\documentclass[aps,superscriptaddress]{revtex4}
\usepackage[T1]{fontenc}
\usepackage[utf8]{inputenc}
\usepackage{mathtools}
\usepackage{graphicx}
\usepackage{caption}
\usepackage{subcaption}
\usepackage{xcolor}
\usepackage{soul}
\usepackage{amsmath,amssymb}
\usepackage[colorlinks=true, pdfstartview=FitV, linkcolor=blue, citecolor=blue, urlcolor=magenta, breaklinks=true]{hyperref}
\usepackage{orcidlink}

\begin{document}

\title{An Undergraduate Approach to the Quantum Hadrodynamics and the Physics of Neutron Stars Part II: Neutron Stars' Exotic Content}

\author{Luiz L. Lopes}

\affiliation{ Centro Federal de Educac\~{a}o Tecnol\'{o}gica de Minas Gerais Campus VIII; CEP 37.022-560, Varginha - MG - Brazil\\}
% \mbox{$^2$Depto de F\'{\i}sica - CFM - Universidade Federal de Santa Catarina  Florian\'opolis - SC - CP. 476 - CEP 88.040 - 900 - Brazil }}

\date{\today}

\begin{abstract}
In this second part, I discuss how to introduce and the role played by non-atomic degrees of freedom in neutron stars' core using the formalism developed in Part I.
\end{abstract}

\maketitle

\section{Introduction}

As a massive star dies with a supernova explosion, it leaves behind its core, called the supernova remnant.  For a supernova remnant below approximately 2.5 $ M_\odot$, the huge density induces electron capture on protons and prevents neutron $\beta$-decay, causing the so-called neutronization of matter; hence the name, neutron star. As neutrons are unstable, a small fraction of them is expected to decay into protons and electrons to maintain chemical equilibrium. This is the classic neutron star, composed of protons, neutrons, and electrons. Here, I called it an atomic neutron star.

However, due to the Pauli exclusion principle, as the density increases towards the neutron star's core, the chemical potential of nucleons and electrons increases as well.
Therefore, it can become energetically favorable to convert some of these particles into non-atomic degrees of freedom. Due to their very small masses, it is expected that some electrons convert into muons even below the saturation density. Muons are easily handled, as it behaves as a free Fermi Gas.

As the density continues to increase, it is also possible that some new hadronic degrees of freedom appear in nuclear matter. Being only 20$\%$ heavier than the nucleon, and feeling a moderately attractive potential, once the density exceeds approximately 2.5 times the nuclear saturation density, the $\Lambda^0$ becomes energetically favorable, and its onset becomes possible. With the density still increasing, other hyperons, such as the $\Xi^-$ and $\Xi^0$, can also appear.  The presence of the $\Sigma$ triplet is unlikely due to the strong repulsive potential. Some studies point out that the presence of hyperons is not only possible but inevitable~\cite{Baldo1998PRC, Dapo2010,lopesnpa}.

More speculative possibilities are the presence of $\Delta's$ resonances and (anti)kaon condensate. Despite being heavier than the $\Lambda^0$ hyperon, due to its strongly attractive potential and negative electric charge, the onset of the $ \Delta^-$ can happen at densities lower than those found for the $\Lambda^0$~\cite{Kauan2022PRC}. In the same sense, despite being literally a thousand times heavier than electrons, antikaons can replace them because not only do they possess a very attractive potential, but due to the G-Partiy, their chemical potential always decreases with the density~\cite{kaonp}. At densities close to three times the saturation point, antikaons can become energetically favorable over electrons as the neutralizing agent in charge-neutral matter.

In Part I of the present work~\cite{LopesUNIVERSE2025}, I describe how to build a realistic neutron star from the quantum hadrodynamics (QHD) in mean field approximation (MFA). The present Part II begins immediately where Part I ends, focusing on the exotic, non-nucleonic degrees of freedom that can be present at the core of massive neutron stars. Such a discussion is not present in Part I. The inclusion of non-nucleonic degrees of freedom is presented as detailed as possible, especially for the hyperons. The inclusion of the $\Delta$'s is analogous, and some passages are purposely omitted and left to the reader as exercises. The same is true for the antikaons. Essential discussions are presented, while those that employ techniques already discussed are omitted. 

This paper is organized as follows: In Sec.~\ref{Sec2}, I made a quick review of the QHD formalism in MFA. The muons are introduced in Sec .~\ref {sec3} and the $\Lambda^0$ in Sec.~\ref{Sec4}, together with the cutoff potential, which plays an essential role in preventing the nucleon mass from vanishing. In Sec.~\ref{baryonoctet}, all the baryon octets are added, as well as the discussion of the constraint related to their interaction. The $\phi$ meson is introduced in this section, along with the 
The Sakurai's proposal. Indeed, the Sakurai's proposal and the use of the SU(3) flavor symmetry, introduced in Sec.~\ref{secCGC}, play the major role in fixing the coupling constants of the hadrons with the vector mesons. Finally, the $\Delta$'s are added in Sec.~\ref{secDelta} and the kaons in Sec.~\ref{secKaon}.

\newpage

%%%%%%%%%%%%%%%%%%%%%%%%%%%%%%%%%%%%%%%%%%
\section{Formalism}~\label{Sec2}

\subsection{QHD formalism in mean field approximation}

In this section, I present only a quick review of the QHD formalism in MFA, which is the main scope of Part I~\cite{LopesUNIVERSE2025}. Additional discussion can also be found in ref.~\cite{Serot_1992,Glenbook,debora-universe,Miyatsu2013,IUFSU} and the references therein. Consider a strongly interacting matter composed, in principle, of $N$ baryons and $J$ leptons in any arbitrary proportion. The Lagrangian density of such a system reads:

\begin{eqnarray}
\mathcal{L}_{\rm QHD} = \sum_B \bar{\psi}_B\left[\gamma^\mu\left(\mbox{i}\partial_\mu  - g_{BB\omega}\omega_\mu -  g_{BB\phi}\phi_\mu  -  \frac{g_{BB\rho}}{2}\vec{\tau} \cdot \vec{\rho}_\mu\right)- (M_B - g_{BB\sigma}\sigma)\right]\psi_B \nonumber \\      \nonumber   \\
+ \frac{1}{2}(\partial_\mu \sigma \partial^\mu \sigma - m_s^2\sigma^2) - \frac{\kappa}{3} M_N(g_{\sigma} \sigma)^3 - \frac{\lambda}{4}(g_{\sigma}\sigma)^4 - \frac{1}{4}\Omega^{\mu \nu}\Omega_{\mu \nu} + \frac{1}{2} m_v^2 \omega_\mu \omega^\mu\nonumber\\
-  \frac{1}{4}{\bm{P^{\mu\nu}\cdot P_{\mu\nu}}}+ \frac{1}{2} m_\rho^2 \vec{\rho}_\mu \cdot \vec{\rho}^{ \; \mu}   + \Lambda_{\omega\rho}(g_{\rho}^2 \vec{\rho^\mu} \cdot \vec{\rho_\mu}) (g_{\omega}^2 \omega^\mu \omega_\mu) +
\sum_l \bar{\psi}_l[(i\gamma^\mu\partial_\mu - m_l)\psi_l, \label{LQCD}
\end{eqnarray}
where the $\psi_B$  represent the Dirac field of the baryon $B$ and with mass $M_B$, while $\psi_l$  represent the Dirac field of the lepton $l$ and with mass $m_l$. The sum for the baryons runs from 1 to $N$, while for the leptons runs from 1 to $J$.

The $i=\sigma, \omega_\mu,,\vec{\rho}_\mu$ are the mesonic fields with mass $m_i$. The $g_{BBi}$ are the Yukawa coupling constants that simulate the strong interaction between the baryon $B$ and the meson $i$, and $\vec{\tau}$ are the Pauli matrices.
 The antisymmetric mesonic field strength tensors are given by their usual expressions as presented in~\cite{Glenbook} and do not contribute in MFA. The $\Lambda_{\omega\rho}$ term represents a non-linear  $\omega$-$\rho$ coupling between the vector mesons, as the one presented in the IUFSU model~\cite{IUFSU}.

Applying  Euler-Lagrange (E-L) equations within MFA approximation, we can obtain the energy eigenvalue for baryons and leptons, which at $T = 0K$ approximation is also the chemical potential ($\mu$): 

\begin{equation}
  E_B = \mu_B = \sqrt{M_B^{*2} +k_{FB}} + g_{BB\omega}\omega_0 + \frac{1}{2}g_{BB\rho}\tau_{3B}\rho_0 \label{beev} ,
\end{equation}

\begin{equation}
  E_l = \mu_l = \sqrt{m_l^2 + k_{Fl}^2}  , \label{leev}
\end{equation}
where $M^{*}_B~\equiv~M_B - g_{BB\sigma}\sigma_0$ is the effective baryon mass, $k_F$ is the Fermi momentum of the particle, and $\tau_{3B}$ is the isospin projection of the baryon $B$.

With the help of Statistical mechanics for the fermions and taking $\langle\mathcal{H}\rangle = -\langle \mathcal{L} \rangle$ for the mesons, we obtain the total equation of state (EOS)~\cite{LopesUNIVERSE2025}:

 \begin{eqnarray}
  \epsilon = \frac{\gamma}{2\pi^2} \sum_B\int_0^{k_{FB}} [\sqrt{M_N^{*2} +k^2}]  k^2 dk \nonumber  + \frac{1}{2}m_s^2\sigma_0^2 + \frac{1}{2}m_\omega^2\omega_0^2  + \frac{1}{2}m_\rho^2\rho_0^2  \nonumber \\ +  \frac{\kappa M_N (g_{\sigma}\sigma)^3}{3} + \frac{\lambda (g_{\sigma}\sigma)^4}{4} +3\Lambda_v\omega_0^2\rho_0^2\nonumber \\ + \frac{1}{\pi^2}\sum_l\int_0^{k_{Fl}}\sqrt{m_l^{2} +k^2} k^2 dk ,\label{EDfinal}
 \end{eqnarray}

 \begin{eqnarray}
  p = \frac{\gamma}{6\pi^2}\sum_B\int_0^{k_{FB}}\frac{k^4dk}{\sqrt{M_B^{*2} +k^2}}   - \frac{1}{2}m_s^2\sigma_0^2 \nonumber  + \frac{1}{2}m_\omega^2\omega_0^2  
+ \frac{1}{2}m_\rho^2\rho_0^2 \\- \frac{\kappa M_N (g_{\sigma}\sigma)^3}{3} -\frac{\lambda (g_{\sigma}\sigma)^4}{4} + \Lambda_v\omega_0^2\rho_0^2  \nonumber \\+  \frac{1}{3\pi^2}\sum_l\int_0^{k_{Fl}}\frac{k^4dk}{\sqrt{m_l^{2} +k^2}} ,\label{Pfinal}
 \end{eqnarray}
and the total number density is given by:

  \begin{equation}
   n = \sum_B n_B = \sum_B \gamma\frac{k_{FB}^3}{6\pi^2} . \label{eosNLWM}
 \end{equation}
 where $\Lambda_v~\equiv~\Lambda_{\omega\rho}g_{\omega}^2g_{\rho}^2$~, $g_{\sigma}~\equiv~g_{NN\sigma}$, $g_{\omega}~\equiv~g_{NN\omega}$, $g_{\rho}~\equiv~g_{NN\rho}$, and $\gamma$ is the degeneracy factor $\gamma = (2S+1$).

 The expected  values of the mesonic fields are:

\begin{equation}
 m_s^2\sigma_0 = \sum_B {g_{BB\sigma}} n_B^S  - {g_{\sigma}}  \bigg [\kappa M_N(g_{\sigma}\sigma_0)^2 + {\lambda}(g_{\sigma}\sigma_0)^3 \bigg ]\label{snlinear},
\end{equation}

\begin{equation}
 (m_\omega^2 +2\Lambda_v\rho_0^2)\omega_0 =  \sum_B g_{BB\omega}n_B  \label{nlomega} ,
\end{equation}

\begin{equation}
 (m_\rho^2 +2\Lambda_v\omega_0^2)\rho_0 =  \sum_B g_{BB\rho}\frac{\tau_{3B}}{2}n_B,  \label{nlrho} 
\end{equation}
with $n_B^S$ being the scalar density of the baryon $B$:

\begin{equation}
   n_B^S = \frac{\gamma}{2\pi^2}\int_0^{k_{FB}} \frac{M_B^{*}k^2 dk}{\sqrt{M_B^{*2} + k^2}} , \label{scalardensity}
  \end{equation}

Now, any realistic EOS must be able to fulfill six physical quantities at the saturation density: the saturation density itself $(n_0)$, the incompressibility $(K_0)$, the binding energy per nucleon $(B/A)$, the effective nucleon mass $M_N^*/M_N$, the symmetry energy $(S_0)$, and its slope $(L)$. To accomplish this task, I use the Enhanced L3$\omega\rho$ (eL3$\omega\rho$)~\cite{lopesPRD}, an improved version of L3$\omega\rho$ originally presented in ref.~\cite{Lopes2022CTP}. The model parameters, the calculated values of the physical quantities at the saturation density, and the phenomenological constraints (taken from Refs.~\cite{Dutra2014,Micaela2017,Essick2021}) are presented in Tab.~\ref{TL1}.  The nucleon mass is $M_N = 939$ MeV, the electron mass is $m_e =0.51$ MeV, and the meson masses are $m_\omega$ = 783 MeV, $m_\rho$ = 770 MeV, and  $m_\sigma$ = 512 MeV. To be consistent with Part I~\cite{LopesUNIVERSE2025}, I also define:

\begin{equation}
(g_{NN\sigma}/m_\sigma)^2 = G_S, \quad  (g_{NN\omega}/m_\omega)^2 = G_V, \quad (g_{NN\rho}/m_\rho) = G_\rho.
\end{equation}

%\begin{widetext}
\begin{center}
\begin{table}%[ht]
\begin{center}
\caption{Parameters of the $eL3\omega\rho$ model utilized in this work and their prediction for the symmetric nuclear matter properties at the saturation density; the parametrization is taken from ref.~\citet{lopesPRD} and the phenomenological constraints are taken from  Refs.~\cite{Dutra2014,Micaela2017,Essick2021}}.
\label{TL1}
\scalebox{0.90}{
\begin{tabular}{|c|c||c|c|c||c|}
\hline 
  & Parameters & &  Constraints  & This model  \\
 \hline
 $G_S$ & 12.108 fm$^2$ &$n_0$ (fm$^{-3}$) & 0.148 - 0.170 & 0.156 \\
 \hline
  $G_V$ & 7.132  fm$^2$ & $M^{*}/M$ & 0.65 - 0.75 & 0.69  \\
  \hline
  $G_\rho$ & 5.85  fm$^2$ & $K$ (MeV)& 220 - 260                                          &  256  \\
 \hline
$\kappa$ & 0.004138 & $S_0$ (MeV) & 31.2 - 35.0 &  32.1  \\
\hline
$\lambda$ &  -0.00390 & $L$ (MeV) & 38 - 67 & 66\\
\hline 
$\Lambda_{\omega\rho}$ &  0.0283 & $B/A$ (MeV) & 15.8 - 16.5  & 16.2  \\ 
\hline
\end{tabular}}
\end{center}
\end{table}
\end{center}
%\end{widetext}
%%%%%%%%%%%%%%%%%%%%%%%%%%%%%%%%%%%%%%%%%%

\subsection{Hydrostatic Equilibrium and astrophysical constraints}

The static structure of spherically symmetric distributions of self-gravitating matter is determined by the condition of hydrostatic equilibrium consistent with General Relativity. Combining Einstein’s field equations with the energy–momentum tensor of a perfect fluid leads to the equations established by Oppenheimer and Volkoff~\cite{TOV}, which encode the balance between pressure gradients and gravitational attraction in the relativistic regime. It reads:

\begin{equation}
\frac{dp(r)}{dr} = - \frac{GM(r)\epsilon(r)}{r}\bigg [1 + \frac{p(r)}{\epsilon(r)}\bigg]\bigg [1 +\frac{4\pi p(r) r^3}{M(r)} \bigg ]\bigg [1 -\frac{2GM(r)}{r} \bigg ]^{-1}, \nonumber
\end{equation}
\begin{equation}
M(r) = \int 4\pi r^2 \rho(r) dr \label{OV}
\end{equation}

As someone may notice, there are two equations but three variables $(M,~\epsilon,~p)$. The third equation, which allows us to solve the so-called OV equations, is the EOS, which will be the main subject of this work. For a single EOS, we have a whole family of solutions, as each initial condition $p(0) = p_0$ yields a different star.

Analogous to the experimental constraints on nuclear physics at saturation density, there are observational constraints from neutron star observations, which allow us to infer the properties of nuclear matter at supranuclear densities. At present, one of the strongest astrophysical constraints on the EOS comes from the pulsar PSR J0740+6620, whose measured mass and radius are $M = 2.08 \pm 0.07\, M_\odot$ and $R = 12.39^{+1.30}_{-0.98}$ km, respectively~\cite{Miller2021,Riley2021}. A realistic EOS must be able to reproduce these observational constraints.
Another key observable is the radius of a canonical neutron star with mass $M = 1.4\,M_\odot$. Reference~\cite{Miller2021} reports a value of $R_{1.4} = 12.45 \pm 0.65$ km, implying an uncertainty of only about 5\% in the radius of the canonical star. Furthermore, to describe the outer and inner crust regions of neutron stars, we employed the Baym-Pethick-Sutherland (BPS) EOS~\cite{BPS} and the Baym-Bethe-Pethick (BBP) EOS~\cite{BBP}, respectively.\\

\section{Atomic neutron stars and the muon} \label{sec3}

\subsection{Chemical equilibrium}

Let us begin with the simplest possible neutron star configuration, composed only of neutrons, protons, and electrons. As these are the same particles that compose the ordinary atoms, we call them here atomic neutron stars. Newborn neutron stars are essentially composed of neutrons, in a process called neutronization~\cite{pizzochero2010}. As neutrons are unstable, some of them must decay into protons+electrons (+anti-neutrinos, which are weakly interacting and therefore will be fully disregarded in this entire work). Furthermore, we also expect neutron stars to be electrically neutral. So, in the simplest neutron star model, the neutron star must be composite by neutrons, protons, and electrons in beta equilibrium with zero net charge:

\begin{eqnarray}
 \mu_n  = \mu_p + \mu_e, \nonumber \\
 n_p = n_e, \label{betaeq} 
\end{eqnarray}

The coupling constants of the protons and neutrons are always the same, nevertheless, the $\tau_{3B}$ assumes +1 for protons and -1 for neutrons, while electrons are treated as a free Fermi gas. Therefore, their chemical potentials are:

\begin{eqnarray}
   \mu_i  = \left\{ \begin{array}{ll}
         \sqrt{M^{*2}_N + k_{Fn}^2} + g_{NN\omega}\omega_0 +\frac{1}{2}g_{NN\rho} {\rho}_0  & \mbox{for protons};\\
      \sqrt{M^{*2}_N + k_{Fp}^2} + g_{NN\omega}\omega_0 -\frac{1}{2}g_{NN\rho} {\rho}_0  & \mbox{for neutrons} \\
      \sqrt{m_e^2 + k_{Fe}^2} & \mbox{for electrons} .\end{array} \right.  \label{energyrho}
\end{eqnarray}

So, for a given neutron density (or neutron Fermi momentum), using Eq.~\ref{betaeq}, we are able to determine the amount of protons and neutrons and ultimately, we obtain the total EOS. \\

\subsection{Muons}

As the density increases, the chemical potential of all particles increases as well. At some point, it can become energetically favorable to convert some electrons into muons. Imposing chemical equilibrium with respect to muons (disregarding the neutrinos), we obtain:

\begin{equation}
  \mu_n = \mu_p +\mu_\mu.  
\end{equation}

Now, comparing with Eq.~\ref{betaeq}, we have:

\begin{eqnarray}
   \mu_\mu =\mu_e \label{muon=electron}
\end{eqnarray}

Muons can easily be handled in beta-stable matter, as they are also introduced as a free Fermi gas. Indeed, their chemical potentials are identical to those of electrons, but their masses; $m_\mu = 106$ MeV.  From Eq.~\ref{muon=electron}, we can obtain an equation for the Fermi momentum of the muons:

\begin{eqnarray}
  k_{F\mu}^2 = \sqrt{m_e^2 +k_{Fe}^2} -m_\mu^2  .
\end{eqnarray}

Once their densities  are related to their Fermi momentum, we have:

\begin{eqnarray}
  n_\mu  = \left\{ \begin{array}{ll}
         0 & \mbox{if $k_{F\mu}^2~\leq~0$ };\\
      {k_{F\mu}^3}/{(3\pi^2)}  & \mbox{if $k_{F\mu}^2~>~0$ }.\end{array} \right.  \label{muond}
\end{eqnarray}

The presence of muons in neutron stars' interiors is considered in the literature since the late 1950s, from the work of A.G.W. Cameron~\cite{Cameron1959APj}. Nowadays, the presence of muons is almost unanimously accepted, and their absence must be strongly justified, or such a work will be treated as only pedagogical.  Calculations with various models suggest that muons appear at densities slightly below the saturation point  (around 0.11 to 0.12 fm$^{-3}$, depending on the model). Indeed, the standard model of neutron stars consists of protons, neutrons, electrons, and muons~(abbreviated as $npe\mu$ matter) in chemical equilibrium and with zero electric charge net. The chemical equilibrium for $npe\mu$ matter reads:

\begin{eqnarray}
    \mu_n = \mu_p +\mu_e , \nonumber \\
    \mu_\mu =\mu_e, \nonumber \\
    n_p = n_e +n_\mu. \label{npeu}
    \end{eqnarray}

\subsection{Numerical results}

    Before explicitly presenting the numerical results, one may ask what the effects of considering muons in neutron stars' interiors. Being able to make qualitative estimations before resorting to numerical calculations can help us to develop valuable insights about dense matter in neutron stars.

    First, the presence of the muons reduces the chemical potential of electrons, which consequently increases the proton fraction, since we now have  $k_{FP}~>k_{Fe}$. As muons are more massive than electrons, their presence softens the EOS by their own. However, the main softening of the EOS comes from reducing the energy associated with the asymmetric parameter $\alpha$: $(n_n - n_p)/(n_n + n_p)$, thereby reducing the contribution of the $\rho$ field. 

\begin{figure*}[ht]
\begin{tabular}{ccc}
\centering % \begin{center}/\end{center} takes some additional vertical space
\includegraphics[scale=.54, angle=270]{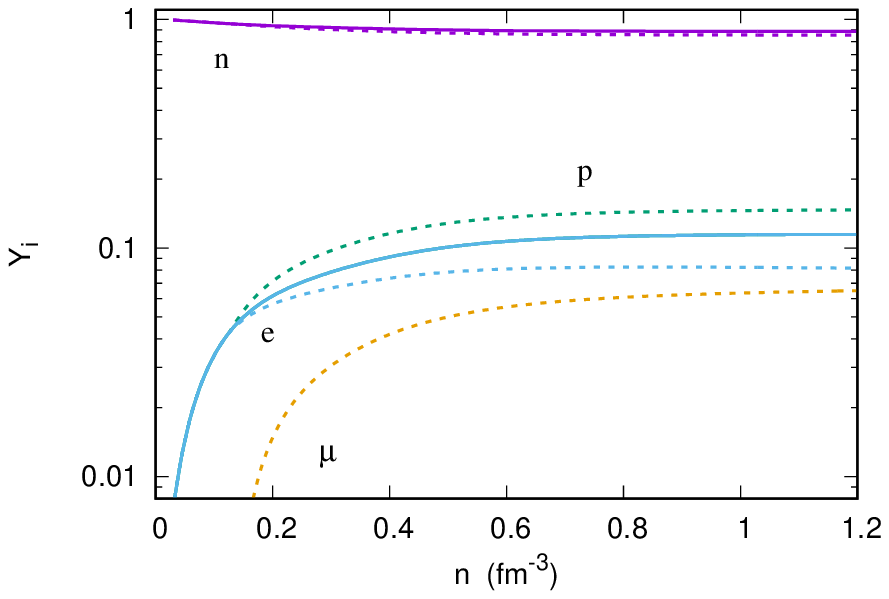} &
\includegraphics[scale=.54, angle=270]{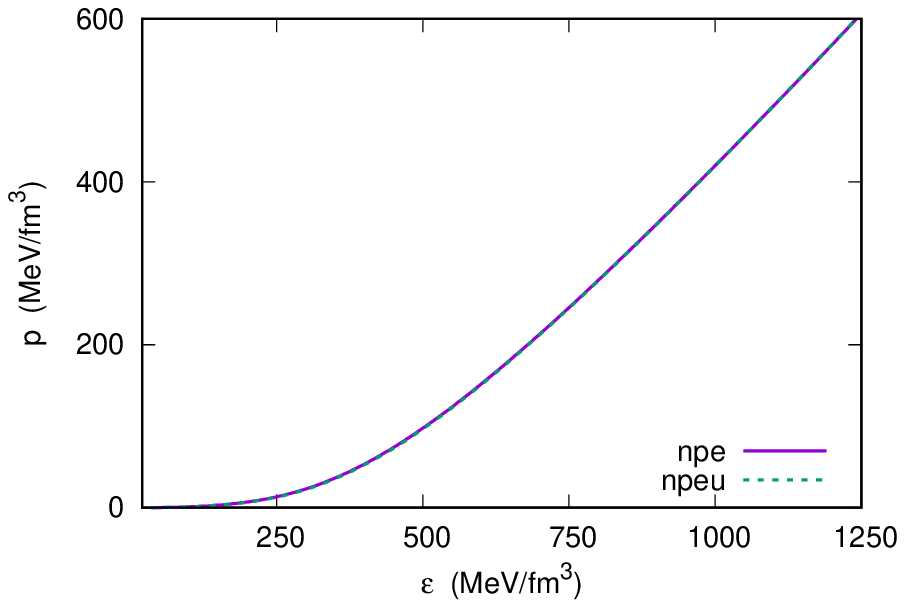} \\
\includegraphics[scale=.54, angle=270]{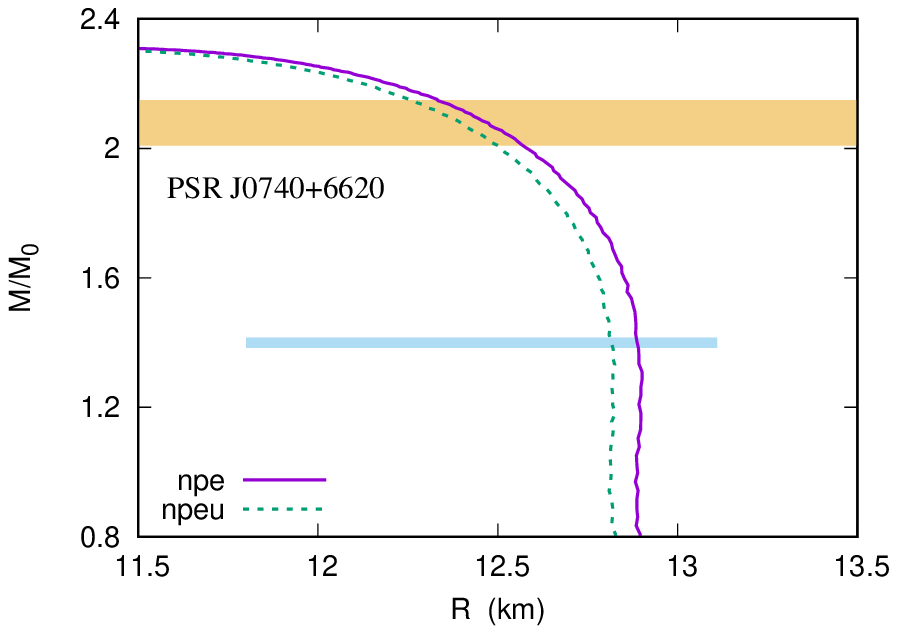} 
\end{tabular}
\caption{ (\textbf{a}) Particle population,  (\textbf{b}) EOSs, and (\textbf{c}) OV solutions. Solid lines represent $npe$ matter, and dashed lines represent $npe\mu$ matter. \label{FL1}}
\end{figure*}

    Therefore, we expected softer EOS when compared with $npe$ matter of atomic stars, with a larger amount of protons. This implies lower radii and a slightly smaller maximum mass. The decrease of the maximum mass must be very small because of the presence of the non-linear term $\Lambda_{\omega\rho}$, which strongly reduces the $\rho$ field at high densities. A detailed discussion about the mesons, their coupling, and their effects can be found in Part I~\cite{LopesUNIVERSE2025} and the references therein. The quantitative results related to the onset of muons are presented in Fig.~\ref{FL1}.

%%%%%%%%%%%%%%%%%

In {\bf (a)}, I present the particle population for $npe$ matter (solid lines) and $npe\mu$ matter (dashed lines). We can see that our qualitative discussion presented above is indeed correct. The proton fraction increases due to the onset of muons reaching over  0.15, while in the case of $npe$ matter, the value stabilize slight above 0.11. However, as presented in {\bf (b)}, the EOSs for $npe$ and $npe\mu$ matter are visually indistinguishable. Finally, in {\bf (c)}, I present the OV solution. The presence of muons causes the radius of the canonical star to drop from 12.90 km to 12.82 km. The reduction of the maximum mass is very subtle, from 2.31 $M_\odot$ to 12.30 $M_\odot$. As can be seen, the eL3$\omega\rho$ model can satisfy both constraints related to neutron stars' observations: the mass and radius of the PSR J0740+6620 pulsar, and the radius of the canonical 1.4$M_\odot$~\cite{Riley2021,Miller2021}.\\

\section{The $\Lambda^0$ hyperon} \label{Sec4}

The lightest baryonic non-nucleonic degree of freedom that can be present on the neutron star is the $\Lambda^0$. With a mass around 1116 MeV~\cite{PDG2020}, the $\Lambda^0$ is less than 20\% heavier than the nucleon. One may wonder whether, as the neutron chemical potential increases, it becomes energetically favorable for some neutrons to convert into $\Lambda^0$ hyperons, analogous to the appearance of muons at high electron chemical potentials.  So, a new generalized chemical equilibrium can be postulated with an additional equation:

\begin{equation}
  \mu_\Lambda = \mu_n.  \label{l=n}
\end{equation}

However, unlike the muons that could be easily handled because they were treated as free Fermi gas, the $\Lambda^0$'s are strongly interacting particles, which in principle, couples to the $\sigma$, $\omega$ and $\rho$ mesons. Now, as the isospin projection of the $\Lambda^0$ is null, we can assume that $g_{\Lambda\Lambda\rho} = 0$. Its chemical potential now reads:

\begin{equation}
  \mu_\Lambda = \sqrt{M_\Lambda^{*2} + k_{F\Lambda}^2} + g_{\Lambda\Lambda\omega} \omega_0. \label{mulambda}
\end{equation}
where $M^*_\Lambda = M_\Lambda - g_{\Lambda\Lambda\sigma}\sigma_0$ is the $\Lambda^0$ effective mass. We can now construct a formulation analogous to the muons to determine the density of $\Lambda^0$ in chemically stable matter:

\begin{eqnarray}
  k_{F\Lambda}^2 = \mu_n^2 -M_\Lambda^{*2} - g_{\Lambda\Lambda\omega}\omega_0 ,
\end{eqnarray}
remembering that the $\mu_n$ explicitly depends on the mesonic  $\sigma$, $\omega$ and $\rho$ fields (Eq.~\ref{energyrho}). We then have:

\begin{eqnarray}
  n_\Lambda  = \left\{ \begin{array}{ll}
         0 & \mbox{if $k_{F\Lambda}^2~\leq~0$ };\\
      {k_{F\Lambda}^3}/{(3\pi^2)}  & \mbox{if $k_{F\Lambda}^2~>~0$ }.\end{array} \right.  \label{mlamb}
\end{eqnarray}

A generalized beta-stable matter containing neutrons, protons, lambdas, electrons, and muons can now be constructed. However, a crucial point remains unresolved. What are the values of $g_{\Lambda\Lambda\sigma}$ and $g_{\Lambda\Lambda\omega}$? For more than 30 years, since the suggestion that hyperons could be present at the core of neutron stars~\cite{Ambar1960} until precise measurements of hyperon potential depths~\cite{Potentials2000}, the values of the hyperon-meson coupling constants were very speculative and almost treated as a full free parameter.

However, in the literature, it is more common to report the value normalized to the nucleon coupling. I therefore define the quantity $\chi_{BBM}$, which represents the coupling strength between the baryon $B$ and the meson $M$, normalized to the corresponding nucleon coupling:

\begin{equation}
\chi_{BBM} = \frac{g_{BBM}}{g_{NNM}}   . \label{chi1}
\end{equation}

Explicitly, for the $\Lambda^0$ we have:

\begin{equation}
 \chi_{\Lambda\Lambda\sigma} =  \frac{g_{\Lambda\Lambda\sigma}}{g_{NN\sigma}} \quad \mbox{and} \quad \chi_{\Lambda\Lambda\omega} =  \frac{g_{\Lambda\Lambda\omega}}{g_{NN\omega}}.  \label{chiL}
\end{equation}

\subsection{Problems with high values of coupling constants and the lack of true maximum masses.}

Until the late 1970s, the theoretical and experimental understanding of hypernuclear matter was extremely rudimentary. A common approach to hyperons was simply consider $\chi_{\Lambda\Lambda\sigma} = \chi_{\Lambda\Lambda\omega}$. Following this approach, several values for the $\chi's$ can be found in the literature. Let us consider here a particular (still arbitrary) case: $\chi_{\Lambda\Lambda\sigma} = \chi_{\Lambda\Lambda\omega} = 0.75$.

\begin{figure*}[ht]
\begin{tabular}{ccc}
\centering % \begin{center}/\end{center} takes some additional vertical space
\includegraphics[scale=.54, angle=270]{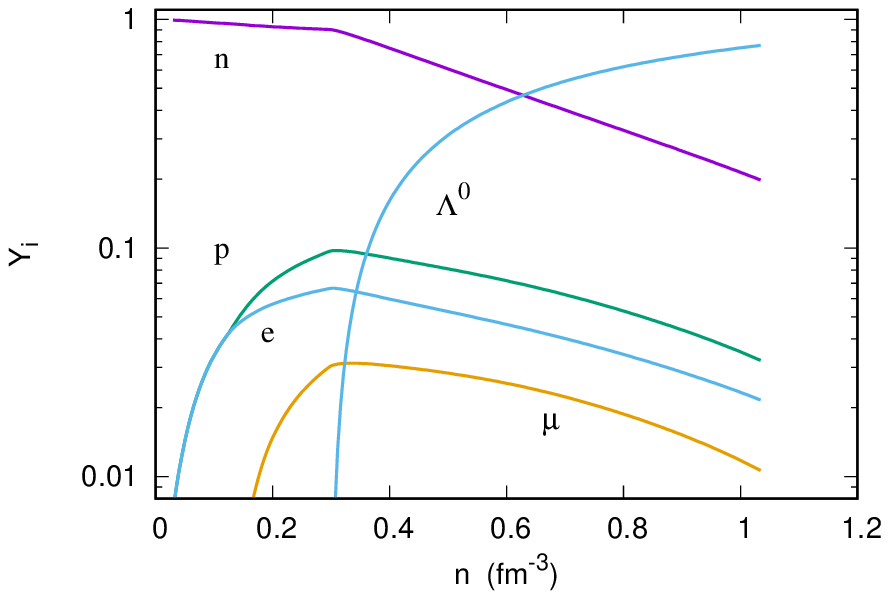} &
\includegraphics[scale=.54, angle=270]{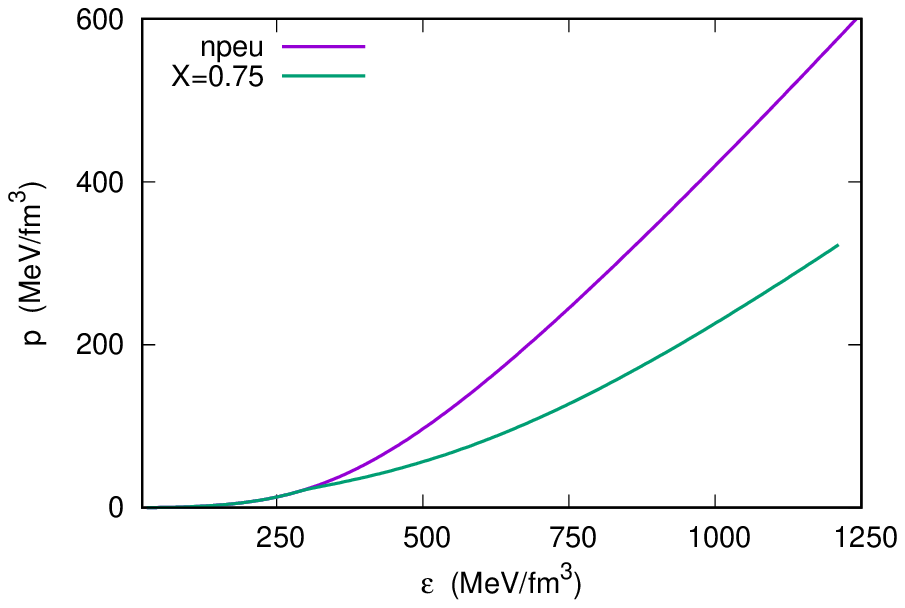} \\
\includegraphics[scale=.54, angle=270]{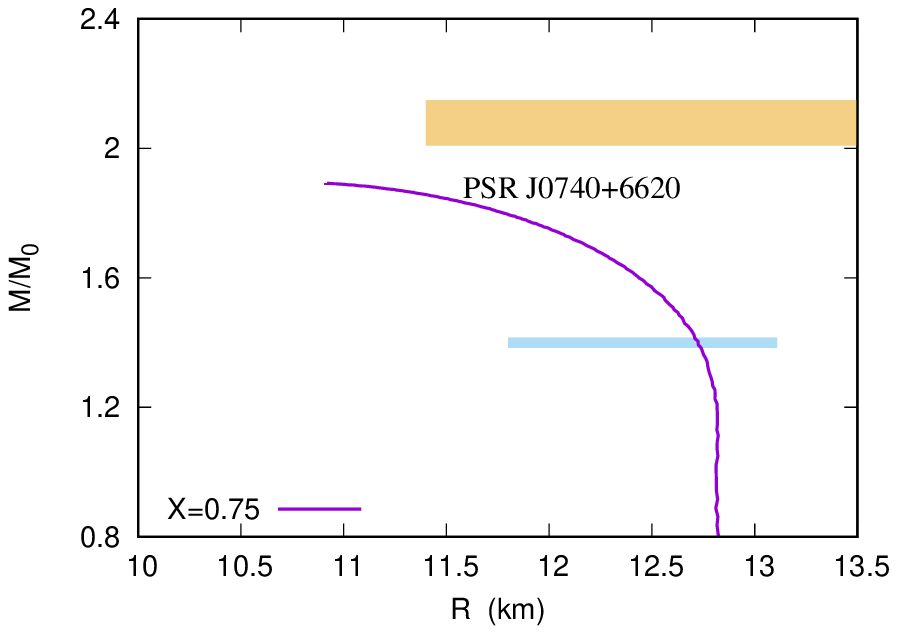} &
\includegraphics[scale=.54, angle=270]{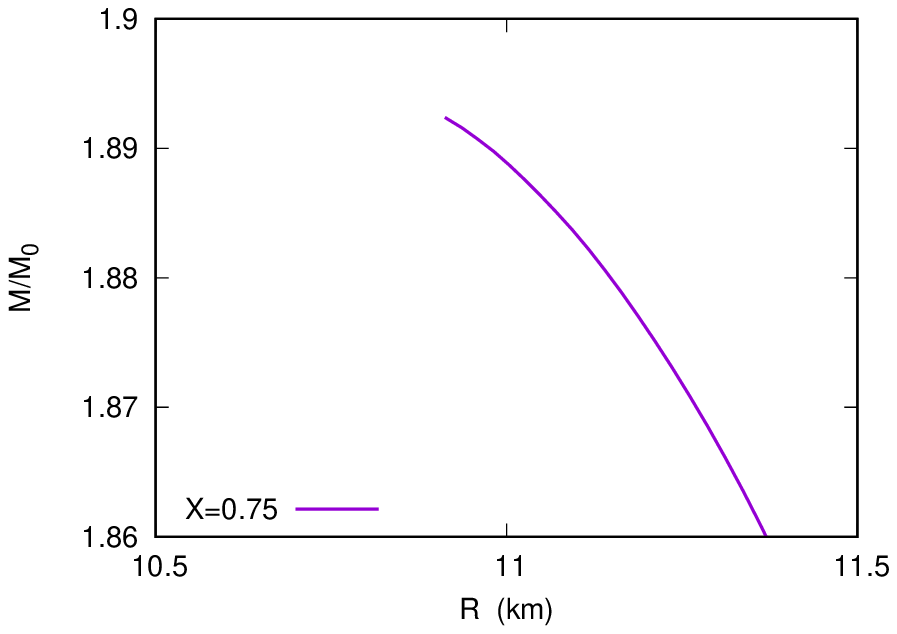}\\
\end{tabular}
\caption{ (\textbf{a}) Particle population.  (\textbf{b}) EOSs for $npe\mu$ and $npe\mu\Lambda$ matter. (\textbf{c}) General OV solution (\textbf{d}) OV solutions zoomed-in close to the maximum mass. The numerical code stops before the true maximum mass is reached.\label{FL2}}
\end{figure*}

In Fig.~\ref{FL2}, I display the particle population, the EOS, and the OV solution for $npe\mu\Lambda$ matter with $\chi_{\Lambda\Lambda\sigma} = \chi_{\Lambda\Lambda\omega} = 0.75$.  At first look, it seems that such parametrization is simply not good enough to reproduce the constraint related to the  PSR J0740+6620 pulsar, as its maximum mass is only 1.89$M_\odot$.

Nevertheless, let us take a closer look. In Fig.~\ref{FL2} {\bf (a)}, I show the particle population. We can see how the $\Lambda^0$ threshold causes a drop in the proton population (followed by a drop in the $e$ and $\mu$ to keep the charge neutrality), and an even larger drop in the neutron population. But the key point here is not the particle population itself. A peculiar feature is the absence of numerical solutions for densities exceeding approximately 1 fm$^{-3}$. Similarly, Fig.~\ref{FL2} {\bf (b)} shows the EOS for $npe\mu\Lambda$ matter alongside the $npe\mu$ case. As expected, the appearance of $\Lambda$ hyperons considerably softens the EOS, which ceases to exist at an energy density of about 1200 MeV/fm$^{3}$.
Analyzing the OV solutions in Fig.~\ref{FL2} {\bf (c))}, it looks like that the maximum mass was reached at 1.89$M_\odot$, however, with a closer look at Fig.~\ref{FL2} {\bf (d)} shows that the true maximum mass was never reached, i.e, the condition $dM/d\epsilon_c~<0$ was not obtained.

%%%%%%%%%%%%%%%%%

One might expect that the numerical calculations could simply be extended to densities above 1 fm$^{-3}$. However, the difficulty is not numerical but physical: no self-consistent solution exists beyond this point. The calculations terminate because the nucleon effective mass vanishes, causing the $\sigma$ field to become imaginary. Although the inclusion of the $\Lambda^0$ hyperon leads to the breakdown of the solution at densities around 1 fm$^{-3}$, incorporating the entire baryon octet with the same coupling constants causes the calculations to terminate at even lower densities, below 0.8 fm$^{-3}$.\\

\subsection{The Cutoff Scheme}

To avoid the nucleon mass vanishing, it is necessary to restrict the $\sigma$ field from taking an arbitrarily large value. It is possible to accomplish such a task by introducing a cutoff potential, as done originally by Maslov and his collaborators in ref~\cite{Maslov2015PRC}.  Consider the following term to be added to the Lagrangian density of Eq~\ref{LQCD}.

\begin{equation}
\mathcal{L}_{cut} =  -U_{cut}(\sigma), \quad \mbox{with} \quad U_{cut}(\sigma) =  \alpha\ln[1 +\exp(\beta(f -f_c))] \label{scut}
\end{equation}
with $\alpha = m_\pi^4~(m_\pi = 130$ MeV), $\beta =120$, as suggested in refs.~\cite{Maslov2015PRC,Thakur2024PRC}, $f =g_\sigma\sigma/M_N$, and $f_c$ is a free parameter that controls where the effect of the cutoff will be relevant.
We can now discuss how it modifies the equation for the $\sigma$ field in MFA by imposing that the energy density is stationary (see Part I~\cite{LopesUNIVERSE2025} for a detailed discussion), the Eq.~\ref{snlinear} must be modified and now reads:

\begin{equation}
 m_s^2\sigma_0 + U_{cut}'(\sigma_0)= \sum_B {g_{BB\sigma}} n_B^S  - {g_{\sigma}}  \bigg [\kappa M_N(g_{\sigma}\sigma_0)^2 + {\lambda}(g_{\sigma}\sigma_0)^3 \bigg ]\label{scut2},
\end{equation}
with

\begin{eqnarray}
 U'_{cut}(\sigma_0) = \frac{\alpha\beta}{M_N}\frac{1}{1 +\exp(-\beta(f - f_c))}   .
\end{eqnarray}

Finally, the $U_{cut}(\sigma_0)$ itself must be taken into account in the total energy density and pressure from its Hamiltonian expected value: $p = -\epsilon = \langle\mathcal{H}\rangle = -U_{cut}(\sigma_0)$ and added to Eq~\ref{EDfinal} and Eq.~\ref{Pfinal}.  The extended EOS now reads:

 \begin{eqnarray}
  \epsilon = \frac{\gamma}{2\pi^2} \sum_B\int_0^{k_{FB}} [\sqrt{M_N^{*2} +k^2}]  k^2 dk \nonumber  + \frac{1}{2}m_s^2\sigma_0^2 + \frac{1}{2}m_\omega^2\omega_0^2  + \frac{1}{2}m_\rho^2\rho_0^2  \nonumber \\ +  \frac{\kappa M_N (g_{\sigma}\sigma)^3}{3} + \frac{\lambda (g_{\sigma}\sigma)^4}{4} +3\Lambda_v\omega_0^2\rho_0^2 + U_{cut}(\sigma_0)\nonumber \\ + \frac{1}{\pi^2}\sum_l\int_0^{k_{Fl}}\sqrt{m_l^{2} +k^2} k^2 dk ,\label{EDfinal2}
 \end{eqnarray}

 \begin{eqnarray}
  p = \frac{\gamma}{6\pi^2}\sum_B\int_0^{k_{FB}}\frac{k^4dk}{\sqrt{M_B^{*2} +k^2}}   - \frac{1}{2}m_s^2\sigma_0^2 \nonumber  + \frac{1}{2}m_\omega^2\omega_0^2  
+ \frac{1}{2}m_\rho^2\rho_0^2 \\- \frac{\kappa M_N (g_{\sigma}\sigma)^3}{3} -\frac{\lambda (g_{\sigma}\sigma)^4}{4} + \Lambda_v\omega_0^2\rho_0^2  - U_{cut}(\sigma_0)  \nonumber \\+  \frac{1}{3\pi^2}\sum_l\int_0^{k_{Fl}}\frac{k^4dk}{\sqrt{m_l^{2} +k^2}} ,\label{Pfinal2}
 \end{eqnarray}

Due to the large value of $\beta$, the effects of $U_{cut}(\sigma_0)$ are numerically null until $f~\approx~f_c$. As $f~>~f_c$, $U_{cut}(\sigma_0)$ quickly diverges, and the $\sigma$ field becomes almost constant. The reader is invited to investigate the behavior of $U_{cut}(\sigma_0)$ and the $\sigma$ field itself as a function of the density.

%%%%%%%%%%%%%%%%%
\begin{figure*}[ht]
\begin{tabular}{ccc}
\centering % \begin{center}/\end{center} takes some additional vertical space
\includegraphics[scale=.58, angle=270]{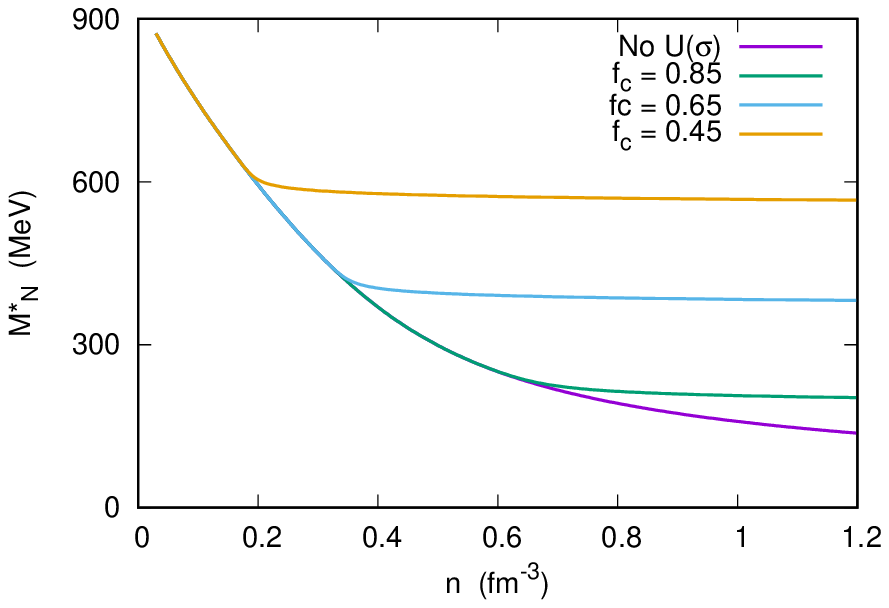} &
\includegraphics[scale=.58, angle=270]{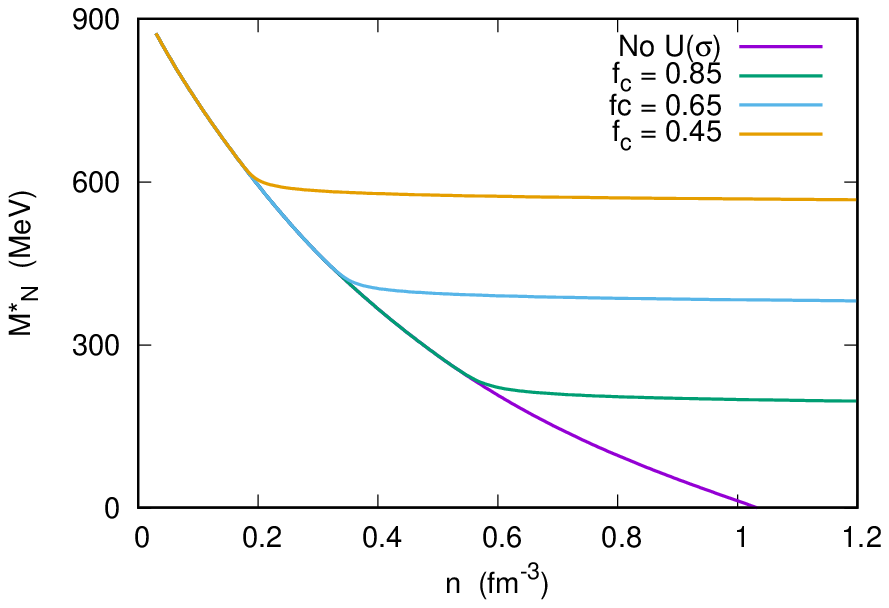} \\
\end{tabular}
\caption{ Effective nucleon mass for $npe\mu$ matter (\textbf{a}), and for $npe\mu\Lambda$ matter (\textbf{b}) with $\chi_{\Lambda\Lambda\sigma} = \chi_{\Lambda\Lambda\omega} = 0.75$ for different values of $f_c$ \label{FL3}. The value $f_c = 0.85$ will be solely used from this point onward.}
\end{figure*}

I display in Fig.~\ref{FL3} the nucleon effective mass for different values of $f_c$ in $npe\mu$~(Fig.~\ref{FL3} {\bf (a)}) and $npe\mu\Lambda$~(Fig.~\ref{FL3} {\bf (b)}) matter. We can notice that for $npe\mu$ matter, the effective nucleon mass does not go to zero in the range analyzed. Nevertheless, using $f_c = 0.85$, the cutoff potential only becomes relevant for densities around 0.64 fm$^{-3}$, i.e, for densities above 4$n_0$. As we reduce $f_c$, the influence of  $U_{cut}(\sigma_0)$ becomes relevant for lower densities; therefore, the effective nucleon mass stabilizes at higher values. 
In the case of $npe\mu\Lambda$, we notice the vanishing of the nucleon mass around  1 fm$^{-3}$. The cutoff potential prevents such an effect, and again, the nucleon mass stabilizes at different values for different values of $f_c$. This allows us to use high values of $\chi_{BB\sigma}$ even when several additional degrees of freedom are present.

The $\sigma$ cutoff potential was introduced to artificially stiffen the EOS as discussed in ref~\cite{Maslov2015PRC} (I call it an artificial stiffening of the EOS because this potential was added only to fulfill this task. There is no experimental or theoretical reason beyond this desirable stiffening to motivate the introduction of $U_{cut}(\sigma)$). Here, however, I will employ it solely to ensure that the nucleon effective mass remains finite at high densities. As the L3$\omega\rho$ model was tested against several constraints up to 4 times the nuclear saturation density (see ref.~\cite{Lopes2022CTP}), I will use from this point onward only $f_c = 0.85$.

\subsection{The adiabatic index}

One of the biggest issues in modern nuclear astrophysics is the fact that the macroscopic properties of neutron stars, such as masses and radii, are largely insensitive to the internal composition and are determined exclusively by the equation of state (EOS).
Nevertheless, from the microscopic point of view, there is a physical parameter very sensitive to the onset of new degrees of freedom, the adiabatic index $\Gamma$. 
For multicomponent matter, the adiabatic index exhibits jumps at densities coincident with density thresholds of individual components, signaling phase
transitions and/or changes in the matter constitution~\cite{lopesPRD,Chamel2008}. It is defined as:

\begin{equation}
 \Gamma = \frac{(p +\epsilon)}{p} \bigg (\frac{\partial p}{\partial \epsilon} \bigg ). \label{adiabatic}   
\end{equation}

The adiabatic index presents information not only on the EOS ($p$ and $\epsilon$), but also on the speed of sound, $v_s^2 = \partial p/\partial \epsilon$. \\

%%%%%%%%%%%%%%%%%
\begin{figure*}[b!]
\begin{tabular}{ccc}
\centering % \begin{center}/\end{center} takes some additional vertical space
\includegraphics[scale=.54, angle=270]{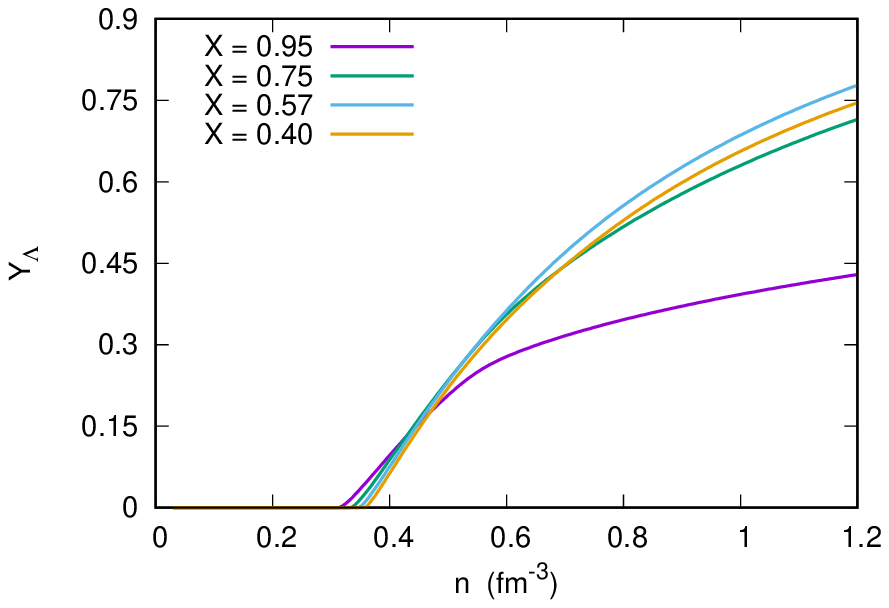} &
\includegraphics[scale=.54, angle=270]{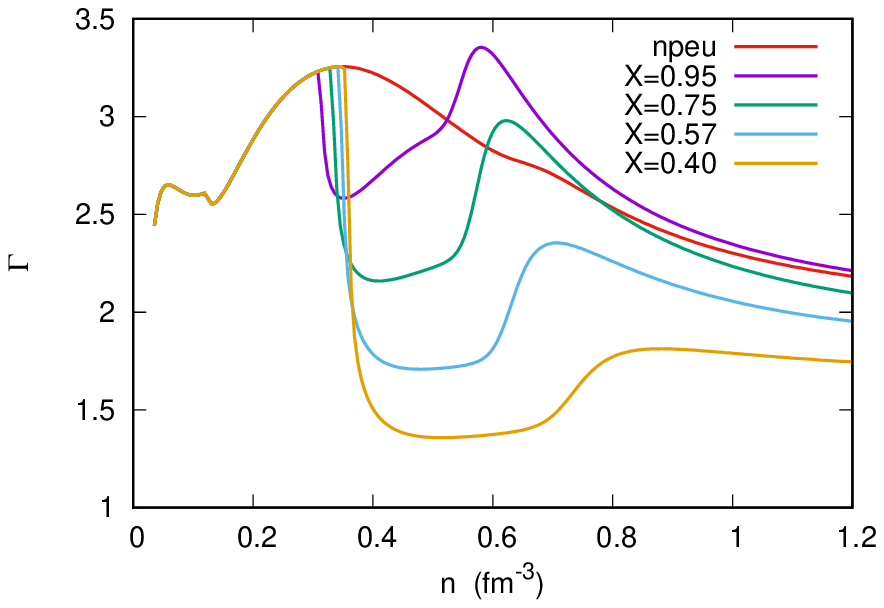} \\
\includegraphics[scale=.54, angle=270]{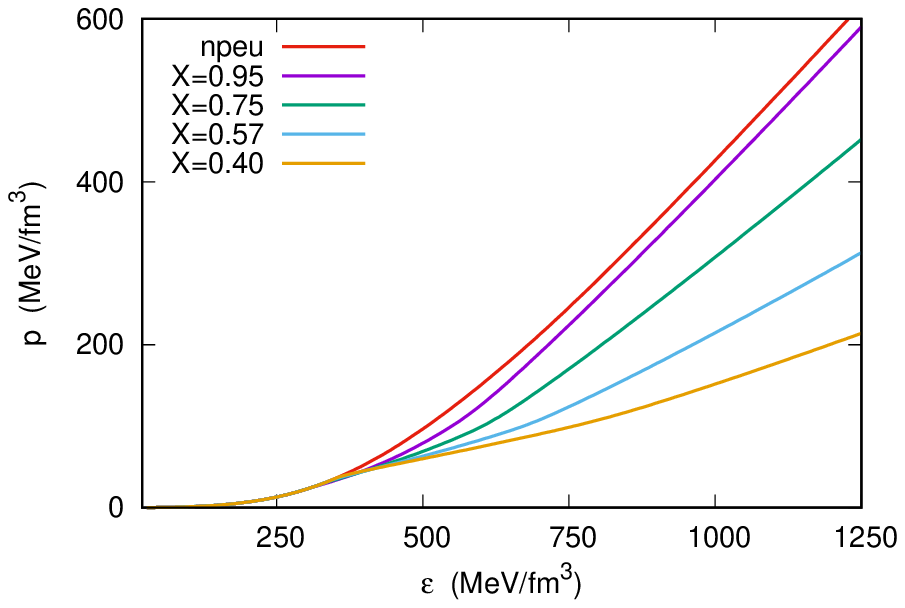} 
\end{tabular}
%} % If the paper is ``preprints'', please uncomment this parenthesis.
\caption{The fraction of $\Lambda^0$'s (\textbf{a}), the adiabatic index (\textbf{b}), the EOSs (\textbf{c}), and the  OV solutions  (\textbf{d})  for different values of $\chi_\Lambda$.~\label{FL4}}
\end{figure*}

\subsection{Numerical Results}

We now analyse the effect of different coupling constants for the $\Lambda^0$ and how it affects some macroscopic and microscopic neutron star properties. In this first analysis limit the study to $\chi_{\Lambda\Lambda\sigma} = \chi_{\Lambda\Lambda\omega} = \chi_\Lambda$. From a historical point of view, values $\chi_\Lambda < 0.5$ were favored by the study of hypernuclei until the late 1980s, in particular the value $\chi_\Lambda = 0.40$~\cite{WALKER1986NPA,Cohen1987PRC}, still values as low as 0.2 can also be found (see a discussion in ref.~\cite{Glenbook}). Here, nevertheless, I use four different values: $\chi_\Lambda = 0.40$, 0.57, 0.75, and 0.95. Afterwards, I compare with $npe\mu$ matter. The relevant results are shown in Fig.~\ref{FL4} for different values of $\chi_\Lambda$.

In Fig.~\ref{FL4} {\bf (a)}, I display the $\Lambda^0$ population. As can be seen, larger values of $\chi_\Lambda$ tend to favor an earlier onset of the $\Lambda^0$ hyperon, while leading to smaller $\Lambda^0$ fractions at higher densities. It should be emphasized, however, that this behavior represents a general tendency rather than a strict rule.
 We can see that the larger $\chi_\Lambda = 0.95$ have the low $\Lambda^0$ threshold, at $n =0.31$ fm$^{-3}$, but reaching values $Y_\Lambda <0.45$ in the high density limit.  On the other hand, $\chi_\Lambda = 0.40$ has the highest density threshold for the $\Lambda^0$, $n = 0.36$ fm$^{-3}$, while at the high densities, we obtain a very high value of $Y_\Lambda$, slightly below 0.75. Yet, surprisingly, the highest $Y_\Lambda$ is obtained not fro $\chi_\Lambda = 0.40$, but for $\chi_\Lambda = 0.57$, reaching $Y_\Lambda~>0.75$.

In Fig.~\ref{FL4} {\bf (b)} it is displayed the adiabatic index $\Gamma$. In the case of $npe\mu$ matter, we can see that $\Gamma$ smoothly grows, reaching a peak around $n = 0.40$ fm$^{-3}$, followed by a smooth decrease.
Such smooth behavior ceases with the onset of new degrees of freedom. The $\Lambda^0$ threshold causes a sharp fall in $\Gamma$ followed by a steep growth until a new peak., leading to a local minimum and a local maximum.  The higher the value of $\chi_\Lambda$, the smaller the drop, but the higher the peak.

In Fig.~\ref{FL4}~{\bf (c)} and {\bf (d)}, the EOSs and the corresponding OV solutions are shown. The relation between these quantities is straightforward: increasing $\chi_{\Lambda}$ results in a stiffer EOS and more massive stars. Such behavior is explained by the $\omega$ dominance, as detailed in Part I~\cite{LopesUNIVERSE2025}: { the repulsive vector-$\omega$ contribution grows with density (is proportional to $n$) and dominates the pressure at high density, so that increasing $g_{BB\omega}$ stiffens the EOS and raises $M_{max}$, overriding the secondary role of the scalar sector.}
Quantitatively, the true maximum mass for $\chi_{\Lambda} = 0.75$ is $1.99,M_{\odot}$, highlighting the importance of the cutoff potential at high densities. Only nucleonic and $\chi_\Lambda = 0.95$ can reproduce the observational constraints.\\

\section{The Baryon Octet and some constraints} \label{baryonoctet}

The proton, neutron, and $\Lambda^0$ belong to a larger family of spin-$1/2$ baryons known as the baryon octet. This multiplet also contains the $\Sigma$ triplet and the $\Xi$ doublet. We, therefore, have a big problem. With six hyperons and three mesons, we have no less than eighteen free parameters to fix. The members of the baryon octet can be rearranged in terms of their hypercharge ($Y$) and isospin projection $(I_3)$, as suggested by Gell-Mann~\cite{GellMann1962PR}. Furthermore, the vector mesons are also grouped in a nonet, composed of an octet plus a singlet state. Their schemes are presented below.

We can now search for experimental and theoretical constraints to reduce the number of free parameters.

\begin{center}

\begin{minipage}{0.45\textwidth}
\centering
\begin{tikzpicture}[scale=2, thick]

% Eixos
\draw[->, gray] (-1.5, 0) -- (1.5, 0) node[right] {$I_3$};
\draw[->, gray] (0, -1.5) -- (0, 1.5) node[above] {$Y$};

% Pontos
\fill (0.5,1) circle (2.2pt);
\fill (-0.5,1) circle (2.2pt);
\fill (1,0) circle (2.2pt);
\fill (-1,0) circle (2.2pt);
\fill (0.5,-1) circle (2.2pt);
\fill (-0.5,-1) circle (2.2pt);
\fill (0,0) circle (2.2pt);

% Conexões
\draw[gray] (0.5,1) -- (-0.5,1);
\draw[gray] (0.5,1) -- (1,0) -- (0.5,-1);
\draw[gray] (-0.5,1) -- (-1,0) -- (-0.5,-1);
\draw[gray] (-0.5,-1) -- (0.5,-1);

% Rótulos
\node at (0.65,1.1) {$p$};
\node at (-0.65,1.1) {$n$};
\node at (1.3,0.2) {$\Sigma^+$};
\node at (-1.3,0.2) {$\Sigma^-$};
\node at (0.65,-1.1) {$\Xi^0$};
\node at (-0.65,-1.1) {$\Xi^-$};

\node at (-0.35,0.25) {$\Sigma^0$};
\node at (0.4,-0.25) {$\Lambda^0$};

\node at (0,1.8) {\textbf{Baryon Octet}};

\end{tikzpicture}
\end{minipage}
\hfill
\begin{minipage}{0.45\textwidth}
\centering
\begin{tikzpicture}[scale=2, thick]

% Eixos
\draw[->, gray] (-1.5, 0) -- (1.5, 0) node[right] {$I_3$};
\draw[->, gray] (0, -1.5) -- (0, 1.5) node[above] {$Y$};

% Pontos
\fill (1,0) circle (2.2pt);
\fill (0,0) circle (2.2pt);
\fill (-1,0) circle (2.2pt);
\fill (0.5,1) circle (2.2pt);
\fill (-0.5,1) circle (2.2pt);
\fill (0.5,-1) circle (2.2pt);
\fill (-0.5,-1) circle (2.2pt);

% ponto extra do singlet
\fill (0,0) circle (2.2pt);

% Conexões
\draw[gray] (1,0) -- (0,0) -- (-1,0);
\draw[gray] (0.5,1) -- (1,0) -- (0.5,-1);
\draw[gray] (-0.5,1) -- (-1,0) -- (-0.5,-1);
\draw[gray] (0.5,1) -- (-0.5,1);
\draw[gray] (0.5,-1) -- (-0.5,-1);

% Rótulos
\node at (1.25,0.15) {$\rho^+$};
\node at (-1.25,0.15) {$\rho^-$};
\node at (-0.20,0.30) {$\rho^0$};

\node at (0.7,1.15) {$K^{*+}$};
\node at (-0.7,1.15) {$K^{*0}$};

\node at (0.7,-1.15) {$\bar K^{*0}$};
\node at (-0.7,-1.15) {$K^{*-}$};

\node at (-0.25,-0.35) {$\omega_8$};
\node at (0.35,-0.35) {$\omega_1$};

\node at (0,1.8) {\textbf{Vector Meson Nonet}};

\end{tikzpicture}
\end{minipage}

\end{center}

\subsection{The potential depth and the quark-isospin counting rule}

The hyperon potential depth is the single-particle potential experienced by a hyperon in symmetric nuclear matter at saturation density. Negative values correspond to an attractive interaction and generally favor the appearance of the hyperon in dense matter, whereas positive values correspond to a repulsive interaction and tend to disfavor its onset. The hyperon potential depth is defined as:

\begin{eqnarray}
  U_Y(n_0) = g_{YY\omega}\omega_0 - g_{YY\sigma}\sigma_0.  
\end{eqnarray}

In the earlier 1990s, the $\Lambda^0$ potential depth was measured with a high degree of precision, assuming $U_\Lambda = -28$ MeV. In the following years, the potential of the $\Sigma$ triplet and the $\Xi$ were also measured, but with a lower degree of precision, lying in the range~ $+15$ MeV $<~U_\Sigma~<$ +40 MeV, and  $-20$ MeV $<~U_\Xi~<$ 0 MeV~\cite{Potentials2000,WEISSENBORN2012NPA}. In the late 2010s, these results were confirmed by studies coming from lattice QCD (LQCD)~\cite{LQCD}.
Here, I present the so-called canonical values of the potential depths:

\begin{equation}
 U_\Lambda = -28~\mbox{MeV}, \quad U_\Sigma = +30~\mbox{MeV}, \quad U_\Xi = -4~\mbox{MeV.}
\end{equation}

However, the potential depth alone cannot fix the coupling constants, as different sets can produce the same values.

Let us introduce here a very naive constraint, based on the nature of the $\omega$ and $\rho$ mesons, the quark-isospin counting rule. 
The $\omega$ meson is a vector-isoscalar meson (a vector meson in the spin space and a scalar in the isospin space). As it does not contain strange quarks, it is assumed that the strength of the coupling of the $\omega$ mesons with a baryon $B$ is proportional to the numbers of $u$ and $d$ quarks. The nucleons contain no strange quarks, the $\Lambda^0$ and the $\Sigma$ triplet contain one, and the $\Xi$ doublet contains two strange quarks. Therefore, the ratio of their coupling constants is: 

\begin{equation}
\frac{g_{\Lambda\Lambda\omega}}{g_{NN\omega}}
=
\frac{g_{\Sigma\Sigma\omega}}{g_{NN\omega}}
=
\frac{2}{3},
\quad
\frac{g_{\Xi\Xi\omega}}{g_{NN\omega}}
=
\frac{1}{3}.
\end{equation}

On the other hand, the $\rho$ meson is a vector-isovector meson (a vector meson in both, the spin and isospin spaces). Therefore, it is assumed to couple to the isospin:

\begin{equation}
\frac{g_{\Lambda\Lambda\rho}}{g_{NN\omega}}
=
0,
\quad 
\frac{g_{\Sigma\Sigma\rho}}{g_{NN\rho}}
=
2
\quad
\frac{g_{\Xi\Xi\rho}}{g_{NN\rho}}
=
1.
\end{equation}

Furthermore, the eigenvalue of $\tau_{3B}$ reads +1 for $p,~\Sigma^+,$ and $\Xi^0$; $-1$ for $n,~\Sigma^-,$ and $\Xi^-$; and zero for the $\Sigma^0$. The quark-isospin counting rule, combined with the potential depths, fully determines all coupling constants.

Now, we introduce hyperons by generalizing chemical equilibrium while maintaining charge neutrality. We can write:

\begin{align}
&\mu_{\Sigma^+} = \mu_p = \mu_n - \mu_e,&  \quad \mu_{\Lambda} = \mu_{\Xi^0} = \mu_{\Sigma^0} = \mu_n, \nonumber \\
&\mu_{\Sigma^-} = \mu_{\Xi^-} = \mu_n + \mu_e,& \quad \mu_\mu = \mu_e, \nonumber \\
&n_{p} + n_{\Sigma^+} = n_e +n_\mu + n_{\Sigma^-} + n_{\Xi^-}. \label{genBE}
\end{align}

%%%%%%%%%%%%%%%%%
\begin{figure*}[ht]
\begin{tabular}{ccc}
\centering % \begin{center}/\end{center} takes some additional vertical space
\includegraphics[scale=.54, angle=270]{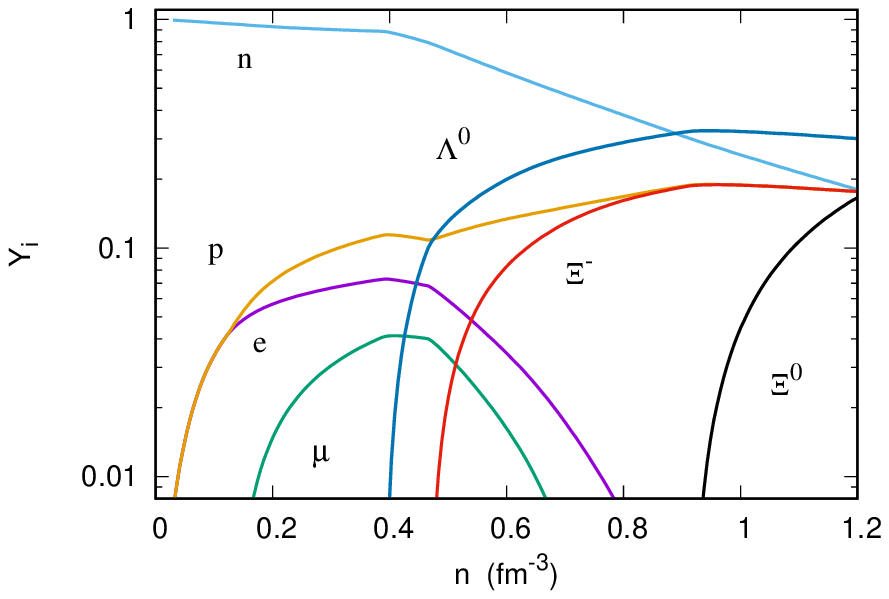} &
\includegraphics[scale=.54, angle=270]{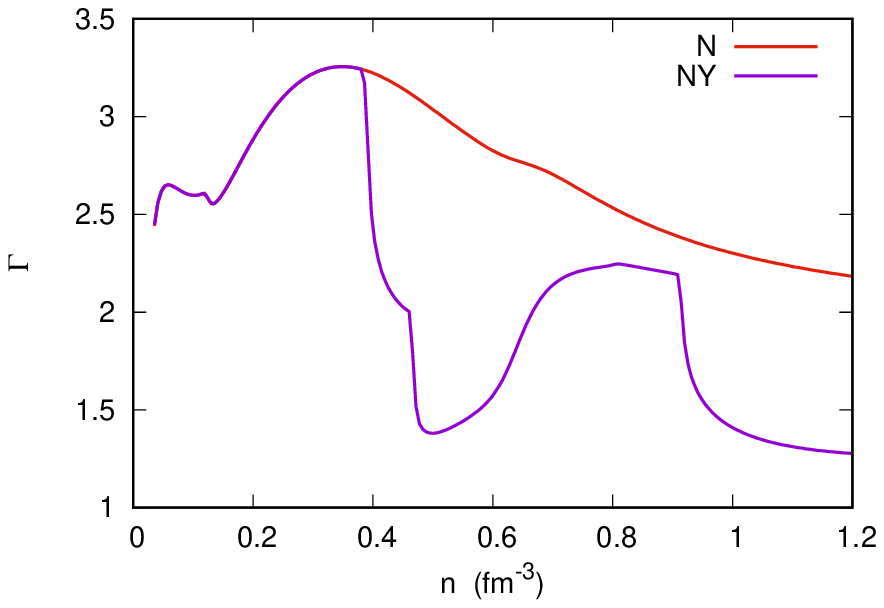} \\
\includegraphics[scale=.54, angle=270]{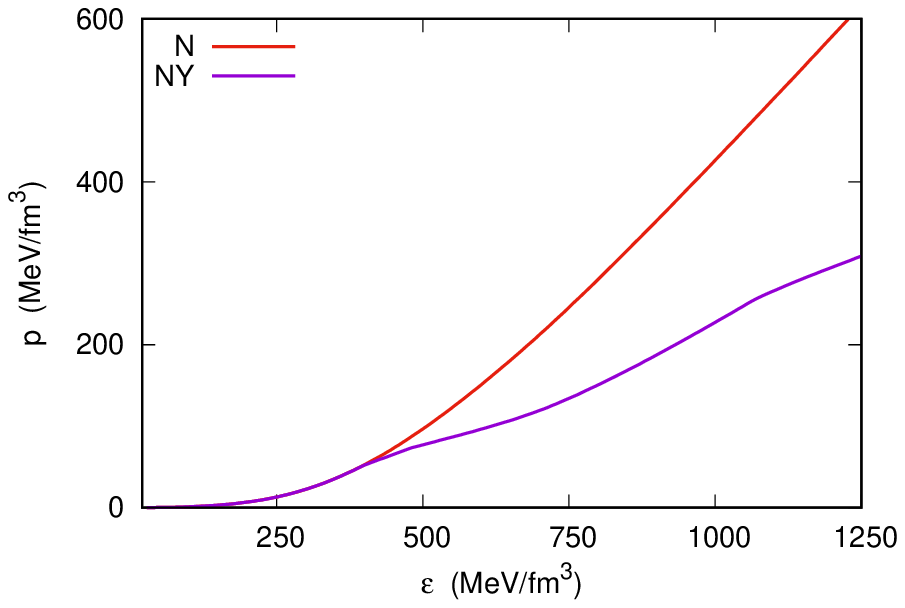} &
\includegraphics[scale=.54, angle=270]{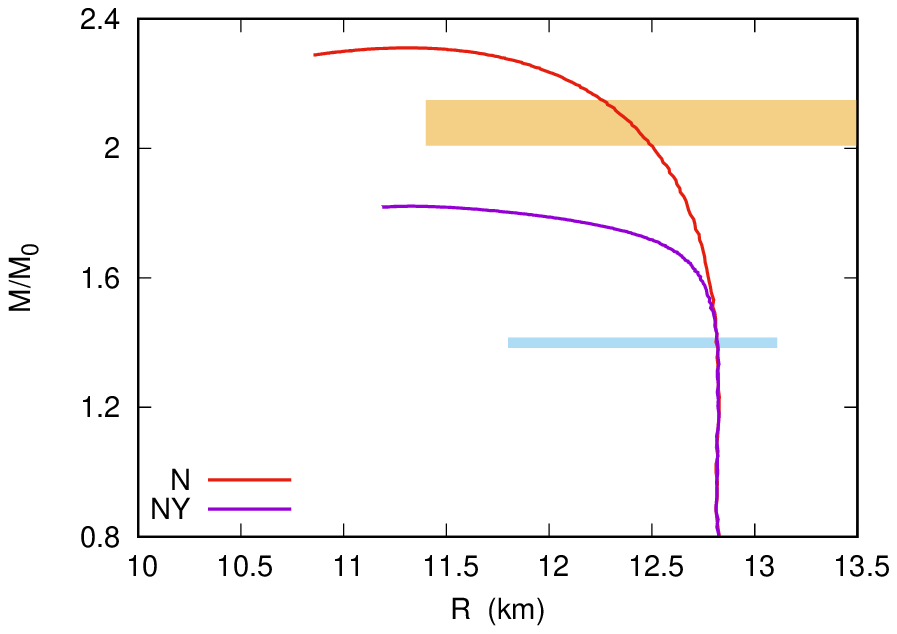}\\
\end{tabular}
\caption{|Particle population (\textbf{a}), adiabatic index (\textbf{b}), EOSs (\textbf{c}), and OV solution (\textbf{d}) for the whole baryon octet within the quark-isospin counting rule. The mass of the PSR J0740+6620 cannot be reproduced with hyperons in the neutron stars' core. \label{FL5}}
\end{figure*}

From this point on, $npe\mu$ matter will be referred simply as nucleonic matter ($N$), while matter that contains hyperons will be referred to as hyperonic matter ($NY$). The results for nucleonic and hyperonic neutron stars within the quark-isospin counting rule are presented in Fig.~\ref{FL5}. 

Form Fig.~\ref{FL5} (\textbf{a}), we can notice that besides the $\Lambda^0$, the $\Xi$ doubled is also present. However, the $\Sigma$ triplet is suppressed due to the strong repulsive potential, $U_\Sigma = + 30$ MeV. One can also observe that, at sufficiently high densities, charge neutrality is maintained entirely by baryons, despite the electron mass being more than 2600 times smaller than that of the $\Xi^-$. The reader is invited to ponder the physical origin of this seemingly surprising behavior.

Figure~\ref{FL5} (\textbf{b}) displays the adiabatic index for both nucleonic and hyperonic matter. The onset of each additional degree of freedom is clearly reflected by a reduction in $\Gamma$. In particular, the appearance of the $\Lambda^0$ produces a pronounced drop, followed by a smaller decrease associated with the onset of the $\Xi^-$. After the second peak in $\Gamma$, a further reduction occurs when the $\Xi^0$ becomes populated. Figures~\ref{FL5} (\textbf{c}) and \textbf{(d)} show the corresponding EOSs and OV solutions, respectively. As can be seen, the inclusion of hyperons dramatically softens the EOS, reducing the maximum neutron-star mass to only $1.82\,M_\odot$. This result is particularly alarming, as it places the maximum mass well below the firmly established two-solar-mass constraint. The apparent irreconcilability between massive neutron stars and { the apparent} inevitability of the presence of hyperons in their cores is known as the hyperon puzzle.

\subsection{Sakurai's proposal and the $\phi$ meson}

As pointed in Part I~\cite{LopesUNIVERSE2025}, until the mid-2000s, an EOS that predicts a maximum mass of 1.65 $M_\odot$ could be considered realistic. However, since the early 2010s, this is not true anymore, and a maximum mass above 2.0 $M_\odot$ nowadays is imperative~\cite{Demorest2010,Antoniadis}. 

Even before the proposal of the quark model, J. J. Sakurai investigated how meson–baryon coupling constants could be constrained using minimal symmetry arguments~\cite{SAKURAI1960AP}. Assuming exact internal symmetries, Sakurai proposed that vector mesons could play the role of gauge bosons associated with hadronic currents. In particular, he argued that different neutral vector mesons couple to distinct conserved currents: one to the isospin current (isospin conservation), another to the hypercharge (hypercharge conservation) current, and an additional singlet current, with a universal coupling (baryon number conservation).

The $\rho^0$ meson is the one that couples to the isospin current, while the $\omega_8$ couples to the hypercharge, and the $\omega_1$ is the singlet with a universal coupling. We must ask now if the physical $\omega$ meson is the singlet or the octet state.

At first glance, something seems inconsistent. The $\omega$ meson cannot correspond to the $\omega_8$, since a pure octet state couples to the hypercharge current. This would imply $\chi_{\Lambda\Lambda\omega} = 0$ and $\chi_{\Xi\Xi\omega} = -1$, instead of the values $2/3$ and $1/3$ predicted by the quark--isospin counting rule. At the same time, the physical $\omega$ is not the $\omega_1$ either, because this would result in $\chi_{BB\omega} =1$ for all baryons.  Could be Sakurai's theory is wrong?

The answer is no. Sakurai's theory is not wrong. The physical $\omega$ meson is a linear combination of the $\omega_8$ and $\omega_1$ states. Because $\omega_8$ and $\omega_1$ are linearly independent, the mixing scheme necessarily generates a second orthogonal physical state with the same quantum numbers, $Y = I_3 = 0$, which is identified with the $\phi$ meson. We therefore have~\cite{DOVER1984}:

\begin{align}
|\omega\rangle = \cos\theta_V|\omega_1\rangle + &\sin\theta_V |\omega_8\rangle   , \nonumber \\
|\phi\rangle =  -\sin\theta_V|\omega_1\rangle + &\cos\theta_V|\omega_8\rangle, \label{mix}    
\end{align}
where $\theta_V$ is the mixing angle between the singlet and octet states. Sakurai himself played a central role in the development of the $\omega-\phi$ mixing scheme, as discussed in Ref.~\cite{Sakurai1963}. The value of $\theta_V$ can be estimated from the Gell-Mann--Okubo mass formula~\cite{GellMann1962PR,GreinerSymmetries,PDG2020}. In the ideal mixing limit, the quark content of the $\omega$ and $\phi$ mesons is given by

\begin{align}
\omega = \frac{1}{\sqrt{2}}(u\bar{u} + d\bar{d}), \quad \mbox{and} \quad \phi = (s\bar{s)}.  
\end{align}
resulting $\theta_V = 35.264^o$. The mass of the physical $\phi$ meson is 1020 MeV.

Following Sakurai's proposal, the $g_{BB\omega}$ for a baryon $B$ can now be write as:
\begin{equation}
g_{BB\omega} = \frac{\sqrt{3}}{3}(g_1\sqrt{2} + g_8Y),
\end{equation}
while at the same time, by just exchange $\cos\theta_V \to -\sin\theta_V$ and $\sin\theta_V \to \cos\theta_V$, we also obtain the coupling with the $\phi$ meson:

\begin{equation}
g_{BB\phi} = \frac{\sqrt{3}}{3}(-g_1 + g_8\sqrt{2}Y),
\end{equation}
without introducing any additional free parameter. The only free parameter to determine the coupling constants of the $\omega$ and $\phi$ mesons with the whole baryon octet is the ratio $g_1/g_8$. 

Assuming $g_1 = \sqrt{2}\, g_8$, Sakurai's theory becomes fully consistent with the quark-isospin counting rule. Moreover, it predicts $g_{NN\phi} = 0$, which is a desirable result since the nucleon contains no strange quarks. The coupling constants of the hyperons with the $\phi$ mesons in relation to the nucleon-$\omega$ coupling are:

\begin{align}
 \frac{g_{\Lambda\Lambda\phi}}{g_{NN\omega}} =  \frac{g_{\Sigma\Sigma\phi}}{g_{NN\omega}}  = - \frac{\sqrt{2}}{3}, \quad     \frac{g_{\Xi\Xi\phi}}{g_{NN\omega}} = - \frac{2\sqrt{2}}{3} .
\end{align}

\subsection{The $\phi$ mesons' Lagrangian density}

I now extend the $\sigma\omega\rho$ model, allowing that baryons can also interact via the exchange of $\phi$ mesons. Therefore, new terms must be added to the QHD Lagrangian (Eq.~\ref{LQCD})~\cite{WEISSENBORN2012NPA,lopesEPJA2020,Lopes2022ApJ}:

\begin{equation}
 \mathcal{L}_\phi = g_{BB\phi}\bar{\psi}_B(\gamma^\mu\phi_\mu)\psi_B  +\frac{1}{2}m_\phi^2\phi_\mu\phi^\mu - \frac{1}{4}\Phi^{\mu\nu}\Phi_{\mu\nu} .
\end{equation}

Using the formalism developed in Part I~\cite{LopesUNIVERSE2025}, in MFA, the baryon energy eigenvalue and chemical potential now read:

\begin{equation}
  E_B = \mu_B = \sqrt{M^{*2}_B + k_{FB}^2} + g_{BB\omega}\omega_0 + g_{BB\phi}\phi_0 + g_{BB\rho}\frac{\tau_{3B}}{2}\rho_0, \label{energyEV2}
\end{equation}
with 
\begin{eqnarray}
 m_\phi^2\phi_0 = \sum_B g_{BB\phi}n_B.   
\end{eqnarray}
Additionally, the $\phi$ field itself contributes to the total energy density and must be summed to the energy density and pressure (Eq.~\ref{EDfinal2} and Eq.~\ref{Pfinal2}):

\begin{equation}
 \epsilon_\phi = p_\phi = \frac{1}{2}m_\phi^2\phi_0^2.   
\end{equation}

%%%%%%%%%%%%%%%%%
\begin{figure*}[ht]
\begin{tabular}{ccc}
\centering % \begin{center}/\end{center} takes some additional vertical space
\includegraphics[scale=.54, angle=270]{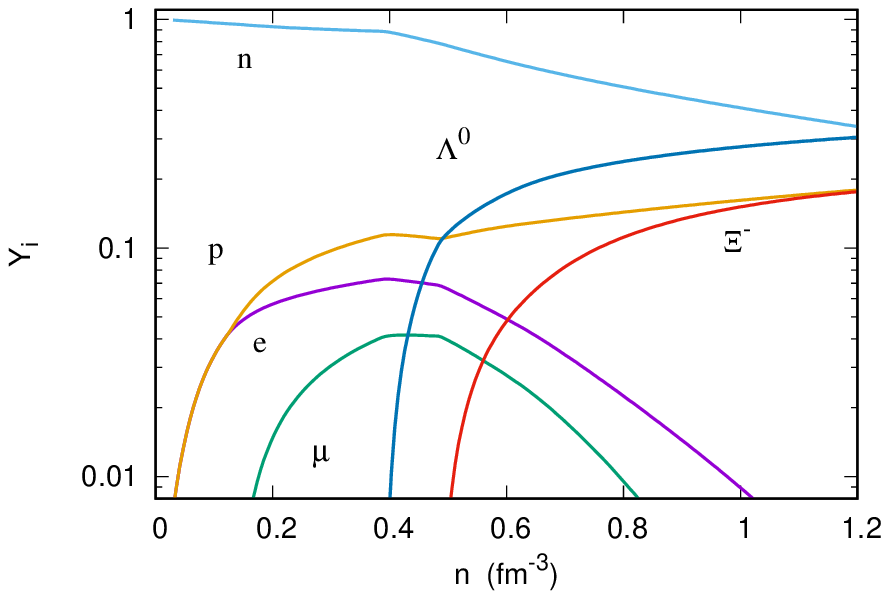} &
\includegraphics[scale=.54, angle=270]{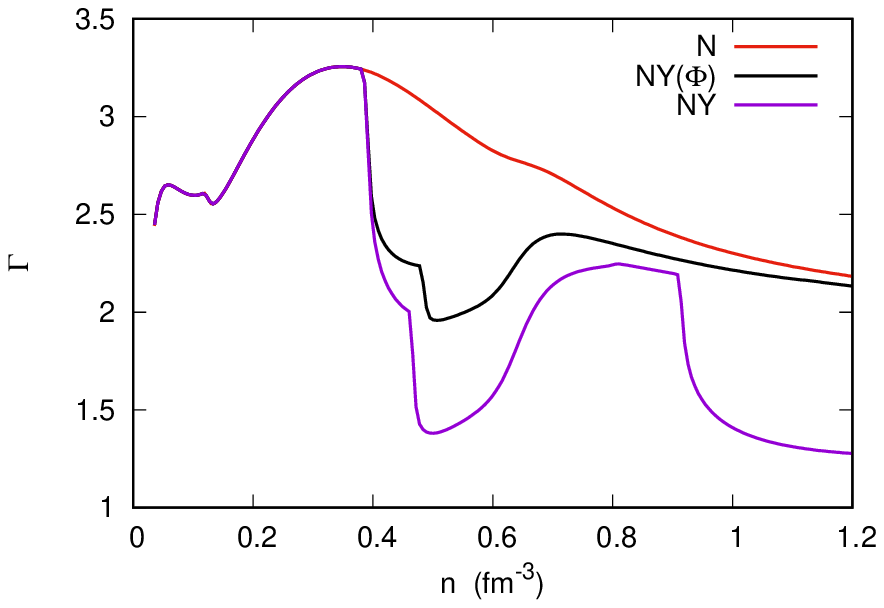} \\
\includegraphics[scale=.54, angle=270]{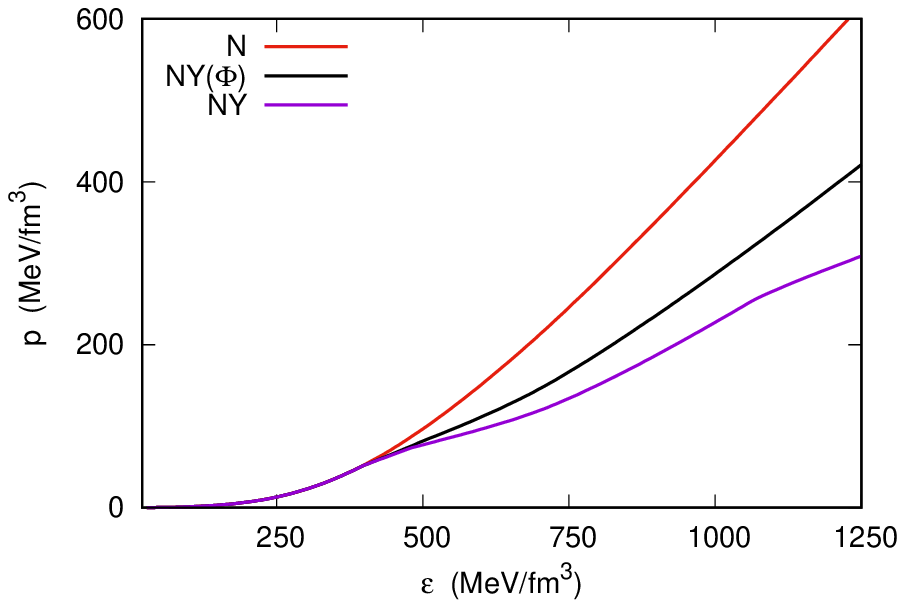} &
\includegraphics[scale=.54, angle=270]{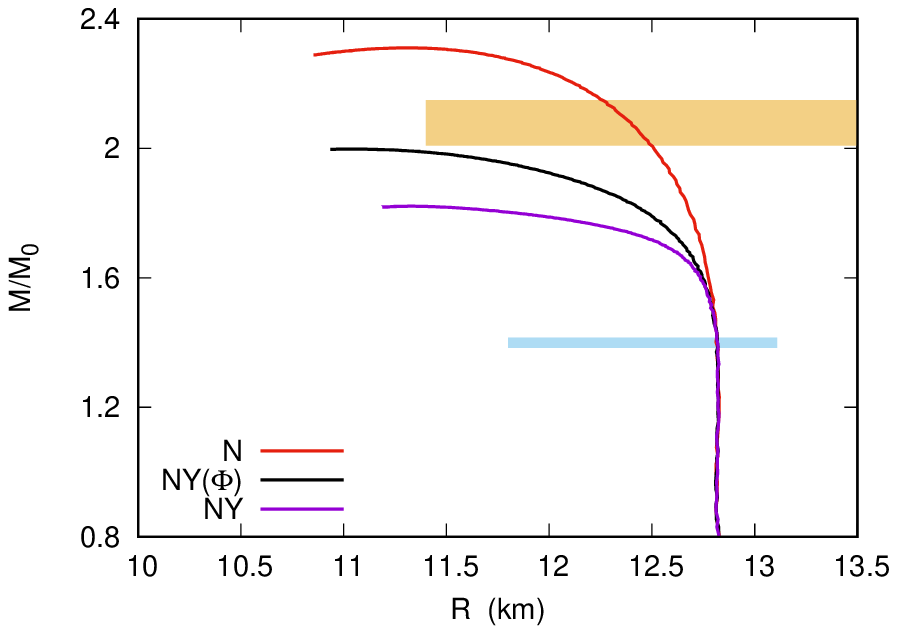}\\
\end{tabular}
\caption{|Particle population (\textbf{a}), adiabatic index (\textbf{b}), EOSs (\textbf{c}), and OV solution (\textbf{d}) for the whole baryon octet within Sakurai's proposal with the $\phi$ meson. Results obtained without the $\phi$ meson are also presented to facilitate comparison.  \label{FL6}}
\end{figure*}

The numerical results are displayed in Fig.~\ref{FL6}. 
The particle population is presented in (\textbf{a}), where we can notice that the $\phi$ meson fully suppresses the $\Xi^0$, being present only the $\Lambda^0$ and the $\Xi^-$. Both, nevertheless, are partially suppressed due to the repulsive nature of the $\phi$ meson, which increases the population of leptons at higher densities. While in the quark-isospin counting rule, the $\Lambda^0$ becomes the most populated particle at higher densities, in Sakurai's model, the $n$ is always the most populated one. Furthermore, the reduction in the $\Xi^-$ population makes the lepton fraction positive even for densities slightly above 1 fm$^{-3}$

In  Fig.~\ref{FL6} (\textbf{b}), I show the adiabatic index for nucleonic and hyperonic matter with and without the $\phi$ meson. We can see that the $\phi$ meson reduces the drop due to the onset of the $\Lambda^0$. Furthermore, the drop related to the $\Xi^0$ is absent for obvious reasons. Finally, in Figs. (\textbf{c}) and \ref{FL6} (\textbf{d}),  I show the EOSs and the OV solutions. As can be seen, the $\phi$  meson causes a stiffening of the EOS, making the maximum mass increase from 1.82 $M_\odot$ to 2.00 $M_\odot$. An almost good result. From this point onward, the $\phi$ meson will always be present.\\

\section{The SU(3) group and its Clebsch-Gordan Coefficients } \label{secCGC}

Soon after J. J. Sakurai’s proposal regarding meson–baryon coupling constants~\cite{SAKURAI1960AP}, Murray Gell-Mann suggested that baryons obey an SU(3) flavor symmetry~\cite{GellMann1962PR}, notably predating his own formulation of the quark model~\cite{GELLMANN1964PL}.

Within the SU(3) group, the baryons, $(\psi_B)$, the anti-baryons $(\bar{\psi})$ and the vector mesons $(M)$ belongs to the irredutible representation IR$\{8\}$, besides the mesonic singlet, which belongs to IR$\{1\}$. Each one of these fields is labeled by its eigenstates: $|N,Y,I,I_3\rangle$, where $N$ is the dimension of the IR, $Y$ is the hypercharge, $I$ is the isospin, and $I_3$ is the isospin projection.  

The Yukawa Lagrangian (sometimes called Dirac-Yukawa Lagrangian):

\begin{equation}
 \mathcal{L}_{Yuk} = -g(\bar{\psi}\psi)M,   
\end{equation}
is a scalar, and thereby, a singlet state: $|N,Y,I,I_3\rangle = |0,0,0,0\rangle$.
This implies that the tensor product $\bar{\psi}_B \otimes\psi_B \otimes M$ must also result in a singlet state.  Each eigenstate field represents a decoupled basis, while the final state is the coupled one. Therefore, we are looking for the Clebsch-Gordan (CG) coefficients.

Here, nevertheless, we are not dealing with the SU(2) CG coefficients found in the theory of angular momentum in quantum mechanics~\cite{Sakuraibook}, but its generalization, the SU(3) CG coefficients. Despite these similarities, the SU(3) Clebsch–Gordan coefficients entail a significantly greater mathematical complexity than their SU(2) counterparts. A complete discussion of the Clebsch–Gordan coefficients of the SU(3) group lies far beyond the scope of the present work. The interested reader is referred to Refs.~\cite{DOVER1984,GreinerSymmetries,Swart1963RMP} and the references therein.  Here, I present only a relatively brief and largely qualitative discussion of the SU(3) Clebsch–Gordan coefficients and their application in determining the relative coupling constants in the Yuaka Lagrangian. A more detailed discussion within the context of hyperons in neutron stars can also be seen in Refs.~\cite{lopesnpa,WeissPRC2012,lopes2023ptep} and especially in the appendix of Ref.~\cite{lopesPRD}.

First, to preserve the unitary symmetry, the direct product $\bar{\psi}_B \otimes\psi_B$  must transform as IR$\{8\}$ when the meson eigenstate ($M$) belongs to IR$\{8\}$, and as IR$\{1\}$ when $M$
belongs to IR$\{1\}$.
By the use of the Speiser method~\cite{Swart1963RMP}, the direct product $\{8\}$ $\otimes$  $\{8\}$  results in $\{27\}$ $\oplus$ $\{10\}$ $\oplus$ $\{\bar{10}\}$ $\oplus$ $\{8\}$ $\oplus$ $\{8'\}$ $\oplus$ $\{1\}$.
Therefore, there are two ways to couple $\{8\}$ $\otimes$  $\{8\}$ to $\{8\}$; typically called symmetric  ($\{8\}$) and antisymmetric ($\{8'\}$) couplings~\cite{DOVER1984}.

The Yukawa Lagrangian density can be rewritten as
\begin{equation}
\mathcal{L}_{\rm Yukawa} = -(gC + g'C' )(\bar{\psi}_B\psi_B)M, \label{A1}
\end{equation}
for the mesons belonging to IR$\{8\}$, and
\begin{equation}
\mathcal{L}_{\rm Yukawa} = -g_1(\bar{\psi}_B\psi_B)M, \label{A2}
\end{equation}
for the mesons belonging to IR$\{1\}$. 

The $g$ ($g'$) is the constant associated to the symmetric (antisymmetric) coupling, while the  $C$ ($C'$) is the SU(3) CG coefficients of the symmetric (antisymmetric) coupling. The calculation of the SU(3) Clebsch–Gordan coefficients lies beyond the scope of the present work; however, they can be systematically derived from the isoscalar factors, as discussed in Ref.~\cite{Swart1963RMP}. Fortunately, their values have been extensively tabulated, and in this work I adopt those provided in Ref.~\cite{McNamee1964}.

Now introducing the new coupling constants $g_8$ and $\alpha$ as done in ref.~\cite{Swart1963RMP}:

\begin{equation}
g_8 =\frac{\sqrt{30}}{40} g + \frac{\sqrt{6}}{24 }g', \quad \mbox{and} \quad \alpha_v = \frac{\sqrt{6}}{24}\frac{g'}{g_8}, \label{A3}
\end{equation}

Which gives us the standard expressions:

\begin{align}
g_{NN\rho} ={} g_8, \quad g_{\Sigma\Sigma\rho}= 2g_8\alpha_v, \nonumber \\   g_{\Xi\Xi\rho}=  -g_8(1 - 2\alpha_v), \quad  
g_{\Lambda\Lambda\rho}={}0,
\end{align}
and considering  that the physical $\omega$ and $\phi$ mesons are a mixture of the theoretical $\omega_8$ and $\omega_1$ states (Eq.~\ref{mix}):
\begin{align}
g_{NN\omega}={}&g_1 \cos\theta_v + g_8\sin\theta_v\frac{1}{3}\sqrt{3}(4\alpha_v - 1), \nonumber \\
g_{\Lambda\Lambda\omega}={}&g_1 \cos\theta_v + g_8\sin\theta_v\frac{2}{3}\sqrt{3}(1 - \alpha_v), \nonumber  \\
g_{\Sigma\Sigma\omega}={}&g_1 \cos\theta_v - g_8\sin\theta_v\frac{2}{3}\sqrt{3}(1 - \alpha_v),  \nonumber \\
g_{\Xi\Xi\omega}={}&g_1 \cos\theta_v -  g_8\sin\theta_v\frac{1}{3}\sqrt{3}(1 + 2\alpha_v). \label{A10}
\end{align}
while for the $\phi$ meson we only must 
 replace $\cos\theta_v$ $\rightarrow$ $-\sin\theta_v$ and 
$\sin\theta_v$ $\rightarrow$ $\cos\theta_v$ in  the equations above \cite{DOVER1984}.
As can be seen, there are three free parameters in the SU(3). $\theta_V$, $\alpha_V$ and $z = g_8/g_1$.

Unlike the quark-isospin counting rule or Sakurai's proposal, the SU(3) symmetry is able to fix all the baryon-vector mesons without invoking any $ad~hoc$ consideration, and yet, as we will see, all the results previously discussed can be recovered.

\subsection{The free parameters, the hybrid SU(6) group, and the missing coupling constant}

The first free parameter of the theory is the angle $\theta_V$. From our quark knowledge, this is not a true free parameter, because the masses and content of the physical $\omega$ and $\phi$ mesons are well known. Here, we simply assume $\theta_V = 35.264^o$, assuming therefore an ideal mixing.

The second free parameter is the strength of the octet mesons in relation to the singlet one,  $z = g_8/g_1$. This value can be fixed by imposing $g_{NN\phi} = 0$, yielding $z = 1/\sqrt{6}$.

The last free parameter, $\alpha_V$ is related to the strength of the antisymmetric coupling ($g'$), relative to the symmetric one ($g$) in Eq.~\ref{A1}. $\alpha_V$ is defined as $\alpha_V = F/(F+D)$ and lies $0~<\alpha_V~<1$. If $\alpha_V$ = 1, we have a pure antisymmetric coupling, or F-mode. On the other hand, $\alpha_V = 0$ means a pure symmetric coupling, or D-mode. Intermediate values implies that both the symmetric and the antisymmetric coupling are non-null.

In order to reproduce Sakurai’s proposal and the quark–isospin counting rule, it is necessary to impose $\alpha_V = 1$, together with $z = 1/\sqrt{6}$ and $\theta_V = 35.264^\circ$. These specific values are not arbitrary; they emerge naturally from the SU(6) symmetry scheme~\cite{DOVER1984,WEISSENBORN2012NPA}. The SU(6) group is a hybrid symmetry that combines both flavor and spin degrees of freedom, namely SU(6)$\supset$ SU(3)$_F \otimes$ SU(2)$_S$~\cite{Gursey1964PRL}.

In this work, I assume that SU(3) flavor symmetry is exact, meaning that the relations dictated by flavor symmetry are fully preserved. On the other hand, the more restrict SU(6) symmetry, which imposes additional constraints by linking spin and flavor degrees of freedom, is allowed to be partially broken. As a consequence, I fix $\theta_V$ to its SU(6) value, since it determines the vector-meson mixing structure. The parameter $z$, which controls the relative strength between singlet and octet couplings, is also fixed to its SU(6) prediction in order to preserve the standard normalization between these sectors. In contrast, $\alpha_V$ is treated as a free parameter, allowing us to explore deviations from the exact SU(6) limit.

The $\chi_{BBM}$'s are, therefore, functions of $\alpha_V$:

\begin{align}
 \frac{g_{\Lambda\Lambda\omega}}{g_{NN\omega}} = \frac{4 + 2\alpha_v}{5 + 4 \alpha_v},\quad
 \frac{g_{\Sigma\Sigma\omega}}{g_{NN\omega}} = \frac{8 - 2\alpha_v}{5 + 4 \alpha_v}, \quad
  \frac{g_{\Xi\Xi\omega}}{g_{NN\omega}} = \frac{5 - 2\alpha_v}{5 + 4 \alpha_v}, \label{e1}
\end{align}
for the $\omega$ meson,
\begin{align}
 \frac{g_{\Lambda\Lambda\phi}}{g_{NN\omega}}= -\sqrt{2} \left(\frac{ 5-2\alpha_v}{5 + 4 \alpha_v} \right),\quad
 \frac{g_{\Sigma\Sigma\phi}}{g_{NN\omega}} =- \sqrt{2} \left(\frac{ 1+ 2\alpha_v }{5 + 4 \alpha_v} \right),\quad
  \frac{g_{\Xi\Xi\phi}}{g_{N,\omega}} ={}&- \sqrt{2} \left(\frac{ 4+ 2\alpha_v}{5 + 4 \alpha_v} \right),\label{e2}
\end{align}
with ${g_{NN\phi}} = 0$ for the $\phi$ meson, and
\begin{align}
 \frac{g_{\Lambda\Lambda\rho}}{g_{NN\rho}} = 0,\quad
 \frac{g_{\Sigma\Sigma\rho}}{g_{N,\rho}} = 2\alpha_v,\quad
  \frac{g_{\Xi\Xi\rho}}{g_{NN\rho}} = -(1 - 2\alpha_v),\label{e3}
\end{align}
for the $\rho$ meson.

The equations above are the set usually found in the literature when we study baryon-meson coupling within SU(3) symmetry formalism~\cite{WeissPRC2012,Miyatsu2013,lopesnpa,lopesPRD}. However, it was noted in Ref.~\cite{lopes2023ptep} that the set above is not complete and there is a missing coupling constant. 

As pointed out earlier, SU(3) imposes that the Yukawa Lagrangian be a singlet, i.e., a tensor product $\bar{\psi}_B\otimes\psi_B\otimes M = |0,0,0,0\rangle$. As discussed in Ref.~\cite{lopes2023ptep}, the coupled channels $\bar{\psi}_\Sigma{^0}\otimes\psi_\Lambda\otimes\rho^0$ and $\bar{\psi}_\Lambda\otimes\psi_\Sigma{^0}\otimes\rho^0$ also produce a singlet state, whose coupling constants are:

\begin{eqnarray}
  g_{\Sigma\Lambda\rho} =   g_{\Lambda\Sigma\rho} = \frac{2}{3}\sqrt{3}g_8(1 - \alpha_V), \quad
  \mbox{implying} \nonumber \\
  \frac{g_{\Sigma\Lambda\rho}}{g_{NN\rho}} = \frac{2}{3}\sqrt{3}(1 - \alpha_V).  \label{ncc}    
\end{eqnarray}

Such a coupling constant is needed to restore the relations of completeness and closure of the SU(3) Clebsch-Gordan coefficients.  From the field theory point of view~\cite{SAKURAI1960AP}, the Eq.~\ref{ncc} indicates that the $\Sigma^0$ and the $\Lambda^0$ interact with each other via $\rho$ meson exchange. However, in the mean-field approximation, the $\Lambda^0$ and the $\Sigma^0$ now interact with the background field of the meson $\rho$. The strength of this interaction depends only on the coupling constant. Following the formalism in Ref.~\cite{lopes2023ptep} ,their energy eigenvalue now reads:

 \begin{equation}
 \begin{array}{c}
      E_\Lambda = \sqrt{M_\Lambda^{*2} + k_{F\Lambda}^2}  + g_{\Lambda\Lambda\omega}\omega_0  + g_{\Lambda\phi}\phi_0 -\frac{1}{2}{g_{\Sigma\Lambda\rho}}\rho_0,\\
    E_{\Sigma^0} = \sqrt{M_\Sigma^{*2} + k_{F\Sigma}^2}  + g_{\Sigma\Sigma\omega}\omega_0  + g_{\Sigma\phi}\phi_0 + \frac{1}{2}{g_{\Sigma\Lambda\rho}}\rho_0.
 \end{array} \label{elambdasigma}
 \end{equation}

Assuming that $\theta_V$ and $z$ are fixed with their SU(6) values, we see that when $\alpha_V = 1$, $g_{\Sigma\Lambda\rho} = 0$, and the quark-isospin couting rule is restored. The complete set for hyperons-mesons coupling constants is presented in Tab.~\ref{T2}.\\

%\begin{widetext}
\begin{center}
\begin{table}%[ht]
\begin{center}
\caption{Complete set of hyperon-mesons coupling constants for different values of $\alpha_v$, within the SU(3) symmetry group. The values for the vector mesons are fully model-independent. The couplings with the scalar $\sigma$ meson are fixed to reproduce the potential depths.\label{T2}}
\scalebox{0.90}{
\begin{tabular}{|c|cccc|}
\hline
$\alpha_V$  &   1.00&0.75&0.50&0.25   \\
\hline
 $g_{\Lambda\Lambda\omega}/g_{NN\omega}$        & 0.667 & 0.687   & 0.714 & 0.75   \\

 $g_{\Sigma\Sigma\omega}/g_{NN\omega}$         & 0.667 & 0.812  & 1.0 & 1.25   \\
 
$g_{\Xi\Xi\omega}/g_{NN\omega}$           & 0.333 & 0.437  & 0.571 & 0.75   \\
\hline
$g_{\Lambda\Lambda\phi}/g_{NN\omega}$  & -0.471 & -0.619  & -0.808 & -1.06   \\

$g_{\Sigma\Sigma\phi}/g_{NN\omega}$           & -0.471 & -0.441  & -0.404 & -0.354   \\

$g_{\Xi\Xi\phi}/g_{NN\omega}$           & -0.943 & -0.972  & -1.01 & -1.06   \\
\hline
$g_{\Lambda\Lambda\rho}/g_{NN\rho}$  &0.0&0.0&0.0&0.0\\
$g_{\Sigma\Sigma\rho}/g_{NN\rho}$           & 2.0 & 1.5  & 1.0 & 0.5   \\

$g_{\Xi\Xi\rho}/g_{NN\rho}$           & 1.0 & 0.5  & 0.0 & -0.5   \\
\hline
$g_{\Sigma^0\Lambda\rho}/g_{NN\rho}$           & 0.0 & 0.288  & 0.577 & 0.866   \\
\hline
$g_{\Lambda\Lambda\sigma}/g_{NN\sigma}$           & 0.610 & 0.625  & 0.646 & 0.674   \\

$g_{\Sigma\Sigma\sigma}/g_{NN\sigma}$           & 0.406 & 0.518  & 0.663 & 0.855   \\

$g_{\Xi\Xi\sigma}/g_{NN\sigma}$           & 0.269 & 0.350 & 0.453 & 0.590   \\
\hline
\end{tabular}}
\end{center}
\end{table}
\end{center}
%\end{widetext}

\subsection{Numerical Results}

\begin{figure*}[ht]
\begin{tabular}{ccc}
\centering % \begin{center}/\end{center} takes some additional vertical space
\includegraphics[scale=.54, angle=270]{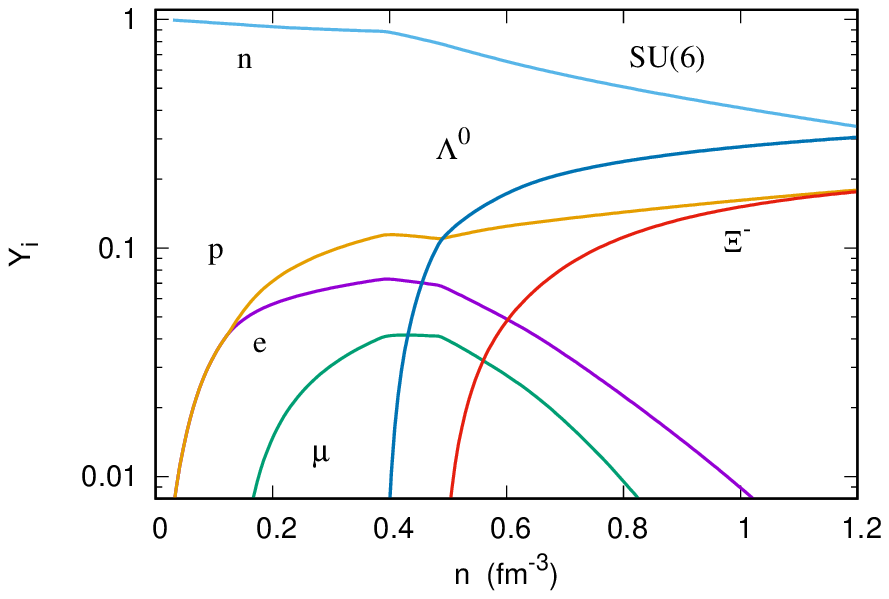} &
\includegraphics[scale=.54, angle=270]{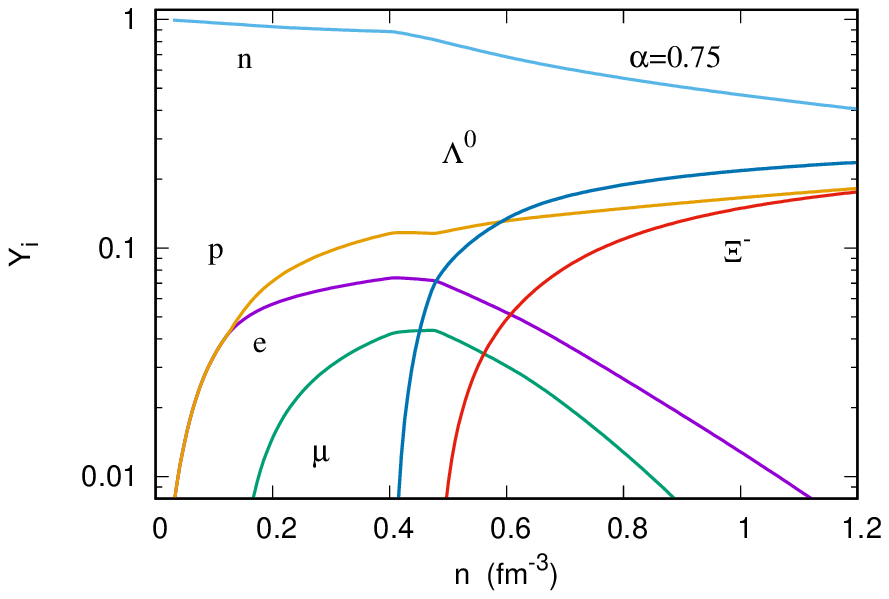} \\
\includegraphics[scale=.54, angle=270]{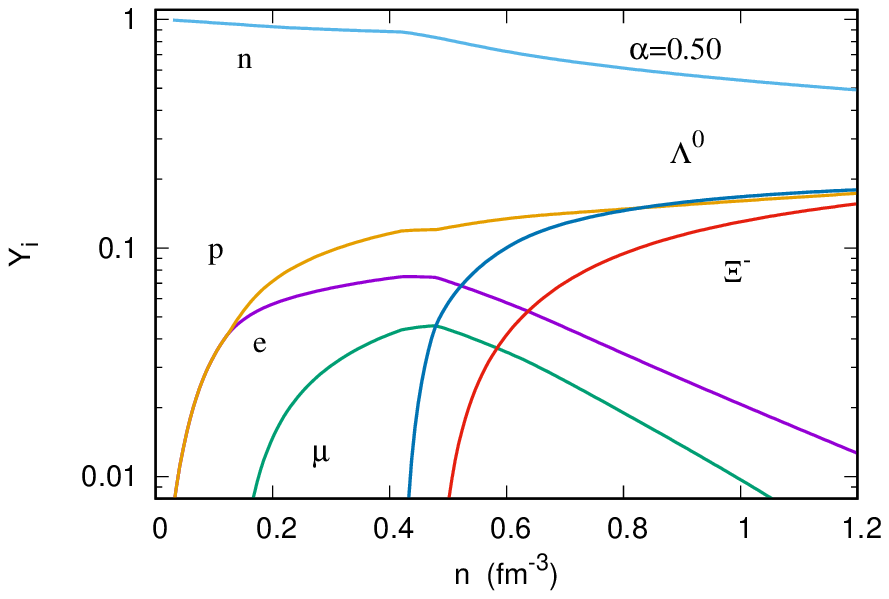} &
\includegraphics[scale=.54, angle=270]{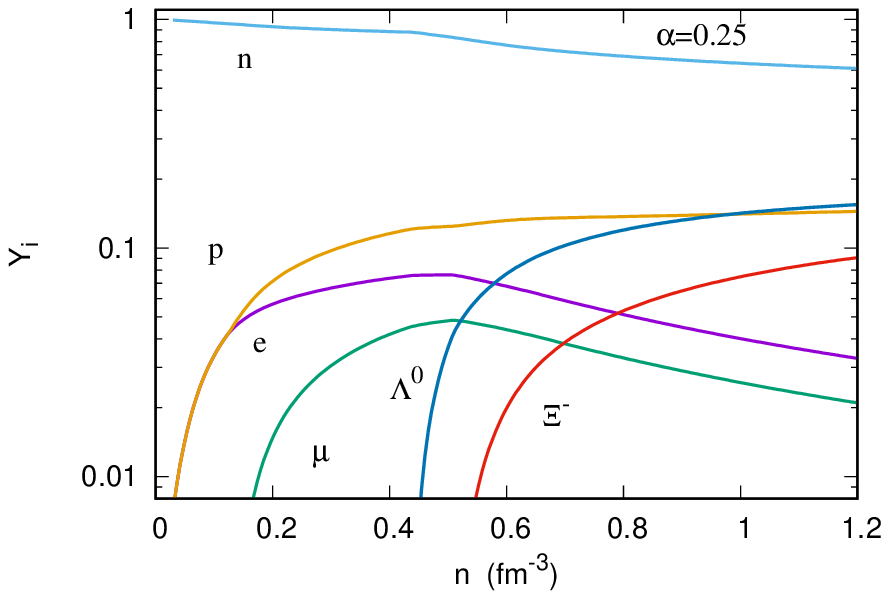}\\
\end{tabular}
\caption{Paticle population for different values of $\alpha_V$.\label{FL7}}
\end{figure*}

In Fig.~\ref{FL7}, we display the particle population for different values of $\alpha_V$. When $\alpha_V =1$, the SU(6) result is recovered, and we also re-obtain the results coming from Sakurai's proposal. As we reduce $\alpha_V$ from 1 to 0.25, we notice that the hyperon population is reduced. This suppression of the hyperons reflects the increase of the $g_{BB\omega}$ for all the hyperons. Furthermore, in general, $g_{BB\phi}$ also increases, causing an additional enhancement of the hyperon repulsion. The $\Sigma$ triplet is an exception, but they are not present. 

The hyperon suppression is also reflected in the total amount of leptons. While for $\alpha_V = 1$ the lepton fraction goes to zero around 1 fm$^{-3}$, when $\alpha_V = 0.25$, electrons and muons are still present even at densities larger than 1.2 fm$^{-3}$.

%%%%%%%%%%%%%%%%%

\begin{figure*}[ht]
\begin{tabular}{ccc}
\centering % \begin{center}/\end{center} takes some additional vertical space
\includegraphics[scale=.54, angle=270]{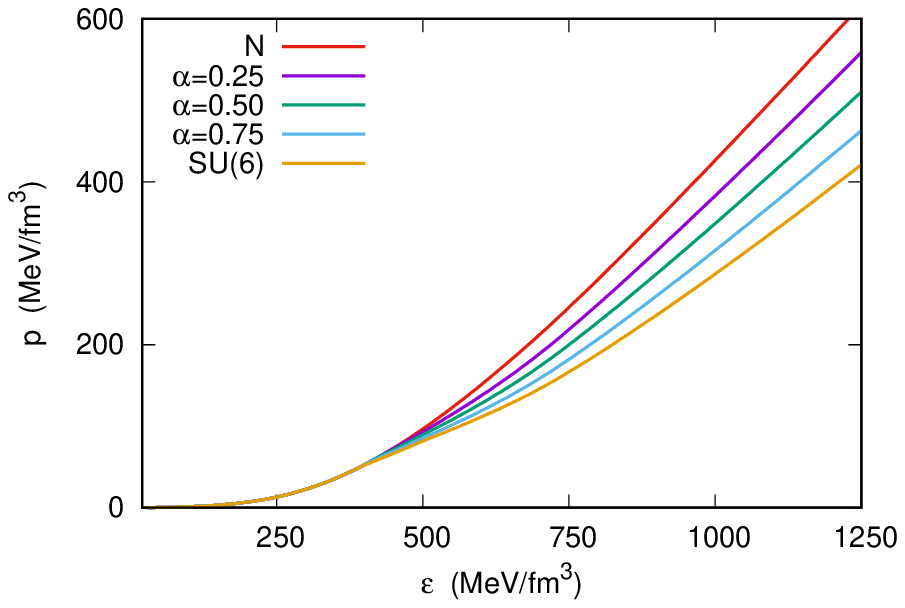} &
\includegraphics[scale=.54, angle=270]{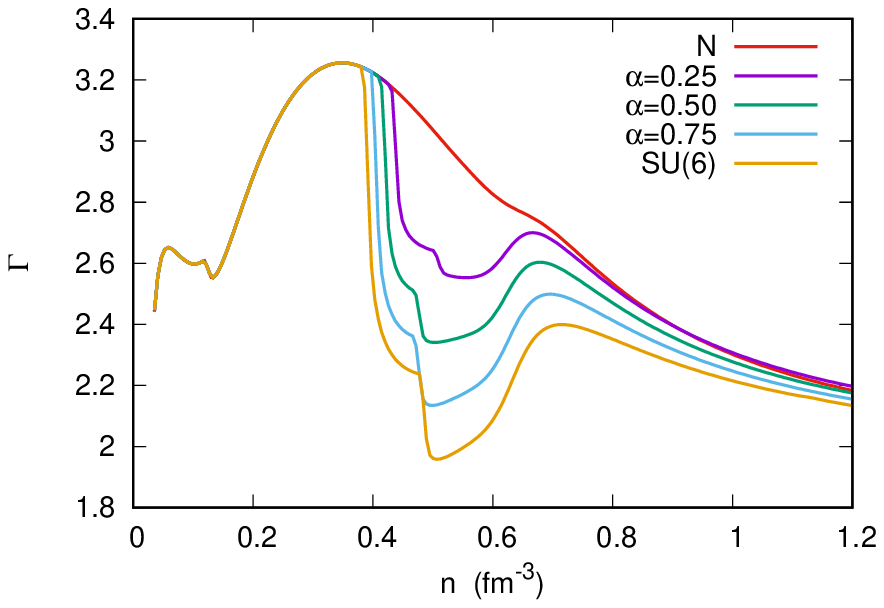} \\
\includegraphics[scale=.54, angle=270]{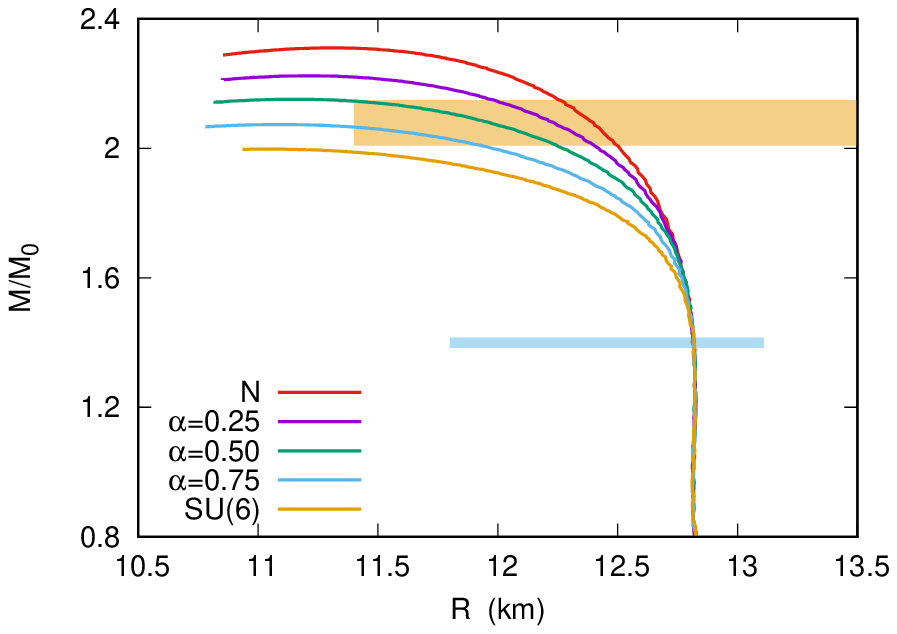} 
\end{tabular}
\caption{ Effects of different values of $\alpha_V$ on microscopic and macroscopic properties of neutron stars: (\textbf{a}) equation of state. (\textbf{b}) Adiabatic index (\textbf{c}) OV solutions.\label{FL8}}
\end{figure*}

The other relevant quantities are displayed in Fig.~\ref{FL8} for different values of $\alpha_V$. In (\textbf{a}), we show the EOSs. Due to $\omega$ dominance, as we decrease the value of $\alpha_V$, the EOS becomes stiffer. As a consequence, the drop in the adiabatic index, related to the onset of new degrees of freedom, and displayed in (\textbf{b}), is reduced. Finally, in (\textbf{c}), the mass-radius relation is obtained by solving the OV equations. Again, due to the $\omega$ dominance, as we reduce $\alpha_V$, the maximum possible mass is raised. Indeed, for $\alpha_V = 0.25$, a maximum mass of 2.22$M_\odot$ is reached. Furthermore, except for the SU(6) parametrization, all values of $\alpha_V$ are in agreement with the constraint related to the PSR J0740+6620 pulsar. Furthermore, as hyperons are not present at 1.4$M_\odot$ neutron stars, all values of $\alpha_V$ predict $R_{1.4} =12.82$ km, in agreement with NICER bounds~\cite{Miller2021}. The relevant quantities are summarized in Tab.~\ref{T3}.

%%%%%%%%%%%%%%%%%

%\begin{widetext}
\begin{center}
\begin{table}[ht]
\begin{center}
\caption{Neutron stars' main properties within NY matter for different values of $\alpha_V$.\label{T3}}
\scalebox{0.90}{
\begin{tabular}{|c|ccc|}
\hline
 $\alpha_V$  &   $M_{max}/M_\odot$ & R (km) & $n_c$ (fm$^{-3}$)   \\
\hline
 Nucleonic       & 2.31 & 11.31   & 0.94   \\

 0.25         & 2.22 & 11.21  & 0.97    \\
 
0.50          & 2.15 & 11.15  & 1.01    \\

0.75  & 2.07 & 11.10  & 1.04   \\

SU(6)      & 2.00 & 11.05  & 1.06    \\
\hline
\end{tabular}}
\end{center}
\end{table}
\end{center}
%\end{widetext}

\subsection{Can we prevent the onset of hyperons in neutron stars' core? The hyperon puzzle}

In the early 2010s, the discovery of neutron stars with mass $M~\gtrsim 2.0M_\odot$, such as the PSR J1614-2230~\cite{Demorest2010}, led to a crisis in the theory of neutron stars. How to reconcile very massive stars with hyperons and yet describe the properties of nuclear matter coming from terrestrial laboratories. At that moment, the IUFSU parametrization was regarded as one of the most realistic QHD-based models. However, with a maximum mass of only 1.94 $M_\odot$, this would imply that when hyperons are also present, the maximum mass would drop to below 1.60 $M_\odot$~\cite{Rafa2011}. This was called the hyperon puzzle. 

Nowadays, the hyperon puzzle has been completely circumvented. The emergence in the literature of models capable of producing very massive stars while still reproducing the properties of nuclear matter (such as the eL3$\omega\rho$ used here~\cite{lopesPRD}), together with the nearly mandatory presence of the $\phi$ meson and the relaxation of the SU(6) group constraints, made stars with masses above $2.2M_\odot$ possible, even with hyperons in their cores.

Nevertheless, it is worth discussing the possibility of avoiding the presence of hyperons without artificially suppressing them by hand. At first glance, we can think of the potential depth. The strong repulsive potential was able to fully suppress the $\Sigma$ triplet. Therefore, it is natural to think that a more repulsive potential can do the same for the $\Lambda^0$ and the $\Xi$ doublet. However, there are some issues related to this approach.

First of all, the effect of the potential depth is only secondary, as can be seen from Fig.~\ref{FL4}. At first glance, one might expect $\chi_\Lambda = 0.95$ to produce the least attractive potential and $\chi_\Lambda = 0.40$ the most attractive one, since the corresponding EOS is significantly stiffer for $\chi_\Lambda = 0.95$. However, a straightforward calculation yields $U_\Lambda = -26.5$ MeV for $\chi_\Lambda = 0.40$ and $U_\Lambda = -62.9$ MeV for $\chi_\Lambda = 0.95$. Therefore, the actual mechanism behind the stiffening of the EOS is the enhanced coupling to the $\omega$ meson through $\omega$ dominance.
Second, the $\Sigma$ triplet is suppressed only because the energetically more favorable $\Xi$ doublet appears first. If the $\Xi$ hyperons are removed, the $\Sigma$ hyperons reappear.
Finally, the potential depth cannot be varied arbitrarily. Only a limited range of values can be considered minimally realistic. Within these values, the differences are minimal.

Another, and one of the earliest approaches after the discovery of the PSR J1614-2230, was the use of three-body forces. As discussed in Refs.~\cite{Lagoteta2012,Chatterjee2016EPJA}, for a sufficiently strong three-body force, a mass above two solar masses can be obtained. However, these same papers argue that the three-body repulsion required to obtain such a high mass is rather unrealistic, since the hyperonic three-body force is far stronger than the nucleonic three-body force.

{ A still incipient possibility is related to the existence of a quarkyonic phase. As the baryon density increases, nucleons overlap, and their constituent quarks fill a common Fermi sea. Deep inside this Fermi sea, the quarks behave as the relevant microscopic degrees of freedom and dominate the bulk thermodynamic properties of the system, such as the pressure and energy density. However, because confinement is assumed to persist, isolated quarks cannot exist. Instead, low-energy excitations near the Fermi surface remain color-singlet hadronic states~\cite{MCLERRAN2007NPA}. It was shown in Ref.~\cite{FujimotoPRC2026} that the possible existence of a quarkyonic phase in pure neutral matter (consisting of only neutrons, $\Lambda^0$ and $\Xi^0$) delays the onset of hyperons, and shifts the threshold density from  $\approx$ 2 - 3$n_0$ to $\approx$ 5 - 6$n_0$. Nevertheless, it is worth emphasizing that this study remains rather incomplete, since it does not include charged baryons, nor does it deal with beta-stable matter. In particular, it does not include the $\Xi^{-}$ hyperon, which appears at low densities, and contains two strange quarks and is, therefore, expected to be less affected by quark Pauli blocking. }

A much more inventive and creative idea, although probably also unrealistic, is to assume the parity-doubled model, where nucleons and hyperons have chiral partners. Some Lattice QCD results suggest that the baryons have a chiral partner, i.e., a baryon with the same quantum numbers and quark content but with a negative parity~\cite{DeTar1989PRD,Jido2001}. In this scenario, the chiral partners are heavier due to the breaking of the chiral symmetry. For larger densities and temperatures, the chiral symmetry is restored, and the masses of a baryon and its partner will converge to the same invariant value. Indeed, we can find in nature baryons with the same quantum numbers and quark content of the members of the baryon octet, but with a negative parity~\cite{PDG2020}. The parity doubled octet is composed by the $N^*(1535)$, $\Lambda^*(1670)$, $\Sigma^*(1750)$, and $\Xi^*(1950)$. 

Ref.~\cite{Gao2026PRD} has explored this possibility and shown that if the chiral invariant mass of the nucleon, $m_0$, is larger than 800 MeV, the hyperons are fully suppressed, once their masses also strong depends of $m_0$. There are, nevertheless, issues with this model.

First, there is no experimental evidence that the negative-parity baryons are indeed the chiral partners of the members of the baryon octet, i.e., there is no indication that their masses become degenerate at high densities and temperatures. Second, the chirally invariant nucleon mass, \(m_0 = 800\) MeV, is significantly larger than what is generally expected in the chiral limit. Indeed, the nucleon effective mass at saturation density is expected to lie in the range of 600--700 MeV~\cite{Dutra2014,Micaela2017}, and even smaller values are anticipated at higher densities. Finally, in order to identify the \(\Lambda^\ast(1670)\) as the chiral partner of the \(\Lambda^0\), one must assume a \(\Lambda\) potential depth of \(U_{\Lambda} = -49\) MeV, which is clearly unrealistic. The same issue also arises for the other hyperons.

Although the calculated results do not correspond to reality, the idea of the parity doubled model to solve the hyperon puzzle is proof of the genius of the human mind to solve complex problems. Ultimately, concerning the question at the title of this section, can we prevent the onset of hyperons in neutron stars' core? An honest answer is: probably not.\\

\section{$\Delta$ resonances} \label{secDelta}

Hyperons are not the only baryons that can potentially be present in the core of neutron stars. $\Delta$ resonances may also play a role here. The principle is the same: as the density increases, it can become energetically favorable the conversion of the nucleons into heavier baryons. The $\Delta^-$ deserves special attention here. Due to its negative charge, the conversion $n +e^- \to \Delta^- +\nu$ becomes more and more energetically favorable as the electron chemical potential increases.

\subsection{Can we prevent the onset of $\Delta$ resonances in neutron stars' core?}

Let us start this section with the question of the last section. Can we prevent the onset of $\Delta$ resonances in the neutron stars' core? Here, { the more honest answer is: we don't know, but there are arguments against $\Delta$ threshold that are not present in the hyperonic sector.}

Beginning with three-body forces, we saw in the previous section that suppressing hyperons requires a three-body interaction that is stronger for hyperons than for nucleons. This may be regarded as unrealistic, since the conventional two-body interaction is known to be stronger in the nucleonic sector. Such an issue is not present for $\Delta$ resonances. Quite the reverse, since two-body forces are stronger in the $\Delta$ sector (as we will see below, symmetry group arguments suggest $\chi_{\Delta\Delta\omega}~\gtrsim 1$, while the highly attractive potential depth also implies $\chi_{\Delta\Delta\sigma}~\gtrsim 1$); one may naturally expect the corresponding three-body forces to be stronger as well. Therefore, three-body forces as a suppression mechanism for $\Delta$ resonances constitute a realistic possibility.

Sawyer, in Ref.~\cite{Sawyer1972APj}, has argued that even in the absence of interactions, the $\Delta$ resonances will suffer a large positive energy shift, varying from 145 to 250 MeV. Therefore, instead of the use of $m_\Delta = 1232$ MeV, a more realistic approach must use at least 1450 MeV.  With such a high mass, the $\Delta$ resonances cannot be present at densities expected in the neutron stars' core.

Using the quark-meson coupling (QMC) model, a more modern study~\cite{MOTTA2020} has also shown that many-body forces generated by the change in the internal quark structure of the baryons in the strong scalar mean fields generated in dense nuclear matter prohibit the appearance of $\Delta$ isobars.

{ The possible existence of a quarkyonic state can also prevent the onset of $\Delta$ resonances. As shown in Ref.~\cite{FujimotoPRC2026}, in matter consisting only of neutral baryons, the $\Lambda^0$ threshold is delayed to 5--6 times the saturation density, while the $\Delta^0$ is fully suppressed in this regime. Although this work is clearly incomplete, since it does not include charged baryons or chemical equilibrium, $\Xi^-$ and $\Delta^-$ are expected to face different scenarios. While the effects of quark Pauli blocking are strongly mitigated for the $\Xi^-$ due to the presence of two strange quarks, the $\Delta^-$ experiences full quark Pauli blocking because its constituent quarks are the same as those of the nucleons. These features disfavor the onset of $\Delta$ resonances.}

Other arguments against the presence of $\Delta$ resonances in dense matter are discussed in Ref.~\cite{BETHE1974}. For instance, it has been argued that $\Delta$ isobars are intrinsically large particles and, therefore, cannot play a role in dense nuclear matter.

The key point here is that, unlike hyperons, there are several plausible arguments against the presence of $\Delta$ resonances in the neutron stars' core, and their presence is significantly more speculative.

We can raise another important question here: Is it possible for neutron stars to contain $\Delta$ resonances but no hyperons? This implies the existence of a mechanism able to suppress the hyperons but not the $\Delta$'s. For all that was discussed here and in the section of hyperons, such a scenario seems very unlikely. Therefore, in this study, I will not consider the possibility of only nucleons and $\Delta$'s (ND matter).

This does not mean, however, that ND matter fails to provide insight into dense matter. In fact, ND matter was investigated in Refs.~\cite{Kauan2022PRC,lopesPRD,Lopes2026EPJA}, where several surprising results were obtained. After reading this section, the reader is encouraged to explore the case of pure ND matter.

\subsection{Formalism}

From a rigorous, formal perspective, the $\Delta$ resonances should be described by the Rarita-Schwinger equations~\cite{Rarita1941,Drago2014}, since they are spin-$3/2$ particles. However, their energy eigenvalue are exactly the same as those from spin-$1/2$, given by Eq.~\ref{energyEV2}. Consequently, we can still use the Lagrangian given in Eq.~\ref{LQCD}, as well as use all the formulations related to the energy eigenvalue, energy density, pressure, and number density from Eq.~\ref{beev} to Eq.~\ref{scalardensity} (as well as the modification introduced by the cutoff potential, $U_{cut}(\sigma_0)$, in Eq.~\ref{scut}).

It is worth emphasizing that the degeneracy factor $\gamma, \gamma = (2S+1)$ is different for the members of the baryon octet and the $\Delta$ resonances. While nucleons and hyperons are spin-$1/2$ particles, and have $\gamma =2$, the $\Delta$ resonances are spin-$3/2$ particles and therefore have $\gamma = 4$. Now, we must generalize the chemical equilibrium and electric charge neutrality to include the $\Delta$ resonances. Indeed, for any baryon $B$ with a charge $Q_B$ (in terms of the proton charge), the chemical equilibrium and the charge neutrality read:

\begin{equation}
 \sum_B \mu_B = \mu_n - Q_B\mu_e, \quad \mbox{with} \quad \sum_BQ_Bn_B +\sum_lQ_ln_l   =0, 
\end{equation}
where $Q_l = -1$ is the electric charge of both leptons, electrons, and muons. 

Finally, we must fix the $\Delta$-mesons coupling constants. The $\Delta$ resonances belong to a large multiplet called the baryon decuplet~\cite{GellMann1962PR,Swart1963RMP}, with $N=10$. They can also be labeled in terms of their hypercharge, isospin, and isospin projection, exactly like the members of the baryon octet. Their representation is presented below:

\begin{center}

\begin{tikzpicture}[scale=2, thick]

% Eixos
\draw[->, gray] (-1.5, 0) -- (1.5, 0) node[right] {$I_3$};
\draw[->, gray] (0, -2.3) -- (0, 1.5) node[above] {$Y$};

% Δ quartet (Y = 1)
\fill (1.5,1) circle (2.2pt);
\fill (0.5,1) circle (2.2pt);
\fill (-0.5,1) circle (2.2pt);
\fill (-1.5,1) circle (2.2pt);

% Σ* triplet (Y = 0)
\fill (1,0) circle (2.2pt);
\fill (0,0) circle (2.2pt);
\fill (-1,0) circle (2.2pt);

% Ξ* doublet (Y = -1)
\fill (0.5,-1) circle (2.2pt);
\fill (-0.5,-1) circle (2.2pt);

% Ω singlet (Y = -2)
\fill (0,-2) circle (2.2pt);

% Conexões horizontais (camadas)
\draw[gray] (1.5,1) -- (0.5,1) -- (-0.5,1) -- (-1.5,1);
\draw[gray] (1,0) -- (0,0) -- (-1,0);
\draw[gray] (0.5,-1) -- (-0.5,-1);

% Conexões verticais corretas (sem diagonais cruzando o centro)
\draw[gray] (1.5,1) -- (1,0);
\draw[gray] (-1.5,1) -- (-1,0);
\draw[gray] (1,0) -- (0.5,-1);
\draw[gray] (-1,0) -- (-0.5,-1);
\draw[gray] (0.5,-1) -- (0,-2);
\draw[gray] (-0.5,-1) -- (0,-2);

% Rótulos
\node at (1.8,1.1) {$\Delta^{++}$};
\node at (0.8,1.1) {$\Delta^{+}$};
\node at (-0.8,1.1) {$\Delta^{0}$};
\node at (-1.8,1.1) {$\Delta^{-}$};

\node at (1.2,0.15) {$\Sigma^{*+}$};
\node at (0.2,0.15) {$\Sigma^{*0}$};
\node at (-1.2,0.15) {$\Sigma^{*-}$};

\node at (0.8,-1.1) {$\Xi^{*0}$};
\node at (-0.8,-1.1) {$\Xi^{*-}$};

\node at (0,-2.2) {$\Omega^{-}$};

\node at (0,1.8) {\textbf{Baryon Decuplet}};

\end{tikzpicture}

\end{center}

Once we know their eigenstate:  $|N,Y,I,I_3\rangle$, we can also use the SU(3) CG coefficients to determine their coupling constants with the members of the vector meson nonet, imposing that the Yukawa Lagrangian is a scalar. A detailed application of the SU(3) formalism to the members of the baryon decuplet can be found in the appendix of Ref.~\cite{lopesPRD}. Here I present a summarized version, focusing on the $\Delta$ resonances. 

To preserve the unitary symmetry, as in the case of the baryon octet, the direct product $(\bar{\psi}_B\otimes\psi_B)$ must transform as IR$\{8\}$ when the meson eigenstate ($M$) belongs to  IR$\{8\}$ and as IR$\{1\}$ when $M$ belongs to  IR$\{1\}$.

Applying the Speiser method, the direct product $\{10\}$ $\otimes$  $\{10^{*}\}$  results in $\{64\}$ $\oplus$ $\{27\}$ $\oplus$  $\{8\}$ $\oplus$ $\{1\}$~\cite{McNamee1964}. Here, unlike the octet case, we have only one coupling resulting in an IR$\{8\}$, the antisymmetric one.

The Yukawa Lagrangian density can be rewritten as
\begin{equation}
\mathcal{L}_{\rm Yukawa} = -(gC)(\bar{\psi}_B\psi_B)M, \label{A12}
\end{equation}
for the mesons belonging to IR$\{8\}$, and
\begin{equation}
\mathcal{L}_{\rm Yukawa} = -g_1(\bar{\psi}_B\psi_B)M, \label{A13}
\end{equation}
for the mesons belonging to IR$\{1\}$, where $C$ are the SU(3) CG coefficients. The CG are listed in Ref.~\cite{McNamee1964}. It is also worth pointing out that the direct product that results in a IR$\{1\}$ must gain a -1 sign, as displayed in Tab. 1 from Ref.~\cite{McNamee1964}. Furthermore, it is useful to divide the four $\Delta$'s into two isospin doublets, once their couplings with the $\rho$ meson are not equal. We  define: ($\Delta^0,\Delta^+$) = $\Delta = |10, 1, 3/2, \pm 1/2 \rangle$, and ($\Delta^-, \Delta^{++}$) = $\Delta^* = |10, 1, 3/2, \pm 3/2 \rangle$, and following Ref.~\cite{lopesPRD} we introduce the coupling $g_{10}$,  in analogy with Eq.~\ref{A3} related to the baryon octet.

\begin{equation}
g_{10} = 10\sqrt{3}g. \label{A15}
\end{equation}

Remembering that the physical $\omega$ and $\phi$ mesons are mixed states of the theoretical $\omega_8$ and $\omega_1$, we can write:

\begin{align}
g_{\Delta^*\Delta^*\rho}={}3g_{10}, \quad g_{\Delta\Delta\rho}={}g_{10}&, \\ \nonumber
g_{\Delta^*\Delta^*\omega}={}g_{\Delta\Delta\omega}=g_1\cos\theta_v +  &g_{10}\sqrt{3}\sin\theta_v, 
\end{align}
and again, the results for the $\phi$ meson are obtained, by  replacing $\cos\theta_v \rightarrow -\sin\theta_v$ and  $\sin\theta_v\rightarrow \cos\theta_v$.

As can be seen, for the baryon decuplet we have only two, rather than three, free parameters. As a consequence, the coupling for the $\rho$ meson is already determined.
Now we employ SU(6) symmetry, and assume an ideal mixing angle ($\theta_v=35.264$). We also determine $z$ in such a way that $g_{\Delta^*\Delta^*\phi} = g_{\Delta\Delta\phi}$ = 0, which implies $z = 1/\sqrt{6}$, exactly as in the octet case. 

To determine the strength of the interaction of the $\Delta$'s in relation to the nucleon, $g_{\Delta\Delta M}/g_{NNM}$, we must define $g_{10}/g_8$. As the octet and the decuplet are independent multiplets, they cannot be fixed by the use of the SU(3) symmetry. 

We can turn again to Sakurai's proposal. We know that, for the members of the baryon octet, in the SU(6) limit, the $\rho$ meson couples to the isospin and the octet part of the $\omega$ meson couples to the hypercharge, restoring the quark-isospin counting rule. Assuming that this is also true for the members of the baryon decuplet, we obtain:

\begin{eqnarray}
\frac{g_{10}}{g_8} = 1.
\end{eqnarray}

Therefore, we can write:
\begin{align}
 &\frac{g_{\Delta^*\Delta^*\omega}}{g_{NN\omega}} = \frac{g_{\Delta\Delta\omega}}{g_{NN\omega}} ={} \frac{9}{5 + 4 \alpha_v},\\
 &\frac{g_{\Delta^*\Delta^*\rho}}{g_{NN\rho}} ={} 3, \quad  \frac{g_{\Delta\Delta\rho}}{g_{NN\rho}} ={} 1,\\
\end{align}
 with  $\chi_{\Delta\Delta\phi} =0$.

 Finally, we can determine the coupling with the scalar $\sigma$ meson from the potential depth. Unlike the hyperons, the potential depths of the $\Delta$'s are still highly uncertain. Modern studies point out that the $U_{\Delta}$ must be, at least, as attractive as the nucleon. Ref.~\cite{Drago2014} discuss the indications of $U_{\Delta} = -75$ MeV and  $U_{\Delta} = -80$ MeV. They also suggest a lower limit of  $U_{\Delta} = U_N - 30$ MeV, which implies  $U_{\Delta} = -95$ MeV for the model I present here.  
 In Ref.~\cite{Bodek2020} a value $U_\Delta = 1.5U_N$ was obtainded, given the value of  $U_{\Delta} = -98$ MeV. This is the value used in Ref.~\cite{lopesPRD}. For all this discussion, we can conclude that the $\Delta$ potential depth must lie $-100<U_\Delta<-70$ MeV.  Here, I use $U_\Delta = -90$ MeV, but the reader is invited to explore the effects of different values of $U_\Delta$. The values of the coupling constants for different values of $\alpha_V$ are presented in Tab.~\ref{T4}.

%\begin{widetext}
\begin{center}
\begin{table}[ht!]
\begin{center}
\caption{Coupling constants for different values of $\alpha_v$, within the SU(3) symmetry group for the $\Delta$ resonances.  The Sakurai's proposal is recovered in the SU(6), and the potential depth is fixed to $U_\Delta = -90$ MeV. Furtheremore, $\chi_{\Delta\Delta\phi} = 0$ for all valeus of $\alpha_V$. \label{T4}}
\scalebox{0.90}{
\begin{tabular}{|c|cccc|}
\hline
$\alpha_V$  &   1.00&0.75&0.50&0.25   \\
\hline
 $g_{\Delta^*\Delta^*\omega}/g_{NN\omega}$        & 1.000 & 1.250   & 1.285 & 1.500   \\

 $g_{\Delta\Delta\omega}/g_{NN\omega}$       & 1.000 & 1.250   & 1.285 & 1.500   \\
$g_{\Delta^*\Delta^*\rho}/g_{NN\rho}$           & 3.0 & 3.0  & 3.0 & 3.0   \\
$g_{\Delta\Delta\rho}/g_{NN\rho}$           & 1.0 & 1.0  & 1.0 & 1.0   \\
 $g_{\Delta^*\Delta^*\sigma}/g_{NN\sigma}$        & 1.080 & 1.178   & 1.300 & 1.466   \\
$g_{\Delta\Delta\sigma}/g_{NN\sigma}$           & 1.080 & 1.178  & 1.300 & 1.466   \\
\hline
\end{tabular}}
\end{center}
\end{table}
\end{center}
%\end{widetext}

\subsection{Numerical Results}

\begin{figure*}[ht]
\begin{tabular}{ccc}
\centering % \begin{center}/\end{center} takes some additional vertical space
\includegraphics[scale=.54, angle=270]{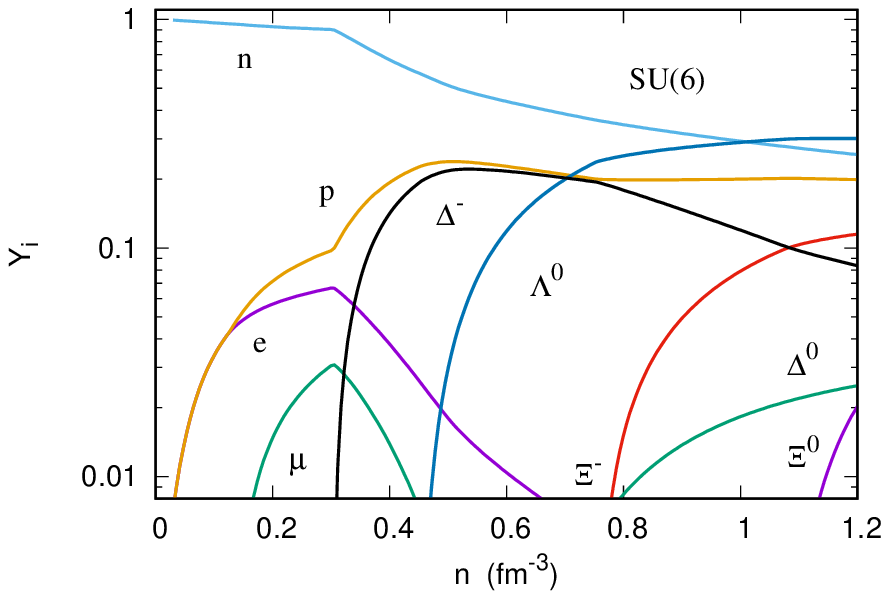} &
\includegraphics[scale=.54, angle=270]{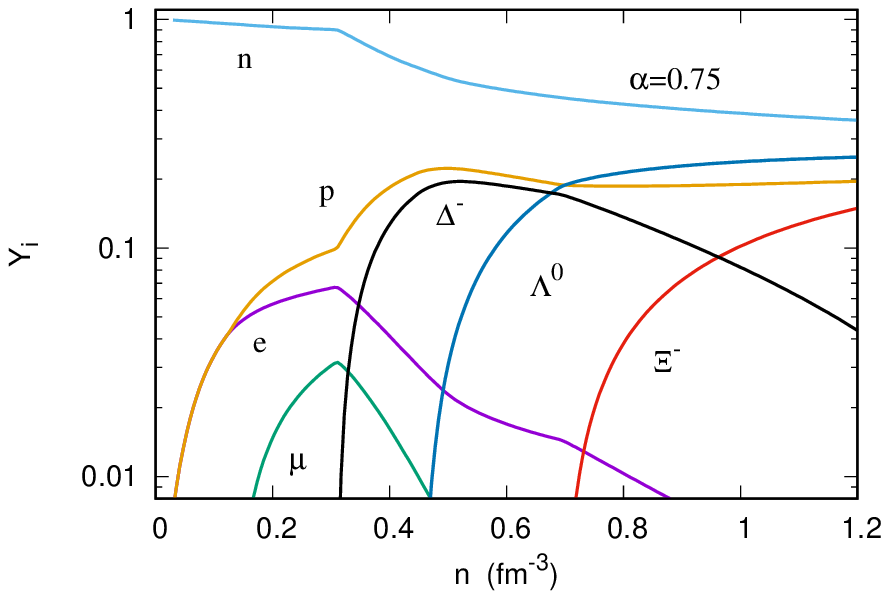} \\
\includegraphics[scale=.54, angle=270]{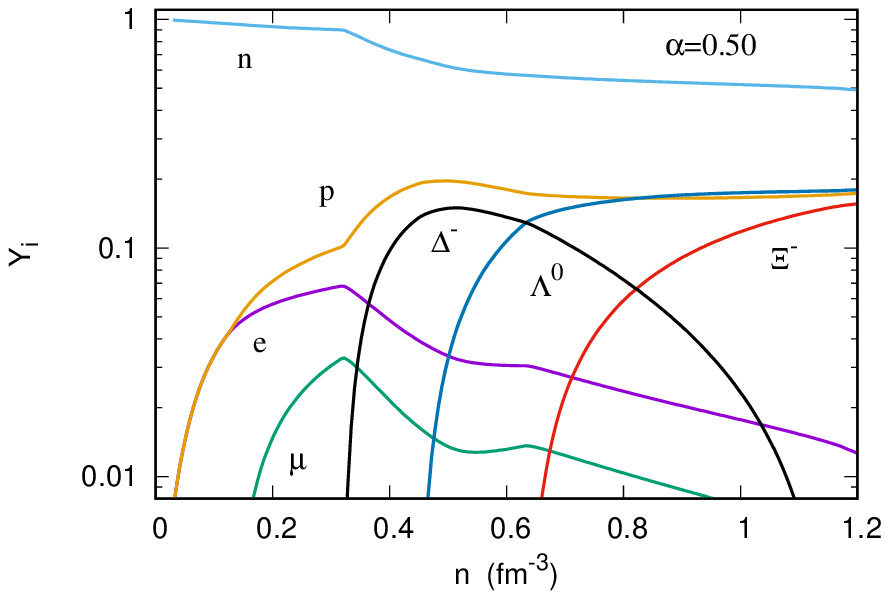} &
\includegraphics[scale=.54, angle=270]{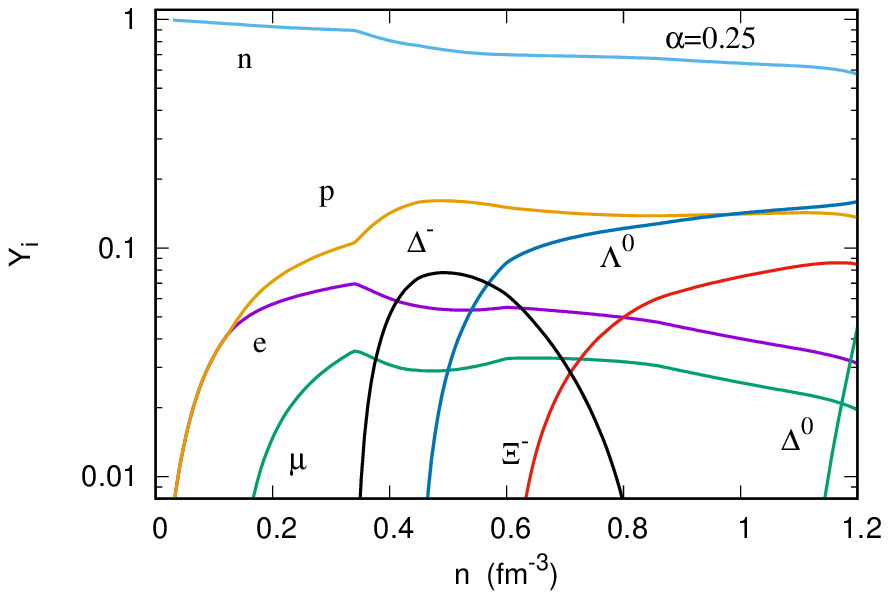}\\
\end{tabular}
\caption{Particle population for different values of $\alpha_V$ for NYD matter.\label{FL9}}
\end{figure*}

We now analyze the effect of $\Delta$ resonances in beta-stable matter. The particle population for NYD matter for different values of $\alpha_V$ is displayed in Fig.~\ref{FL9}.

%%%%%%%%%%%%%%%%%

Let me start the analysis with the SU(6) case. Due to the high attractive potential, the $\Delta^-$ appears at relatively low densities, below two times the saturation density, and due to its value of the degeneracy factor $(\gamma =4)$, its population quickly increases. This increase is followed by a quick decrease in the lepton fraction, leading to an increase in the proton population (the reader is invited to reflect on this assertion). The $\Lambda^0$ threshold happens around 0.44 fm$^{-3}$, and they become the most populated baryon for densities $n~\gtrsim 1$ fm$^{-3}$. The $\Delta^0$ and the $\Xi^-$ appear almost simultaneously around 0.8 fm$^{-3}$, but due to the $\omega$ dominance, the $\Delta^0$ is strongly supressed. In the same sense, the competition between $\Delta^-$ and $\Xi^-$ leads to a decrease in the $\Delta^-$ population. The $\Xi^0$ also appears at high densities but in an insignificant quantity.

As we move away from the SU(6) symmetry, we increase the value of $\chi_{BB\omega}$ for all exotic baryons, and consequently, all of them are suppressed at high densities. This can be noticed by the increase in the neutron population as we reduce $\alpha_V$. For $\alpha_V = 0.75$ and $\alpha_V = 0.25$, the $\Delta^0$ and the $\Xi^0$ are fully suppressed. A more important feature is the strong suppression of the $\Delta^-$ at high densities. As we reduce $\alpha_V$, we obtain $\chi_{\Delta\Delta\omega}~>1$. The $\Delta^-$ population begins to drop not due to the competition with the $\Xi^-$, but mainly due to the $\omega$ dominance.  In the extreme case of $\alpha_V = 0.25$, the existence of $\Delta^-$ particles is restricted to a small range of the neutron stars' core.

We also have the opportunity to discuss the effect of the cutoff potential $U_{cut}(\sigma_0)$. Except for a small difference in the $U_\Delta$, the model and parameters used here are the same as those used in Ref.~\cite{lopesPRD}, but the cutoff potential. We can compare our Fig.~\ref{FL9} with Fig. 9 of Ref.~\cite{lopesPRD}. For the SU(6), we can notice that the absence of the cutoff potential leads to a much higher value of $\Delta^0$'s and the absence of the $\Xi$ doubled. More importantly, without the cutoff potential, the numerical code stops below 1.0 fm$^{-3}$, once the nucleon mass goes to zero. It can also be noted that without the cutoff potential, the $\Delta^0$ is still present at $\alpha_V = 0.75$, while it is fully suppressed when $U_{cut}(\sigma_0)$ is present. A general feature is that the suppression of the $\Delta^-$ at large densities is weakened without the cutoff potential, once the attractive $\sigma$ field grows with the density.

%%%%%%%%%%%%%%%%%
\begin{figure*}[b!]
\begin{tabular}{ccc}
\centering % \begin{center}/\end{center} takes some additional vertical space
\includegraphics[scale=.54, angle=270]{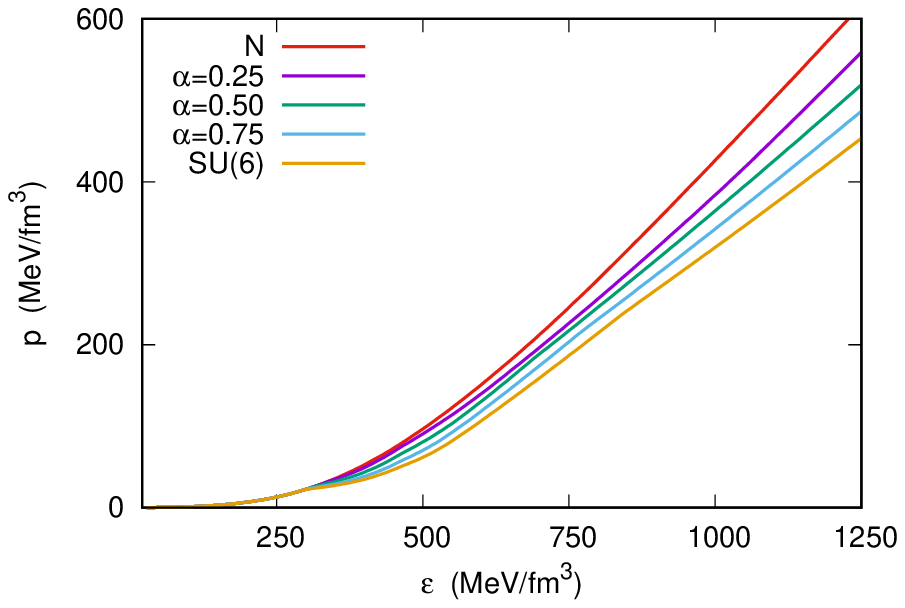} &
\includegraphics[scale=.54, angle=270]{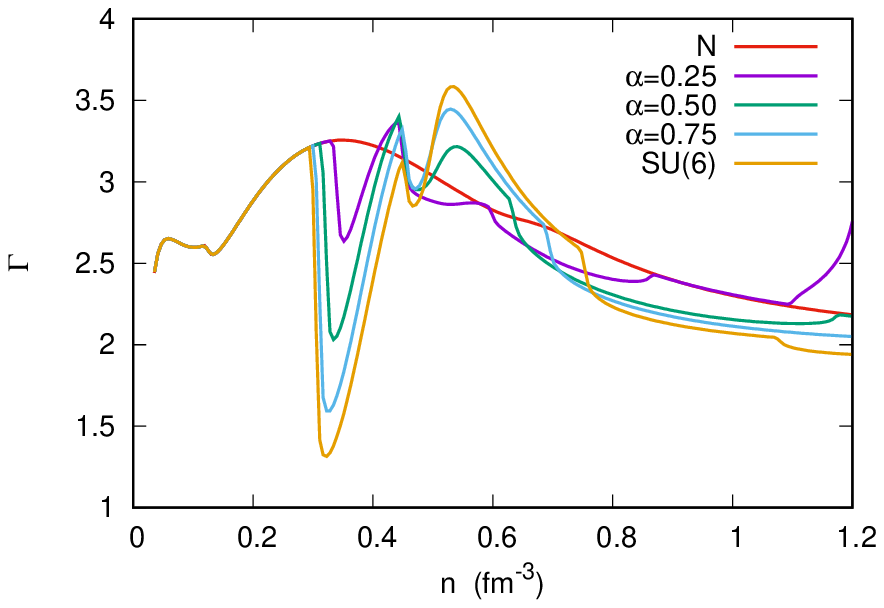} \\
\includegraphics[scale=.54, angle=270]{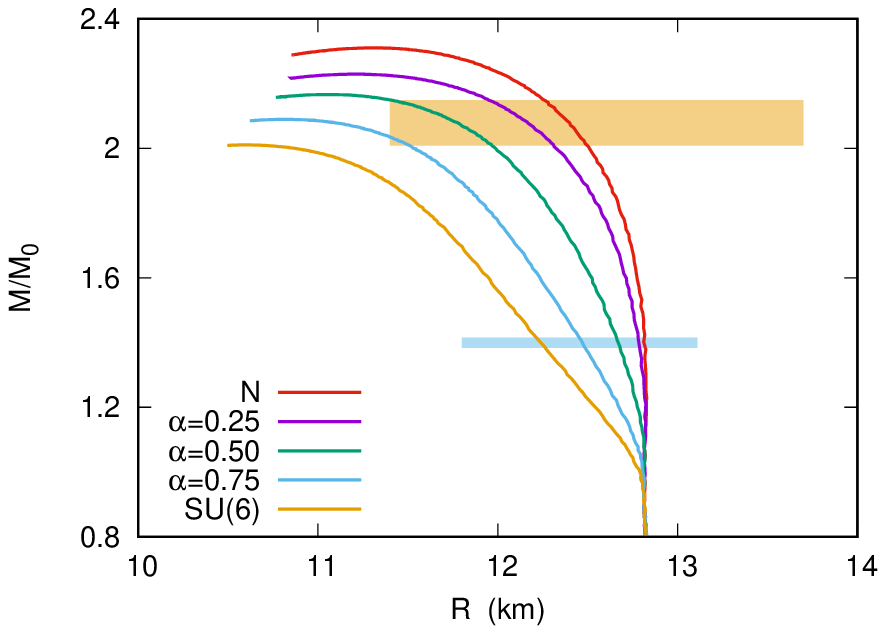} 
\end{tabular}
\caption{Influence of $\alpha_V$ in NYD matter: (\textbf{a}) EOSs; (\textbf{b}) Adiabatic index $(\Gamma)$; (\textbf{c}) OV solutions and astrophysical constraints related to the PSR J0740+6620 and the canonical star.\label{FL10}}
\end{figure*}

We now turn our attention to the EOS-related quantities. The numerical results are presented in Fig.~\ref{FL10}. We begin with the EOSs, presented in (\textbf{a}). In this case, the effects are straightforward. The presence of $\Delta$ resonances softens the EOSs at low densities, while decreasing $\alpha_V$ increases the $\chi_{NN\omega}$ for all non-nucleonic baryons, stiffening the EOS due to the $\omega$ dominance. At high density limit, the softening of EOS due to the onset of the $\Delta^-$ is fully neutralized due to the strong coupling with the $\omega$ meson. The presence of hyperons also causes the well-known softening of the EOS. Unsurprisingly, the stiffer EOS is for pure nucleonic matter.   More interesting are the effects related to the adiabatic index $\Gamma$, presented in (\textbf{b}). The onset of $\Delta^-$ causes a steeper drop in $\Gamma$, followed by an equally steeper growth in $\Gamma$ which produces a local maximum. A second drop is caused by the onset of the $\Lambda^0$, followed by a new growth.  This causes a new point of maximum in $\Gamma$. The maximums in $\Gamma$ are larger in NYD matter than in a pure nucleonic one. In relation to $\alpha_V$, we can see that the effects of the onset of new degrees of freedom are much more exacerbated for larger values of $\alpha_V$. For instance, in the SU(6) ($\alpha_V = 1$), the drop due to the onset of the $\Delta^-$ leads $\Gamma$ to values lower than 1.5, while for $\alpha_V = 0.25$ $\Gamma$ still larger than 2.5. In the same sense, the second maximum is larger than 3.5 in the SU(6), but is essentially the same as the nucleonic matter for $\alpha_V = 0.25$.

We finally analyse the mass-radius relation from the OV solution. The results are presented in (\textbf{c}). The very onset of $\Delta^-$ makes that all stars with masses around 1.0 $M_\odot$ already have $\Delta^-$ in their core. This implies that, unlike the hyperons, the $\Delta^-$ can affect the canonical 1.4$M_\odot$. Larger values of $\alpha_V$ produce softer EOSs, and therefore, the larger the value of $\alpha_V$, the smaller the radius of the canonical star. In varies from 12.24 km for $\alpha_V = 1$ up to 12.79 for $\alpha_V = 0.25$. Again, $\alpha_V$ also reflects in the maximum mass. As can be seen, as in the case of the NY matter, the constraint coming from PSR J0740+6620 is satisfied by all values of $\alpha_V$ but 1. Furthermore, the presence of $\Delta$'s causes the maximum masses to be higher for NYD than for NY for all values of $\alpha_V$. This was already noted in Ref.~\cite{lopesPRD}, and it is related to the $\omega$ dominance. { The fact that $\Delta$ threshold reduces the radius of the canonical star while keeping the maximum mass virtually unchanged is phenomenologically attractive.}

{ Conclusions similar to those presented here were drawn in Ref.~\cite{JIAJIELI2018PLB,Sedrakian2023}, where the authors show that the appearance of $\Delta$’s softens the EoS in the low to intermediate-density region and stiffens it at high densities; pointing out that the maximum mass of a compact star increases by a small amount and, more importantly, the radius of compact stars at low-to-intermediate densities decreases considerably.}

Finally, we can also discuss the effect of the cutoff potential. For SU(6) parametrization, a true maximum mass of 2.01$M_\odot$ was obtained, in contrast with the ``fake" maximum mass of 1.84$M_\odot$ found in Ref.~\cite{lopesPRD} (as the numerical code stops before the condition $\partial M/\partial\epsilon_c~<0$ was reached). For other values of $\alpha_V$ we obtain only a small increase in the mass. All the relevant features for NYD neutron stars are presented in Tab.~\ref{T5}.

%\begin{widetext}
\begin{center}
\begin{table}[ht]
\begin{center}
\caption{Neutron stars' main properties within NYD matter for different values of $\alpha_V$.\label{T5}}
\scalebox{0.90}{
\begin{tabular}{|c|cccc|}
\hline
 $\alpha_V$  &  $R_{1.4}$ (km) & $M_{max}/M_\odot$ & R (km) & $n_c$ (fm$^{-3}$)   \\
\hline
Nucleonic & 12.82      & 2.31 & 11.31   & 0.94   \\

 0.25  & 12.79     & 2.23 & 11.22  & 0.97    \\
 
0.50   &12.66&     2.17 & 11.07  & 1.01    \\

0.75 & 12.47 & 2.09 & 10.82  & 1.07   \\

SU(6)  &12.24 &   2.01 & 10.60  & 1.13    \\
\hline
\end{tabular}}
\end{center}
\end{table}
\end{center}
%\end{widetext}

We conclude this section by pointing out that our current knowledge about nuclear physics and astrophysical observation cannot rule out the possibility of $\Delta$'s in the core of neutron stars.\\

\section{(Anti)kaon condensate} \label{secKaon}

We now discuss the possibility of boson condensation in neutron stars. The most straightforward approach is to consider the appearance of negatively charged pions, $\pi^-$. Since their mass is only slightly larger than that of the muon, the onset of $\pi^-$ condensation may occur even below nuclear saturation density, as negatively charged pions can replace leptons in maintaining chemical equilibrium:

\begin{equation}
\mu_n = \mu_p + \mu_\pi.
\end{equation}

Furthermore, as bosons, the pions can condense with zero momentum (S-wave condensate), which becomes strongly energetically favorable. Study of pion condensate in neutron stars was a hot topic in the 1970s~\cite{Baym1973Pion,MIGDAL1979PLB,BAYM1974NPA}, but since the early 1980s, we have known that it is prevented due to a strong repulsive interaction~\cite{DICKHOFF1983}. Modern studies point out that the pion potential depth can be as high as +100 MeV~\cite{OhnishiPRC2009}.

A viable alternative is the condensation of (anti)kaons. Although their masses are literally about a thousand times larger than that of the electron, studies indicate the presence of a strongly attractive potential. Additionally, different approaches suggest a negative value for the antikaon-omega coupling constant. Due to the $\omega$ dominance, this implies that the antikaon chemical potential monotonically decreases with the density.

\subsection{Formalism}

The Lagrangian for a pseudoscalar-mesont reads~\cite{Glenbook, kaonp,Banik2001,Thapa2021,Thakur2025,Lopes2026PRD}:

\begin{equation}
\mathcal{L}_M = D_\mu^{*}\bar{M}D^\mu M - m_M^{*2}\bar{M}M , \label{kaonLag}
\end{equation}
where $M$ is the mesonic field, $D_\mu = \partial_\mu + ig_{MM\omega}\omega_\mu + \frac{i}{2}g_{MM\rho}\vec{\tau} \cdot \vec{\rho}_\mu + i g_{MM\phi}\phi_\mu$ the covariant derivative, and $m_M^{*} = m_M - g_{MM\sigma}\sigma$ is the effective mass of the meson $M$. By applying the Euler-Lagrange equations, the energy eigenvalue, also known as the dispersion relation~\cite{Banik2001}, is obtained. In MFA, for the S-wave condensate (zero momentum), we have: 

\begin{equation}
  \omega_M = m^*_M + g_{MM\omega}\omega_0 + \frac{\tau_{3}}{2}g_{MM\rho}\rho_0 + g_{MM\phi}\phi_0  . \label{disprel}
\end{equation}

The expected values of the virtual mesonic fields in the presence of baryons and meson condensate are given by:

\begin{eqnarray}
 m_s^2\sigma_0 +  U_{cut}'(\sigma_0) +  g_{NN\sigma}[\kappa M_N (g_{NN\sigma}\sigma_0)^2  +\lambda(g_{NN\sigma}\sigma_0)^3] \\ \nonumber  = \sum_B g_{BB\sigma}n^S_B + \sum_M g_{MM\sigma}n_M ,
\end{eqnarray}
\begin{equation}
 (m_\omega^2 + 2\Lambda_v\rho_0^2)\omega_0 = \sum_B g_{BB\omega}n_B + \sum_M g_{MM\omega}n_M  ,
\end{equation}

\begin{equation}
 (m_\rho^2 + 2\Lambda_v\omega_0^2)\rho_0 = \sum_B g_{BB\rho}\frac{\tau_3}{2}n_B + \sum_M g_{MM\rho}\frac{\tau_3}{2}n_M  
\end{equation}

\begin{equation}
  m_\phi^2\phi_0 = \sum_B g_{BB\phi}n_B + \sum_M g_{MM\phi}n_M ,
\end{equation}
where  $n_M$ is the meson density, given by~\cite{kaonp}:

\begin{equation}
n_M = 2m_M^*\bar{M}M.   
\end{equation}

 The mesonic energy density is

 \begin{equation}
    \epsilon_M =  \sum_M m^{*}_Mn_M ,
 \end{equation}
and the total energy density is the sum of the baryonic, leptonic, and mesonic energy densities. On the other hand, S-wave condensates generate no pressure.

Now, considering a chemically stable matter with zero net charge, composed of nucleons, hyperons, $\Delta$'s, leptons, and antikaon condensate, we have:

\begin{equation}
 \mu_B = \mu_n - Q_B\mu_e; \quad \mu_e =\mu_\mu = \mu_{K^-} = \omega_{K^-};    \nonumber
\end{equation}

\begin{equation}
    \omega_{\bar{K^0}} = 0; \quad \sum_Bq_Bn_B -\sum_ln_l -n_{k^-} = 0, \label{criterias}
\end{equation}
where $Q_B = \pm 1$ is the electric charge of the baryon $B$ in terms of the proton charge.

It is worthwhile to discuss the antikaon density in greater detail, as such discussions are rare in the modern literature. Let us consider a matter composed only of nucleons, leptons, and antikaons (NK matter). In the case of $K^-$, the condensate density is determined by the requirements of chemical equilibrium and charge neutrality. 
Before the onset of the $K^-$, the electron chemical potential is a variable that must be numerically determined. However, after the condition $\mu_{K^-} = \omega_{K^-} = \mu_e$ is satisfied, the electron chemical potential is not a variable anymore, as it is coupled to the kaon chemical potential. As $\omega_{K^-}$ decreases, so does the electron chemical potential. However, a decrease in the electron chemical potential leads to an increase in the proton fraction, due to Eq.~\ref{betaeq} (see also Ref.~\cite{Glenbook}).
Therefore, the value of $n_{K^-}$ is the difference between the density of protons and leptons.
For a matter composed of N different baryons (for instance, hyperons and $\Delta$'s), we can generalize this expression:

\begin{equation}
n_{K^-} = \sum_BQ_Bn_B - \sum_l n_l.  \label{negkaondens}
\end{equation}

Although Eq.~\ref{negkaondens} is simple in form, its numerical implementation can be challenging, since the lepton densities, $n_l$, are independent variables before the onset of kaon condensation but become determined quantities afterward.

In the case of the $\bar{K}^0$, the straightforward approach is to define the plane wave equation for the $K$ field~\cite{Muto2025PRC}:

\begin{equation}
 |K\rangle = \frac{f}{\sqrt{2}}\theta e^{i\omega t}
\end{equation}
where $\theta$ is the chiral angle and $f = 93$ MeV is the meson decay constant~\cite{PDG2020}. To obtain a second-order phase transition: 

\begin{equation}
\theta^2 = - \frac{1}{m_k^{*}}(\omega_{\bar{K}^0} - \omega_c)  \label{theta}
\end{equation}
where $\omega_c$ is the critical energy eigenvalue, which following Eq.~\ref{criterias} is zero. Furthermore, assuming that $\theta$ must be real, Eq.~\ref{theta} assures that no $\bar{K}^0$ condensation occurs when $\omega_{\bar{K}^0} > 0$. Such an approach is fully consistent with the Landau theory of continuous phase transition (see Ref.~\cite{LANDAU1980PT} for a complete discussion).

Finally, the antikaon potential depth at the saturation density is:

\begin{equation}
U_M(n_0) =  g_{\bar{K}\bar{K}\omega}\omega_0 - g_{\bar{K}\bar{K}\sigma}\sigma_0. \label{mesonpd}
\end{equation}

\subsection{Limitation of the theory}

It is worth discussing the uncertainties related to the possibility of antikaon condensate in dense nuclear matter, and comparing with those related to the possibility of hyperon and $\Delta$'s threshold.

From the experimental point of view, the differences are huge. Hyperons in nuclear matter are a well-established field. The existence of hypernuclei is unambiguous. The uncertainties in their potential depth are small. $\Delta$ resonances are well known to NOT be present in nuclear matter at normal densities, but the knowledge of the $\Delta$ width is good enough to provide information about the behaviour of a possible $\Delta$ onset~\cite{OSET1987NPA}. The uncertainties in the potential depth are moderate. On the other hand, our present level of technology does not allow us to have any experimental evidence about the possible kaon condensation and its behaviour. The uncertainties in the potential depth are huge. We can found in the literature values in the range $-180~\mbox{MeV}~<U_{\bar{K}}~<-60$ MeV~\cite{Glenbook,kaonp,Banik2001,Banik2002}.

Additional limitations of the theory are related to the MFA approach to the antikaon condensation. As pointed out in ref.~\cite{Glenbook}, the in-medium interaction of the kaons is certainly a coupled channel problem, and therefore cannot be properly treated at the mean-field level. As mean-field is already an approximation, assuming an MFA for kaons implicates an approximation within an approximation (some kind of inception of approximations), and therefore, a theory based on the chiral effective theory should be mandatory~\cite{THORSSON1994NPA}.  Moreover, the kaon properties are strongly dependent on the $\Lambda(1405)$~\cite{LEE1996PRep,SAVAGE1994PLB}, which is fully disregarded here (as well as in all supra-cited papers about antikaon condensation). This implies that the behaviour of the kaon chemical potential can become significantly different when the baryon chemical potential reaches values around $\mu_B~\gtrsim~1400$ MeV.

Other issues can be related to the presence of the meson decay constant $f$. Although this approach can potentially produce reliable results, once it links MFA with chiral effective fields, this term is not present in the Lagrangian of Eq.~\ref{kaonLag}, indicating that the formalism is not fully self-consistent. Moreover, older studies about antikaon condensation use first-order phase transition~\cite{Glenbook,Banik2001,Banik2002} , while more recent ones use a second-order phase transition~\cite{Thapa2021,Thakur2025,Lopes2026PRD}. In general, the drop in the maximum mass for a potential depth $U_{\bar{K}} = -140$ MeV is around 0.5$M_\odot$ for the first-order phase transition, but only around 0.2$M_\odot$ for a second-order phase transition, or even smaller (See the supra-cited references).

Finally, the possible existence of a quarkyonic phase can produce results even more drastic for kaons than for heavy baryons. While for hyperons and $\Delta$'s the quarkyonic phase can shift their threshold, rendering them energetically unfavorable, in the case of kaons, the quarkyonic phase implies a quark-antiquark interaction that ultimately results in annihilation. The kaon condensate in a quarkyonic phase is therefore not merely suppressed but effectively prohibited.\\

\subsection{Symmetry group arguments for the pseudo-scalar mesons: two different approaches with two different results }

Now that we are aware of the limitations related to the possible onset of the (anti)kaons in dense stellar matter, let us assume that the MFA is still valid for all densities relevant in the neutron stars' core. Our task is now to determine the coupling constants of the kaons with the virtual mesons of the QHD. Mostly of the works uses the quark-isospin counting rule to fix the values of $g_{KK\omega}$ and $g_{KK\rho}$. However, it has no results for the value of $g_{KK\phi}$. To overcome this difficult some authors uses $\sqrt{2}g_{KK\phi} = 6.04$~\cite{Banik2001,Thapa2021}, a resulting coming from $\rho\to2\pi$ decay. 

However, this result leads to problems related to self-consistency. The $g_{KK\omega}$ and $g_{KK\rho}$ are linked to the value of $g_{NN\omega}$ and $g_{NN\rho}$, and are, thereby, model-dependent, while the $g_{KK\phi}$ is fully model-independent. An alternative is to realize that the kaons are members of an octet, with eigenvalues $M = |N~Y~I~I_3\rangle$, totally analogous to the baryon octet. Their representation is displayed below:

\begin{center}
\begin{tikzpicture}[scale=2, thick]

% Eixos
\draw[->, gray] (-1.5, 0) -- (1.5, 0) node[right] {$I_3$};
\draw[->, gray] (0, -1.5) -- (0, 1.5) node[above] {$Y$};

% Pontos
\fill (0.5,1) circle (2.2pt);    % p
\fill (-0.5,1) circle (2.2pt);   % n
\fill (1,0.0) circle (2.2pt);    % Σ⁺ (levemente acima)
\fill (-1,0.0) circle (2.2pt);   % Σ⁻ (levemente acima)
\fill (0.5,-1) circle (2.2pt);   % Ξ⁰
\fill (-0.5,-1) circle (2.2pt);  % Ξ⁻
\fill (0,0) circle (2.2pt);      % Σ⁰ e Λ⁰ no mesmo ponto

% Conexões corretas
\draw[gray] (0.5,1) -- (-0.5,1);          % p <-> n
\draw[gray] (0.5,1) -- (1,0.0) -- (0.5,-1);  % p -> Σ⁺ -> Ξ⁰
\draw[gray] (-0.5,1) -- (-1,0.0) -- (-0.5,-1); % n -> Σ⁻ -> Ξ⁻
%\draw[gray] (-1,0.1) -- (1,0.1);         % Σ⁻ <-> Σ⁺ (acima do centro)
\draw[gray] (-0.5,-1) -- (0.5,-1);       % Ξ⁻ <-> Ξ⁰

% Rótulos
\node at (0.70,1.15) {$K^+$};
\node at (-0.70,1.15) {$K^0$};
\node at (1.3,0.2) {$\pi^+$};
\node at (-1.3,0.2) {$\pi^-$};
\node at (0.65,-1.1) {$\bar{K^0}$};
\node at (-0.65,-1.1) {$K^-$};

% Rótulos no centro
\node at (-0.35,0.25) {$\pi^0$};
\node at (0.4,-0.25) {$\eta_8$};

% Título
\node at (0,1.8) {\textbf{Pseudoscalar Octet (SU(3))}};

\end{tikzpicture}
\end{center}

As can be seen, for each member of the baryon octet, there is a meson with the same eigenvalue. Therefore, the CG coefficients are the same. Two recent works have attempted to fix the CG coefficients using symmetry arguments~\cite{Lopes2026PRD,AthiraJCAP2025}.
Both came with the same general expressions for kaons, antikaons, and pions:

\begin{equation}
  g_{KK\rho} = g_{8(PS)},   \label{krho}
\end{equation}

\begin{equation}
  g_{\bar{K}\bar{K}\rho} =- g_{8(PS)}(1 - 2\alpha_{V(PS)}),  \label{kbarrho}
\end{equation}

\begin{equation}
  g_{\pi\pi\rho} = 2g_{8(PS)}\alpha_{V(PS)},  
\end{equation}

\begin{eqnarray}
  g_{KK\omega} = g_{1(PS)}\cos\theta_{V} + g_{8(PS)}\sin\theta_{V} \frac{1}{3}\sqrt{3}(4\alpha_{V(PS)} -1),   \label{komega}
\end{eqnarray}

\begin{eqnarray}
  g_{\bar{K}\bar{K}\omega} = g_{1(PS)}\cos\theta_V   - g_{8(PS)}\sin\theta_V \frac{1}{3}\sqrt{3}(1 + 2\alpha_{V(PS)}).  \label{kbaromega}
\end{eqnarray}

\begin{eqnarray}
  g_{\pi\pi\omega} =  g_{1(PS)}\cos\theta_{V} + g_{8(PS)}\sin\theta_{V} \frac{2}{3}\sqrt{3}(1 - \alpha_{V(PS)}),
\end{eqnarray}
where the subscript (PS) was added to differentiate the coupling with the pseudo-scalar meson octet with the baryon octet. Both papers also agree with the value of $\theta_V = 35.264$º, since it is related to the nature of the $\omega$ and $\phi$ mesons, leaving only $z_{(PS)}$ and $\alpha_{V(PS)}$ as free parameters. 

The similarities end here, and Ref.~\cite{Lopes2026PRD} and Ref.~\cite{AthiraJCAP2025} follow different paths to fix the values of  $z_{(PS)}$ and $\alpha_{V(PS)}$. To understand both paths, let us define the energy eigenvalue of S-wave kaons $(K)$ and antikaons $(\bar{K})$: 

\begin{eqnarray}
\omega_{0(K)} = m^{*}_K + g_{KK\omega}\omega_0+ \frac{\tau_3}{2}g_{KK\rho}\rho_0 + g_{KK\phi}\phi_0, \nonumber \\
\omega_{0(\bar{K})} = m^{*}_{\bar{K}} - g_{\bar{K}\bar{K}\omega}\omega_0+ \frac{\tau_3}{2}g_{\bar{K}\bar{K}\rho}\rho_0 - g_{\bar{K}\bar{K}\phi}\phi_0, \label{kaonantikaoneigen}
\end{eqnarray}

In Ref.~\cite{AthiraJCAP2025}, the authors assume that the minus sign for the antikaons is independent of the value of $g_{\bar{K}\bar{K}\omega}$, coming exclusively from the solution of Eq.~\ref{kaonLag} (indeed, the reader is invited to solve Eq.~\ref{kaonLag} following the techniques developed in Part I~\cite{LopesUNIVERSE2025},  and obtain both expressions in Eq.~\ref{kaonantikaoneigen}). Therefore, in the SU(6) they imposed:

\begin{equation}
g_{KK\omega} = g_{\bar{K}\bar{K}\omega}.
\end{equation}

Furthermore, as pions have no strange quarks, it is expected that $g_{\pi\pi\phi} =0$ in the SU(6) limit. Within these considerations, the authors propose that in the SU(6) limit.

\begin{equation}
 \alpha_{V(PS)} = 0, \quad \mbox{and} \quad z_{(PS)} = \frac{\sqrt{3}}{2\sqrt{2}},
\end{equation}
and the strength in relation to the coupling of the nucleon reads:

\begin{equation}
\frac{g_{K\bar{K}\omega}}{g_{NN\omega}}
=
\frac{\sqrt{6}-z_{(PS)}(1+2\alpha_{(PS)})}
{\sqrt{6}+z(4\alpha_V-1)} \, .
\end{equation}

Finally, the authors of Ref.~\cite{AthiraJCAP2025} fix $z$ and $z_{(PS)}$ with their SU(6) values and assumes that $\alpha_{V}$ is a multiple of $\alpha_{V(PS)}$. The authors use $\alpha_V = 1.0$, 2.0, and 0.5 times the value of $\alpha_V{(PS)}$, with $\alpha_{V(PS)}$ being the only true free parameter.  The consequences of this approach are the following:\\

\begin{itemize}
    \item The quark-isospin counting rule is never recovered, even in the SU(6) limit. The same is true for Sakurai's proposal.

    \item Moving from SU(6) implies that $|g_{KK\omega}|~\neq~|g_{\bar{K}\bar{K}\omega}|$. As kaons and antikaons form a particle-antiparticle pair, this is a very unorthodox result. The analogy in quantum electrodynamics is that the electric charge of the positron presents a different magnitude of the electric charge of the electron.

    \item Even in the SU(6) limit, the strength of the $g_{\pi\pi\omega}$ is twice the strength of $g_{\bar{K}\bar{K}\omega}$. This results in a strong attractive channel for the $\pi^-$, in disagreement with the results pointed in Ref.~\cite{DICKHOFF1983,OhnishiPRC2009}. Indeed, the combined effect of the low mass of the pions with the $\omega$ dominance would make the presence of $\pi^-$ far more favorable than the presence of $K^-$. 
\end{itemize}

On the other hand, Ref.~\cite{Lopes2026PRD} combines the SU(3) flavor symmetry with the G-parity. G-Parity is a multiplicative quantum number conserved in strong interactions. A detailed discussion about it is beyond the scope of the present work, but the interested reader is referred to Section 5.6 of Ref.~\cite{GreinerSymmetries}.

The key point is that the G-parity gives the interaction of an antiparticle with
a determined meson as a function of the interaction of the
particle with the same meson. For a mesonic field M, we have:

\begin{equation}
G \, |M\rangle = \pm |M\rangle .  \label{Gparity}
\end{equation}

The $\rho$ meson have a positive G-parity~\cite{PDG2020,GreinerSymmetries}, therefore:

\begin{equation}
g_{\bar{K}\bar{K}\rho} =  g_{KK\rho}. 
\end{equation}

As the G-parity is conserved in all strong interactions, this implies that $\alpha_{V(PS)} = 1$.

Now, the $\omega$ and $\phi$ mesons have a negative G-Parity. Implying:

\begin{equation}
 g_{\bar{K}\bar{K}\omega} =  - g_{KK\omega}; \quad    g_{\bar{K}\bar{K}\phi} = - g_{KK\phi}. \label{Gomega}
\end{equation}

As we fixed $\alpha_{V(PS)}$, comparing with Eq.~\ref{komega} and Eq.~\ref{kbaromega}, we obtain $g_{1(PS)} =0$, or equivalently $z_{(PS)}^{-1} = 0 $. This result also implies $g{\pi\pi\omega} = 0$. Another desirable result obtained with this approach is $g_{\pi\pi\phi} = 0$, as pions do not contain strange quarks. Notice that all these results are general, and not restricted to the SU(6) limit.

With those definitions, the strength of the kaon interaction concerning the nucleonic one is:

\begin{equation}
    \frac{g_{KK\omega}}{g_{NN\omega}} =  \frac{g_{8(PS)}}{g_8} \bigg (\frac{3}{5 +4\alpha_V} \bigg ); \quad \frac{g_{KK\rho}}{g_{NN\rho}} =  \frac{g_{8(PS)}}{g_8}.
\end{equation}

Now, in the SU(6) limit ($\alpha_V \to 1.00)$, we have already seen that the quark-isospin counting rule is restored for the members of the baryon octet and baryon decuplet. Assuming this is also true for the members of the pseudoscalar meson octet implies:

\begin{equation}
 \frac{g_{8(PS)}}{g_8} = 1.00.
 \end{equation}

Thereby, the coupling with the pseudoscalar mesons, the baryon octet, and the baryon decuplet with the vector meson octet, depends on a single parameter, $\alpha_V$, in a fully unified scheme. The consequences of this approach are:\\

\begin{itemize}
    \item The quark-isospin counting rule and Sakurai's proposal are restored in the SU(6) limit.

    \item The magnitude of the strength of kaons and antikaons is always equal:  $|g_{KK\omega}| = |g_{\bar{K}\bar{K}\omega}|$, consistent with the concept of particle-antiparticle pair.

    \item $g_{\pi\pi\omega} =0$ for all values of $\alpha_V$, preventing an attractive channel, remaining consistent with the results presented in Refs.~\cite{DICKHOFF1983,OhnishiPRC2009}. The $g_{\pi\pi\phi}$ is also zero for all values of $\alpha_V$, a desirable result.
\end{itemize}

%\begin{widetext}
\begin{center}
\begin{table}[b!]
\begin{center}
\caption{Coupling constants for different values of $\alpha_v$, within the SU(3) symmetry group for the antikaons.  The Sakurai's proposal is recovered in the SU(6), and the potential depth is fixed to $U_{\bar{K}} = -140$ MeV.  \label{TL6}}
\scalebox{0.90}{
\begin{tabular}{|c|cccc|}
\hline
$\alpha_V$  &   1.00&0.75&0.50&0.25   \\
\hline
 $g_{\bar{K}\bar{K}\omega}/g_{NN\omega}$ & -0.333 & -0.375 & -0.429 & -0.50\\ 
 $g_{\bar{K}\bar{K}\phi}/g_{NN\omega}$  & -0.471 & -0.530 & -0.607 & -0.707 \\
$g_{\bar{K}\bar{K}\sigma}/g_{NN\sigma}$ & 0.234 & 0.201 & 0.160  & 0.105\\ 
\hline
\end{tabular}}
\end{center}
\end{table}
\end{center}
%\end{widetext}

Due to the consequences of each approach, I choose to follow Ref.~\cite{Lopes2026PRD} instead of Ref.~\cite{AthiraJCAP2025}. Nevertheless, it is worth emphasizing that both parametrizations are fully consistent with the SU(3) flavor symmetry, as the differences lie only in the free parameters, $\alpha_{V(PS)}$ and $z_{(PS)}$.

The last question is related to the potential depth. As pointed out earlier, unlike hyperons and even the $\Delta$'s, antikaons have a huge uncertainty about its value, lying in $-180~\mbox{MeV}~<U_{\bar{K}}<-60$ MeV. Such uncertainty is high enough to make quantitative and qualitative changes in the microscopic and macroscopic properties of neutron stars. Here I use $U_{\bar{K}} = -140$ MeV, a value well-accepted in the literature. The reader is, nevertheless, invited to investigate the effects of changing this value. Therefore, following Ref.~\cite{Lopes2026PRD}, the coupling constants for the antikaons for different values of $\alpha_V$ are presented in Tab.~\ref{TL6}.

\subsection{Role of the virtual mesons in the (anti)kaons condensate}

The negative value of the $g_{\bar{K}\bar{K}\omega}$ can lead an inattentive reader to assume that the interaction between antikaons mediated by the exchange of the $\omega$ meson is attractive. However, this is not what actually occurs.
The QHD, in its full glory, simulates the strong interaction via the exchange of massive bosons; however, in MFA, the particles behave like free particles within a background classical field. To avoid possible sources of confusion, we shall discuss the role played by each meson in matter containing an antikaon condensate.

\begin{itemize}
    \item The  $\omega$ meson: The interaction of the antikaons via the $\omega$ meson channel is actually repulsive. If we remove the baryons from the equations, the value of the $\omega$ field becomes negative, and therefore the energy eigenvalue increases. This is analogous to the $\phi$ meson and the hyperons. Although the couplings $g_{YY\phi}$ are all negative, the $\phi_0$ field is also negative, and we have a repulsion, and consequently a stiffening of the EOS.  In MFA with nucleons and potentially hyperons and $\Delta$'s, all having positive values for the $g_{BB\omega}$, therefore the value of the $\omega$ field is positive. The effect of a negative $g_{\bar{K}\bar{K}\omega}$ is to weaken the positive $\omega$ field. Now, the lowered $\omega$ field increases the potential chemical of the baryons, but reduces the potential chemical of the antikaons.   

    \item The $\phi$ meson: The interaction of the antikaons vai the $\phi$ meson is also repulsive. However, the $\phi$ meson does not couple with the nucleon, and therefore in matter containing only nucleons and antikaons (NK matter) the $\phi$ field is negative. The presence of hyperons only strengthens the repulsion because the couplings  $g_{YY\phi}$ are also negative. The role of the $\phi$ meson is to strengthen the repulsive interaction among all hadrons, thereby increasing the values of their chemical potentials.

    \item The $\rho$ meson: The $\rho$ meson has a positive G-parity, and couples to the isospin. It is repulsive for all kaons, but the interaction also depends on the sign of the Pauli matrix $\tau_3$ ($+1$ for $\bar{K}^0$ and $-1$ for $K^-$). 

    \item The $\sigma$ field. It is assumed that the G-parity of the $\sigma$ meson is positive because it reflects two-pion exchange (see a discussion in part I~\cite{LopesUNIVERSE2025}), but the true nature of the $\sigma$ meson is not well understood yet, and it is only loosely based on the $f(500)$ resonance~\cite{PDG2020}. Nevertheless, the $\sigma$ field is always attractive. 
        
\end{itemize}

\subsection{Numerical Results}

\begin{figure*}[ht]
\begin{tabular}{ccc}
\centering % \begin{center}/\end{center} takes some additional vertical space
\includegraphics[scale=.54, angle=270]{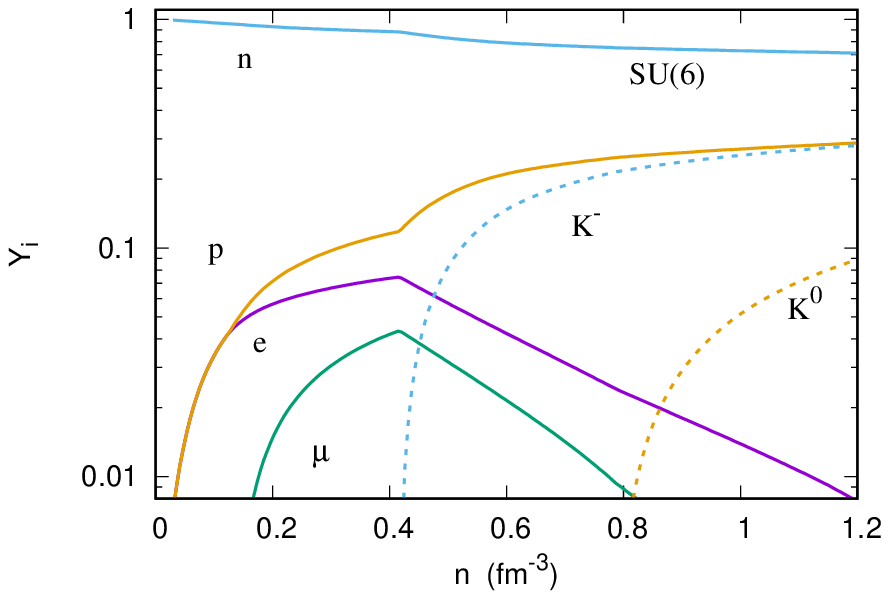} &
\includegraphics[scale=.54, angle=270]{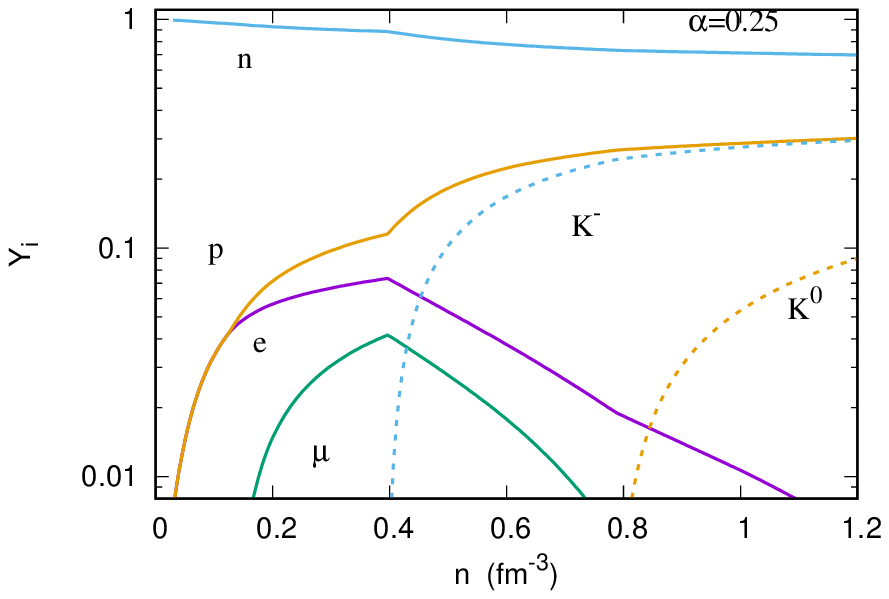} \\
\end{tabular}
\caption{Particle population for NK matter. (\textbf{a}) $\alpha_V = 1.0$ (\textbf{b}) $\alpha_V = 0.25$.\label{FL11}}
\end{figure*}

We now discuss the microscopic and macroscopic properties of neutron stars with kaon condensation. There are three possibilities: neutron stars with nucleons and kaons (NK matter), nucleons, hyperons, and kaons (NYK) and finally, nucleons, hyperons, $\Delta$'s and kaons (NYDK matter). Let us begin with the particle population for NK matter. I present in Fig.~\ref{FL11} for the extreme cases,  $\alpha_V = 1$ (SU(6)), and $\alpha_V = 0.25$.

%%%%%%%%%%%%%%%%%

%%%%%%%%%%%%%%%%%
\begin{figure*}[h!]
\begin{tabular}{ccc}
\centering % \begin{center}/\end{center} takes some additional vertical space
\includegraphics[scale=.468, angle=270]{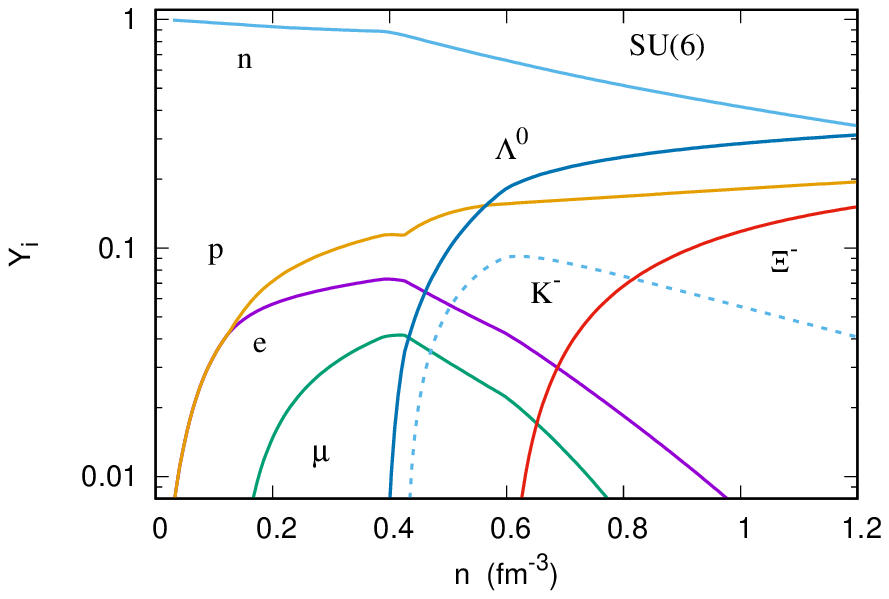} &
\includegraphics[scale=.468, angle=270]{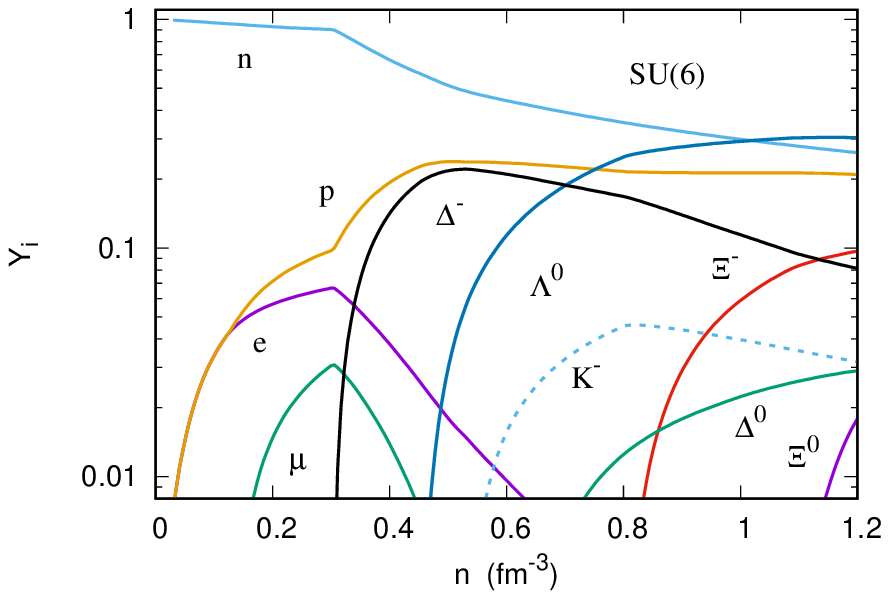} \\
\includegraphics[scale=.468, angle=270]{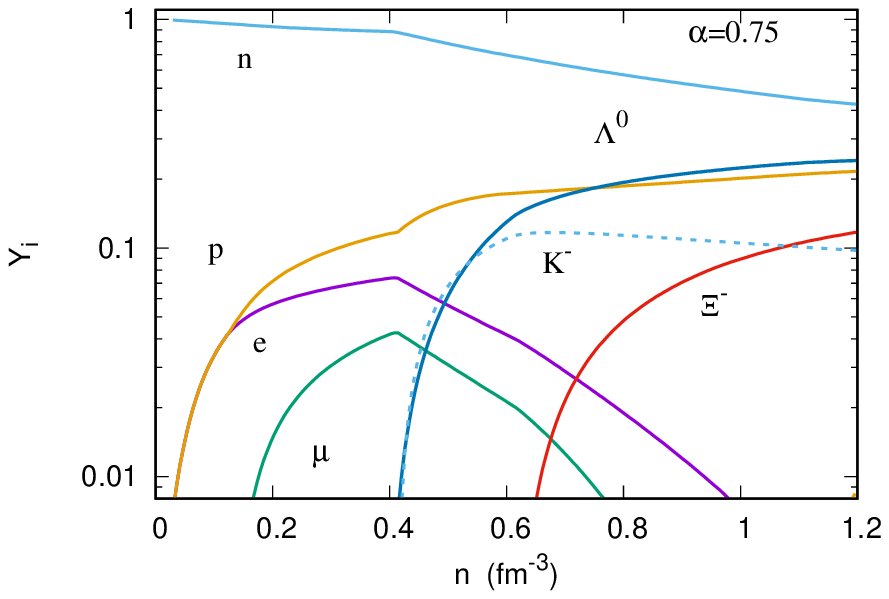} &
\includegraphics[scale=.468, angle=270]{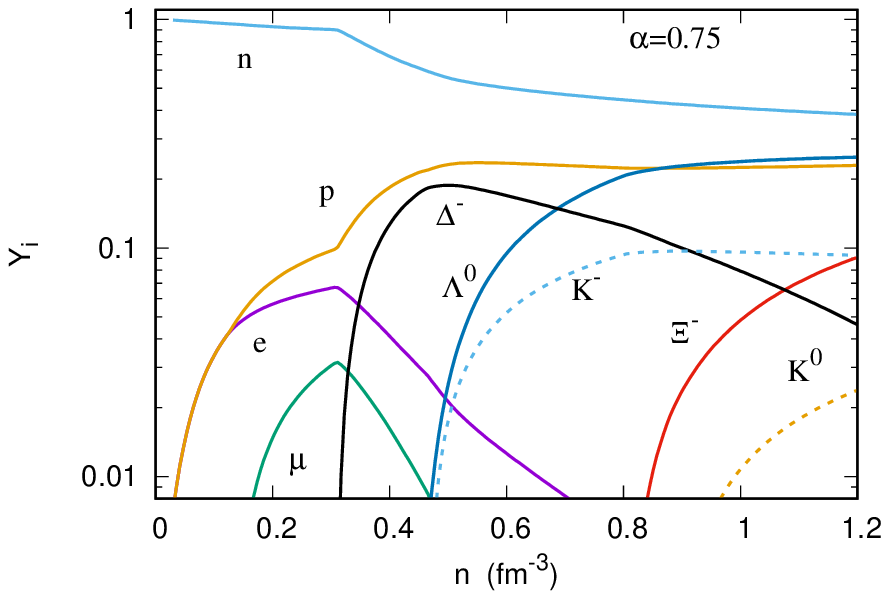}\\
\includegraphics[scale=.468, angle=270]{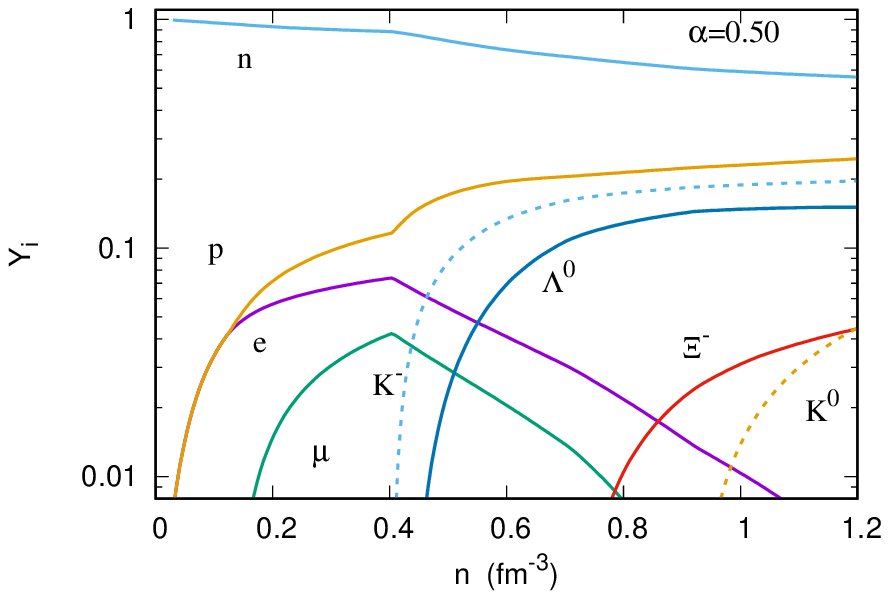} &
\includegraphics[scale=.468, angle=270]{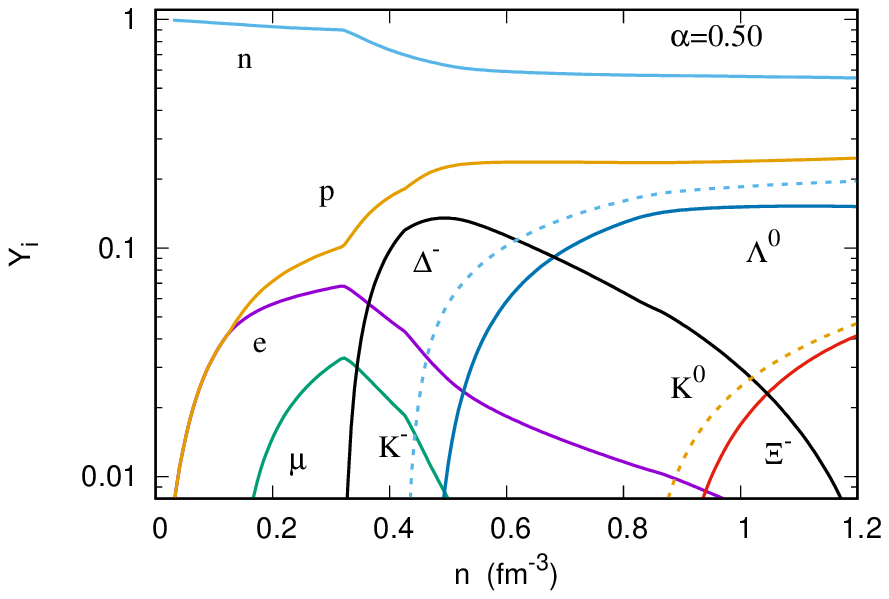}\\
\includegraphics[scale=.468, angle=270]{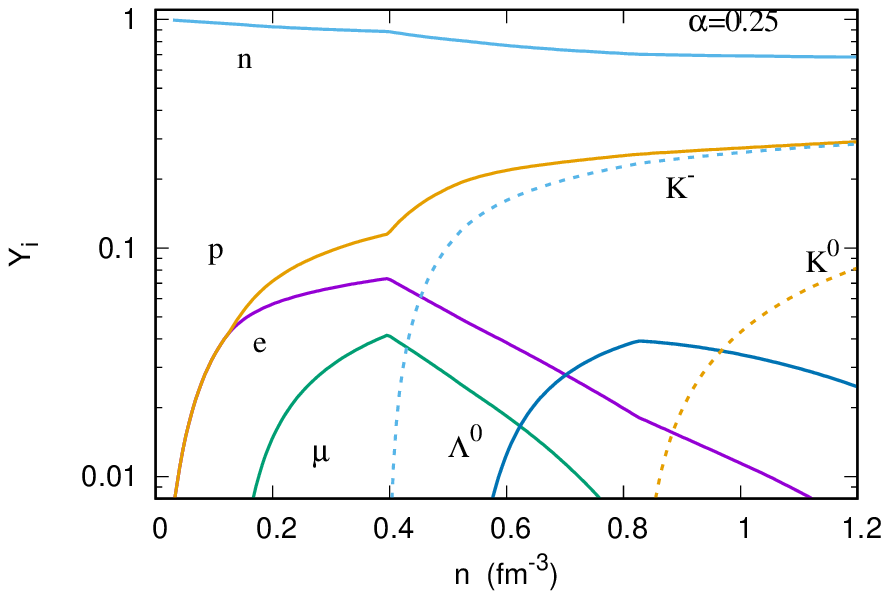} &
\includegraphics[scale=.468, angle=270]{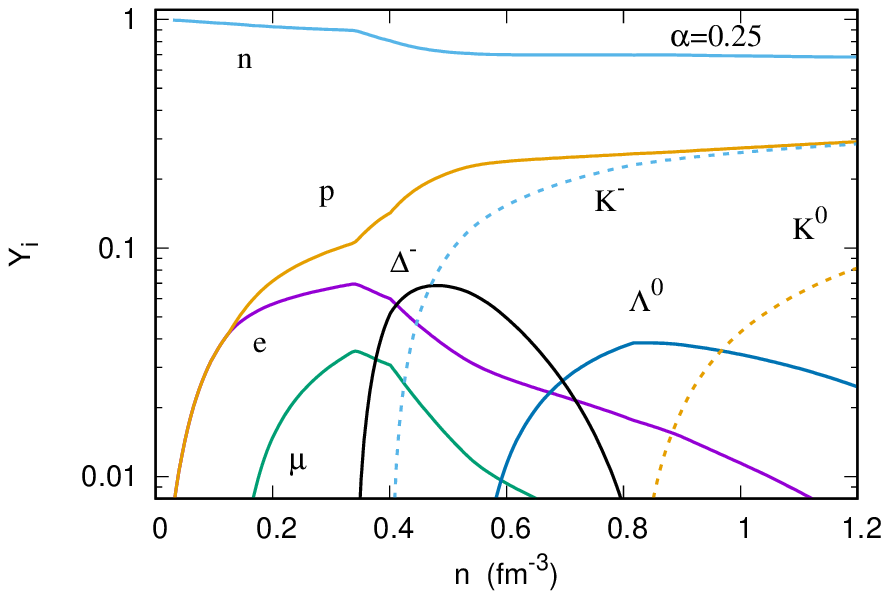}\\
\end{tabular}
\caption{Particle population for different values of $\alpha_V$ for NYK matter (left) and NYDK matter (right).\label{FL12}}
\end{figure*}

The main effect of the kaons is to reduce the lepton population and increase the proton population, due to the lowering of the electron chemical potential. In relation to changes in $\alpha_V$, we can notice that the differences are small. Since the $\omega$ field reduces the (anti)kaons' chemical potentials as we reduce the value of $\alpha_V$, we increase the population of kaons. However, this effect is minimized once the reduction of $\alpha_V$ also increases the magnitude of the repulsive $g_{\bar{K}\bar{K}\phi}$, and reduces the coupling with the attractive $\sigma$ meson. Visually, we observe only a small increase in the kaons, followed by a reduction in the lepton population.

%%%%%%%%%%%%%%%%%
\begin{figure*}[b!]
\begin{tabular}{ccc}
\centering % \begin{center}/\end{center} takes some additional vertical space
\includegraphics[scale=.502, angle=270]{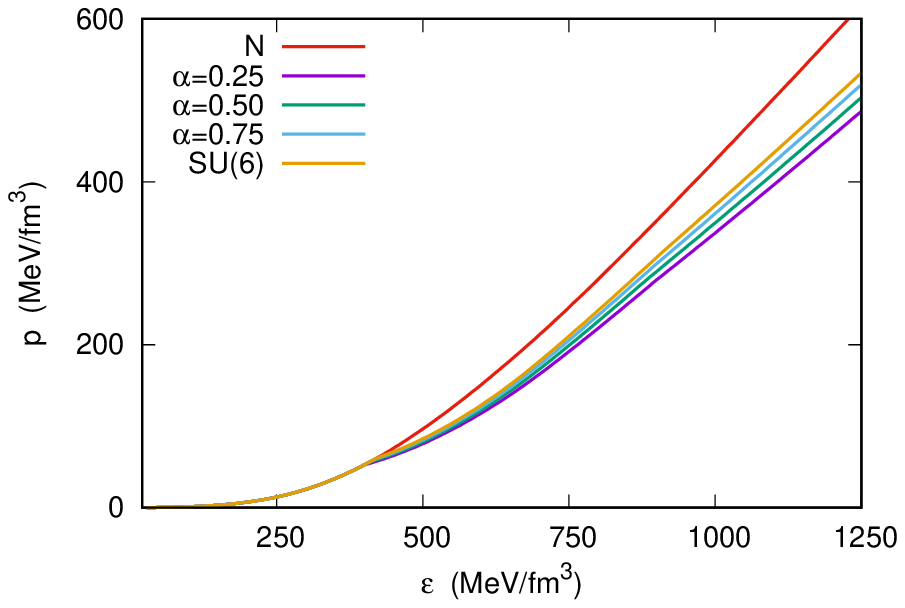} &
\includegraphics[scale=.502, angle=270]{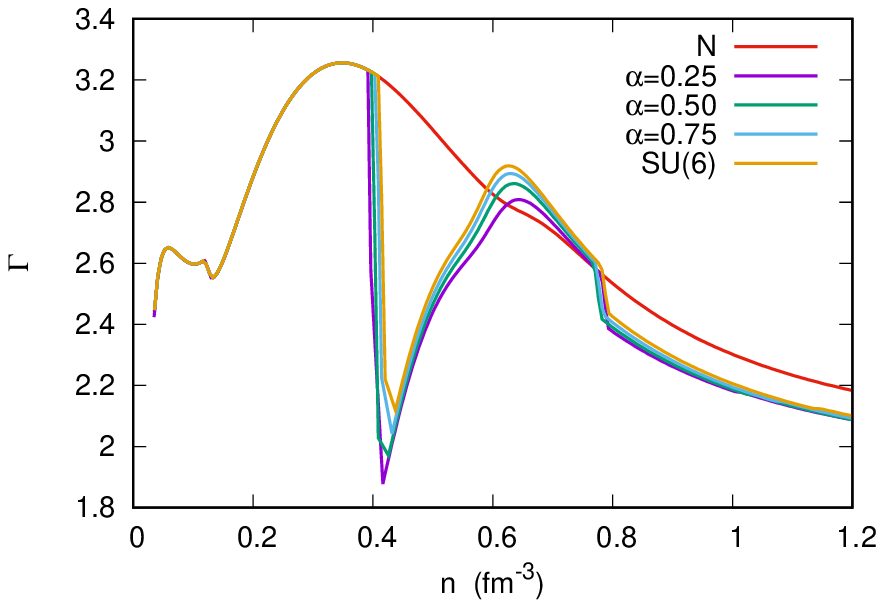} \\
\includegraphics[scale=.502, angle=270]{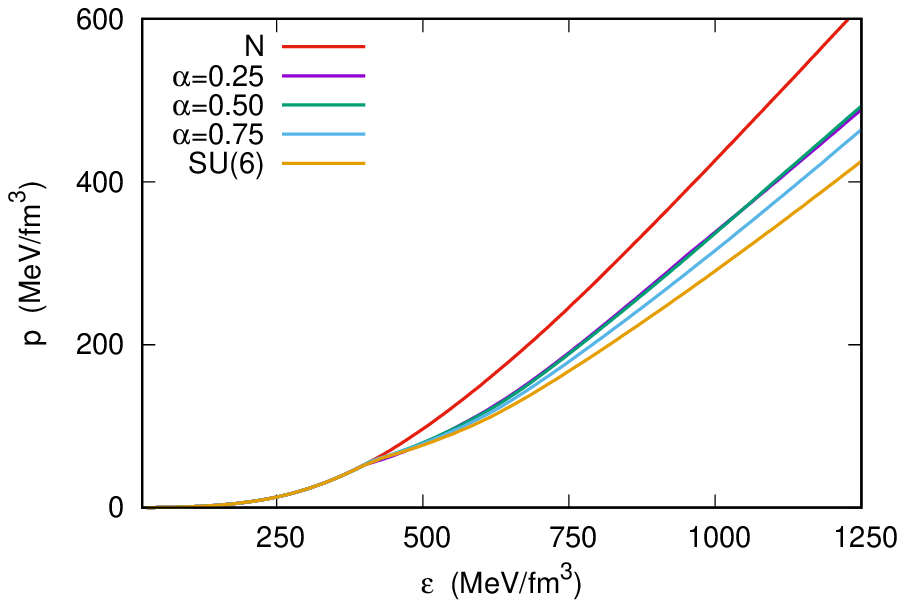} &
\includegraphics[scale=.502, angle=270]{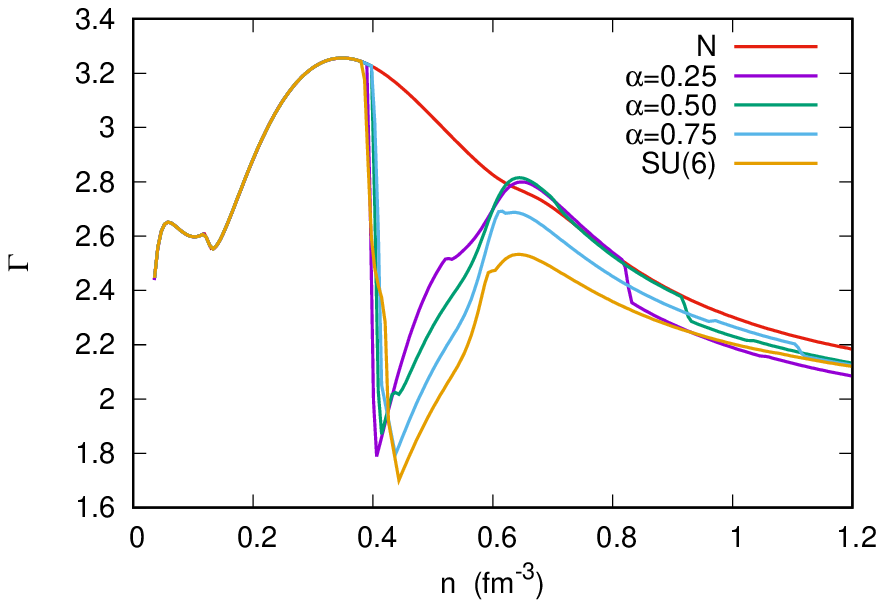}\\
\includegraphics[scale=.502, angle=270]{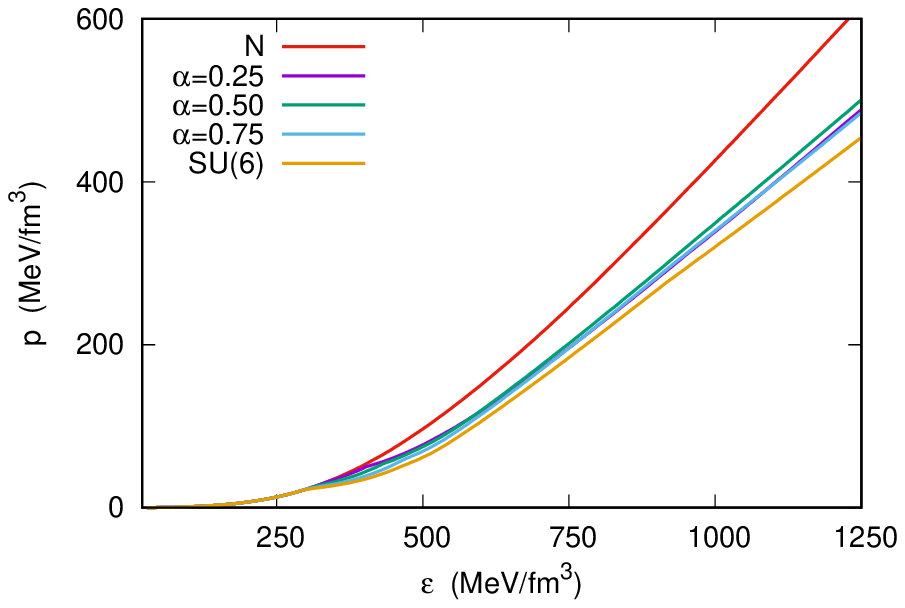} &
\includegraphics[scale=.502, angle=270]{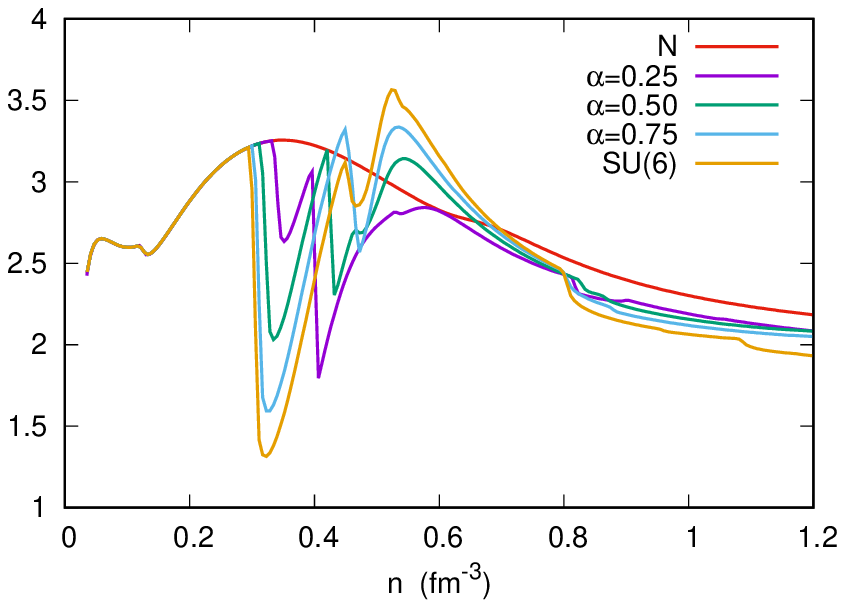}\\
\end{tabular}
\caption{EOSs (left) and the adiabatic index (right) for the NK (top), NYK (middle), and NYDK (bottom) matter within different values of $\alpha_V$.\label{FL13}}
\end{figure*}

The situation becomes more intricate in the NYK and NYDK matter, where the exotic degrees of freedom compete with one another, as illustrated in Fig.~\ref{FL12}. In this case, antikaons, hyperons, and $\Delta$'s compete for the composition of the matter. Nonetheless, as a direct consequence of the dependence of their vector couplings on the value of $\alpha_V$, as it decreases, the coupling $g_{YY\omega}$ increases, while the coupling $g_{\bar{K}\bar{K}\omega}$ becomes progressively more negative. Consequently, two concurrent effects act in the same direction: the baryons become less energetically favored, whereas the antikaons become increasingly favored. It can be noticed that the kaon population is small for SU(6) parametrization, but it dominates over the hyperons and $\Delta$'s for $\alpha_V = 0.25$.

We can also notice that the $\Xi^-$ is the most sensitive exotic baryon to the presence of the kaons. The $\Lambda^0$ only suffers a significant decrease for $\alpha_V = 0.25$, while the population of $\Delta^-$ is essentially the same as presented in Fig.~\ref{FL9}.
The reader is encouraged to contemplate the implications of this result based on the nature and values of $\chi_\omega$ and $\chi_\phi$.

Still about the microscopic properties, I now discuss in Fig.~\ref{FL13} the EOSs (left) and the adiabatic index (right) for the NK (top), NYK (middle), and NYDK (bottom) matter. 

For NK matter, we notice, as expected, an inverted role played by  $\alpha_V$. As we move away from the SU(6), the EOSs becomes progressivaly soften, once the value of $g_{\bar{K}\bar{K}\omega}$ is negative. Although this effect is minimized due to the strong $g_{\bar{K}\bar{K}\phi}$ coupling, it is not enough to prevent the soffetning of the EOS.  In the case of the adiabatic index, the onset of $K^-$ causes a steep drop in  $\Gamma$ around 0.4 fm$^{-3}$. Lower the value of $\alpha_V$, deeper is the drop. A second drop around 0.8 fm$^{-3}$ indicates the onset of $\bar{K}^0$'s.

For the NYK matter, a competition arises between antikaons and hyperons. Although the EOS becomes progressively stiffer as $\alpha_V$ is reduced, this stiffening is significantly mitigated by the presence of antikaons. Indeed, the EOSs obtained for $\alpha_V = 0.50$ and 0.25 are remarkably similar. The adiabatic index reflects the onset of each new degree of freedom. Since the thresholds for $K^-$ condensation and $\Lambda^0$ appearance are very close, the corresponding drops in the adiabatic index are nearly indistinguishable. The only exception occurs for $\alpha_V = 0.25$, for which the $\Lambda^0$ population is strongly suppressed.

Concerning the NYDK matter, a similar situation is observed. However, although the SU(6) parametrization still yields the softest EOS, the stiffest EOS is obtained for $\alpha_V = 0.50$ rather than for $\alpha_V = 0.25$. Since the presence of $\Delta$ baryons has little impact on the maximum mass, this behavior can be attributed to a smaller antikaon content found for $\alpha_V = 0.50$. The adiabatic index exhibits an initial drop associated with the onset of the $\Delta^-$, followed by a second drop related to the nearly simultaneous appearance of the $\Lambda^0$ and $K^-$. As $\alpha_V$ decreases away from the SU(6) limit, the first drop becomes progressively shallower, whereas the second becomes more pronounced. This behavior reflects the suppression of the $\Delta^-$ population together with the increased favorability of antikaon condensation.

%%%%%%%%%%%%%%%%%
\begin{figure*}[b!]
\begin{tabular}{ccc}
\centering % \begin{center}/\end{center} takes some additional vertical space
\includegraphics[scale=.54, angle=270]{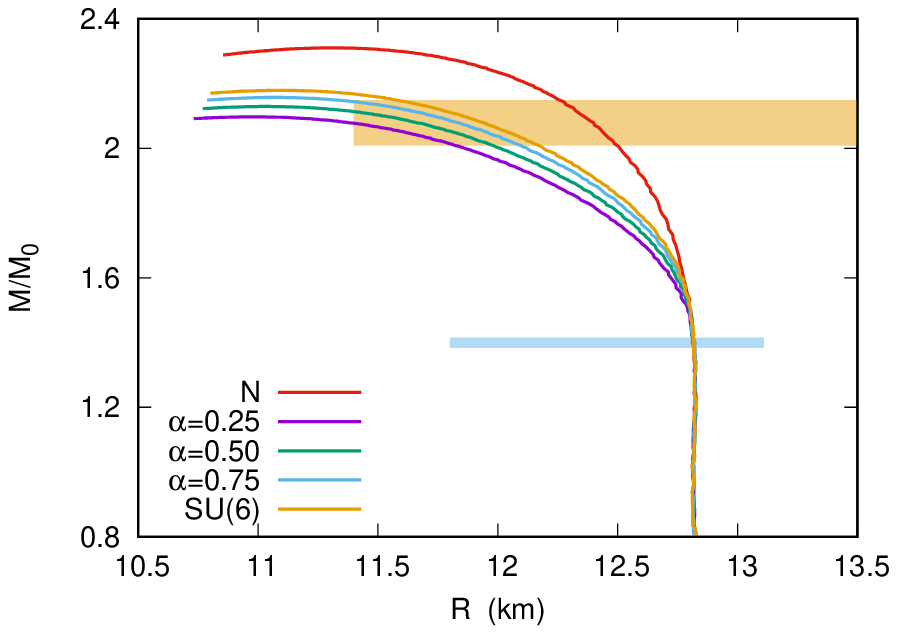} &
\includegraphics[scale=.54, angle=270]{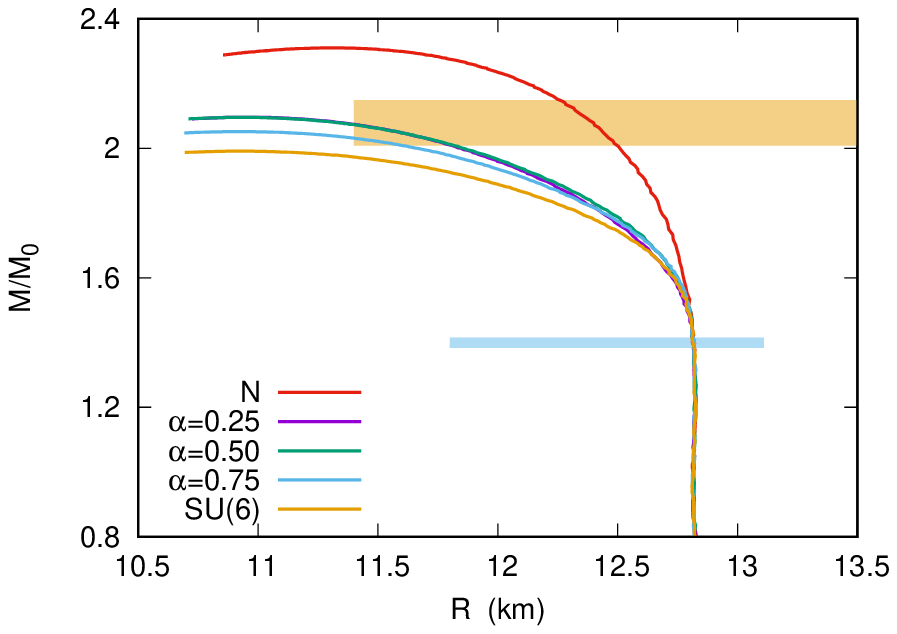} \\
\includegraphics[scale=.54, angle=270]{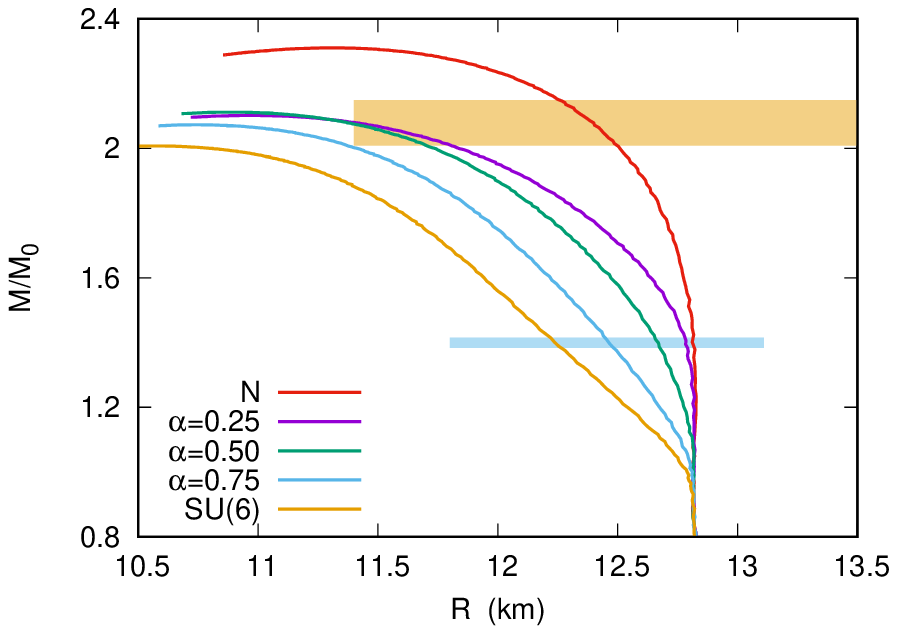} 
\end{tabular}
\caption{OV solutions and astrophysical constraints related to the PSR J0740+6620 and the canonical star for different values of $\alpha_V$ for: (\textbf{a}) NK matter, (\textbf{b}) NYK matter, (\textbf{c}) NYDK matter.\label{FL14}}
\end{figure*}

Ultimately, to assess the impact of antikaon condensation on the macroscopic properties of neutron stars, the OV equations are solved. I present in Fig.~\ref{FL14} the results for (\textbf{a}) NK matter, (\textbf{b}) NYK matter, (\textbf{c}) NYDK matter.

In the case of $NK$ matter, we have the peculiar case where the  SU(6) predicts the highest neutron star mass, $M =2.18M_\odot$, corresponding to a reduction of $0.13,M_\odot$ relative to the purely nucleonic case. For comparison, using the same antikaon potential depth, $U_{\bar{K}}=-140$ MeV, Ref.~\cite{Thakur2025} reported a reduction of $0.19,M_\odot$ for the TM1e model in the absence of the cutoff potential, and only $0.06,M_\odot$ for $f_c=0.6$ (see Eq.~\ref{scut}). It is worth recalling that the calculations presented here employ $f_c=0.85$. For $\alpha_V = 0.25$, the maximum mass is 2.10$M_\odot$. The different path followed by Ref.~\cite{AthiraJCAP2025} leads to a decrease varying from 0.22$M_\odot$ to 0.02$M_\odot$. In relation to observational constraints, we can see that all values of $\alpha_V$ are in agreement with the PSR J0740+6620 pulsar.

For NYK matter, we return to the standard situation, where the SU(6) has the softest EOS, and therefore, the lower maximum mass. A maximum mass of 1.99$M_\odot$ is obtained, virtually equal to the NY matter. This is in agreement with other results found in the literature. For instance, in Ref.~\cite{Thapa2021}, a reduction inferior to 0.01$M_\odot$ was obtained for the same potential depth, $U_{\bar{K}}=-140$ MeV, but with a density-dependent model. It can also be noticed that $\alpha_V = 0.50$ and $\alpha_V = 0.25$ have the same maximum mass of 2.10$M_\odot$. Unsurprisingly, this is also the maximum mass of NK matter for $\alpha_V = 0.25$, due to the strong hyperon suppression. Except for the SU(6) symmetry, all values of $\alpha_V$ can describe the PSR J0740+6620 pulsar.

Last but not least, in the NYDK matter, in the SU(6) we obtain the same maximum mass of 2.01$M_\odot$, reflecting the very small amount of $K^-$. This is again consistent with Ref.~\cite{Thapa2021}, which found a drop of only 0.01$M_\odot$. As we move away from SU(6), the maximum mass increases to 2.11$M_\odot$ and 2.10$M_\odot$ for $\alpha_V$ equals to 0.50 and 0.25, respectively, reflecting the dominance of the $K^-$ for lower values of $\alpha_V$. Here, only $\alpha_V =0.50$ and $\alpha_V = 0.25$ can fulfill the PSR J0740+6620 constraint.

An important feature that must be emphasized in this discussion is how the presence of the kaons strongly compromises the stiffening of the EOS by simply reducing the $\alpha_V$. While for NY and NYD matter, it was possible to reproduce stars with masses above 2.20$M_\odot$, the presence of kaons reduces this limit to around 2.10$M_\odot$, which is essentially the same value for NK matter within $\alpha_V = 0.25$. Additionally, kaons do not affect 1.4$M_\odot$ neutron stars. The main results of this section are summarized in Tab.~\ref{T7}.

%\begin{widetext}
\begin{center}
\begin{table}[ht]
\begin{center}
\caption{Neutron stars' main properties within antikaon condensation for different values of $\alpha_V$. The presence of kaons does not affect the radius of the canonical star.\label{T7}}
\scalebox{0.90}{
\begin{tabular}{|c|cccc|}
\hline
$\alpha_V$  & Composition  &$M_{max}/M_\odot$ & R (km) & $n_c$ (fm$^{-3}$)   \\
\hline
 Nucleonic & N      & 2.31 & 11.31   & 0.94   \\

 0.25  & NK     & 2.10 & 10.98  & 1.06    \\
 
0.50   &NK     & 2.13 & 11.03  & 1.05   \\

0.75 & NK & 2.15 & 11.07  & 1.03   \\

SU(6)  &NK  &  2.18 & 11.11  & 1.01    \\
\hline
 0.25  & NYK    & 2.10 & 10.96  & 1.07    \\
 
0.50   &NYK &     2.10 & 10.94  & 1.07    \\

0.75 & NYK & 2.05 & 10.93  & 1.07   \\

SU(6)  &NYK &   1.99 & 10.93  & 1.08    \\
\hline
 0.25  & NYDK     & 2.10 & 10.98  & 1.06    \\
 
0.50   &NYDK&     2.11 & 10.89  & 1.06    \\

0.75 & NYDK & 2.07 & 10.75  & 1.09   \\

SU(6)  &NYDK &   2.01 & 10.58  & 1.14    \\
\hline
\end{tabular}}
\end{center}
\end{table}
\end{center}
%\end{widetext}

\section{QHD in mean field approximation: extensions and beyond}

In the present work, I discuss the results within QHD in MFA. In the present model, non-linear terms are added to make corrections and ensure a better bound between the theoretical nuclear matter and the experimental one.  It is worth, nevertheless, pointing out that this is not the only possible path to be followed. \\

Still within the MFA, an alternative route is to include coupling constants that are not constants at all, but exhibit density-dependent behaviour. In this case, an {\it ad hoc} dependence of the coupling constants on the density is introduced. The complexity of the calculations increases in these models, since it is mandatory to introduce rearrangement terms to ensure internal consistency (see Refs.~\cite{TYPEL1999,Lala2005PRC}).

A step further is to go beyond MFA. The first step toward this is called the relativistic Hartree approximation (RHA). In this formulation, the contribution of the Dirac sea is explicitly taken into account (sometimes called vacuum fluctuation corrections). In this approach, the contribution from the Dirac sea is divergent, and renormalization is needed~\cite{CHIN1977AP,HEIDE1991PLB,FURNSTAHL1997NPA}.  In general, we have $\mbox{RHA}~\supset~\mbox{MFA}$. Furthermore, in this context, the MFA is sometimes referred as the no-sea approximation~\cite{FURNSTAHL1997NPA}. As shown in Ref.~\cite{Serot_1992}, we have:

\begin{equation}
 \mathcal{L}_{RHA} =  \mathcal{L}_{MFA} + \delta\mathcal{L}   ,
\end{equation}
where $\delta\mathcal{L}$ is the renormalized Dirac sea contribution.

The second step beyond MFA is the so-called Hartree-Fock formalism. This is obtained by including the meson emission and reabsorption (‘exchange’) graphs in the baryon self-energy~\cite{Serot_1992}. The Fock terms are explicitly momentum-dependent and admit tensor couplings. 

In relation to neutron star matter, it was shown in Ref.~\cite{JiaJieLiEPJA2018} that the Hartree-Fock terms lead to a soft EOS when compared with RHA and MFA. The softening of the EOS can be attributed to a negative contribution to the pressure due to the exchange terms, and to the meson-hyperon tensor couplings mediating additional attraction among hyperons compared to the models based on Hartree self-energies. Furthermore, it was shown in Ref.~\cite{Zhao_2015JPG} that the inclusion of the Fock terms reduces the kinetic part of the symmetry energy and may lead to negative values in the supranuclear density region.

The story is not over yet. A third step is called Dirac-Brueckner-Hartree-Fock theory (see Ref.~\cite{Serot_1992}  and the references therein). Such an approach gives us two-body correlations, resulting in a relativistic ladder-approximated Bethe-Salpeter equation~\cite{Bethe-Salpeter1951PR}. However, this theory lies far, far beyond the scope of the present work and is mentioned here solely to stimulate the curiosity of those readers interested in pursuing the subject further.

Beyond QHD models can be seen in Ref.~\cite{BURGIO2021PPNP} and the references therein.\\

%%%%%%%%%%%%%%%%%%%%%%%%%%%%%%%%%%%%%%%%%%
\section{Final Remarks}

In this work, I extend the results presented in Part I~\cite{LopesUNIVERSE2025}, discussing how to introduce new exotic degrees of freedom. The main results can be summarized as follows:

\begin{itemize}
    \item After a quick review of the QHD formalism, I show how to introduce the muon in a generalized chemical equilibrium. The results of neutron stars with $npe$ and $npe\mu$ matter are discussed. 

    \item Using the same arguments for the muon, I then show how to introduce the $\Lambda^0$ in chemically stable matter. However, unlike the muons that are easily handled because they are introduced as a free Fermi gas, the properties of a matter containing $\Lambda^0$ are strongly dependent on the coupling constants of the $\Lambda^0$ with the mesons of the QHD.

    \item I showed that for high values of $\chi_{\Lambda\Lambda\sigma}$, the nucleon mass vanishes, preventing the system from reaching a true maximum mass in the OV solution. I also showed how to solve this problem by adding a cutoff potential.

    \item I showed how changing the values of $\chi_{\Lambda\Lambda M}$ affected the particle population, EOS, $\Gamma$ and the OV solution. 

    \item Constraints related to the coupling constants of the hyperons were introduced. I also generalized the chemical equilibrium to take into account the whole baryon octet. With a maximum mass of only 1.82$M_\odot$, the result was disastrous.

    \item Sakurai's proposal was discussed, leading to another vector meson. The $\phi$ meson. A new maximum mass of 2.00$M_\odot$ was obtained. 

    \item By using symmetry group arguments, it was possible to write the coupling constants of the vector mesons with all the baryon octet with only three free parameters. Fixing $\theta_V$ and $z$ leads to only $\alpha_V$ as a free parameter. 
    Reducing $\alpha_V$ can produce hyperonic neutron stars with masses above 2.20$M_\odot$.
    
    \item Due to the $\omega$ dominance, I showed that the effects of the hyperon potential depth are largely secondary.

    \item I discuss the possibility of preventing the hyperon threshold in dense matter. Within our current knowledge, { the presence of hyperons in neutron-star cores appears to be inevitable.}

    \item The $\Delta$ resonances are introduced analogously to the hyperons. However, unlike hyperons, there are relevant arguments against the onset of $\Delta$ resonances in dense matter. Ref.~\cite{MOTTA2020} deserves special attention. { The possible existence of a quarkyonic phase also disfavors the $ \ Delta$'s more than the hyperons}.

    \item The coupling constants of the $\Delta$'s with the vector mesons are also fixed within SU(3) flavor symmetry. The values of $\chi_{\Delta\Delta\ M}$ can be fully determined by combining symmetry group arguments with Sakurai's proposal.

    \item As in the case of the NY matter, in the NYD matter, reducing $\alpha_V$ stiffens the EOS, suppresses the exotic degrees of freedom, and produces higher values of maximum masses. 

    \item The possible, but speculative presence of (anti)kaons was presented. The limitations of the theory were also discussed. Furthermore, the antikaon potential depth remains highly uncertain, and this uncertainty is large enough to induce not only quantitative differences but also qualitative changes in neutron star properties.

    \item Group theory arguments were also used to fix the coupling constant of the kaons with the vector mesons. Furthermore, with the help of the G-Parity, Sakurai's proposal can also be restored for the members of the pseudo-scalar meson octet.

    \item Due to the G-Parity, an unorthodox, although expected, behaviour appears. As we reduce the value of $\alpha_V$, we favor the presence of antikaons and soften the EOS. 
    For NYK and NYDK matter, the EOS still becomes stiffer as we move away from the SU(6) parametrization.

    \item An important consequence of the presence of kaons is that it strongly compromises the stiffening of the EOS by reducing the parameter $\alpha_V$. While for NY and NYD matter we can produce stars with masses above 2.20$M_\odot$, with kaon condensation, the maximum mass is limited to values around 2.10$M_\odot$.
    
\end{itemize}

The work here is done, but the history is not over yet. There are some aspects of neutron stars within the context of the QHD that were not covered here. Perhaps the most remarkable is the possible presence of deconfined quark matter in the neutron stars' core. Such a star, composed of a deconfined quark core surrounded by hadronic matter, is called a hybrid star~\cite{Acharya1999IJMPE,lopes2024PRC}. Recent studies have pointed that the presence of deconfined quark matter is not only possible but provable~\cite{nature_2020}. A more exotic scenario suggests that the ordinary matter as we know it, composed of protons and neutrons, is only metastable.
In this case, the true ground state of matter is the so-called strange matter, composed of deconfined quarks. A supernova explosion could trigger the deconfinement, and the observed pulsars would actually be strange stars, formed by deconfined $uds$ quarks~\cite{Olinto,lopesEPJC2025}. Moving into more speculative territory, it is also possible that the strong gravitational field of neutron stars captures dark matter. Dark matter-admixed neutron stars exhibit significant changes in the mass-radius relation~\cite{Dasfmode_2021,Grippa2025Universe,Lopes2025ApSc}.

{ There are also additional constraints not mentioned here. One of the more important is the so-called tidal deformation, and the dimensionless tidal parameter $\Lambda$, which is related to gravitational wave observations. The tidal deformability of a compact object is a single parameter that quantifies how easily the object is deformed when subjected to an external tidal field. A larger tidal deformability indicates that the object is easily deformable. On the opposite side, a compact object with a smaller tidal deformability parameter is smaller, more compact, and it is more difficult to deform it. Since its calculation relies on advanced tensor calculus, it is beyond the scope of the present work. The interested reader is referred to Refs.~\cite{Hinderer_2008,Chat2020,Flores2020,lopescesar} and the references therein. The key point here is that the GW170817 event detected by LIGO/VIRGO gravitational wave observatories has constrained the dimensionless tidal parameter of the canonical star in the range $70~<\Lambda_{1.4}<~580$~\cite{AbbottPRL}. The eL3$\omega\rho$ parametrization utilized in this work has $\Lambda_{1.4} = 516$, as shown in Ref.~\cite{lopesPRD}, and therefore is in agreement with this constraint.}

{ In the present work, the model-dependence was explored by varying the parameter $\alpha_V$ and verifying if a determined value fits or not a determined constraint. Modern approaches also apply Bayesian/multi-messenger inference, in which nuclear properties (around saturation, $\chi$EFT), NICER’s mass–radius measurements, and tidal deformabilities are jointly imposed to yield posteriors over the EOS/coupling parameters.
The interested reader is referred to Refs.~\cite{Raithel_2017APJ,Drischler2020PRL,Imam2025PRC} and the references therein. }

Other aspects not covered here include the effects of finite temperature~\cite{Sedrakian2023,Alexandre2004BJP,Tsiopelas2024EPJA}, rotation~\cite{Weber1992,Pattersons2021,Lopes2024ApJ}, and strong magnetic fields~\cite{Prakash_2000,Mallick2014PRC,Lopes2026Universe} on microscopic and macroscopic properties of neutron stars.

A detailed discussion of these cases is left for further studies, perhaps in a future Part III of this work.

\acknowledgments{ L.L.L.  was partially supported by CNPq (Brazil)
under Grant No 305347/2024-1. }

\newpage

%%%%%%%%%%%%%%%%%%%%%%%%%
\bibliography{aref}
%%%%%%%%%%%%%%%%%%%%%%%%%%%%%

\end{document}